\documentclass{lmcs} %
\pdfoutput=1

\usepackage{lastpage}
\lmcsdoi{22}{1}{1}
\lmcsheading{}{\pageref{LastPage}}{}{}%
{Jan.~27,~2025}{Jan.~01,~2026}{}


\usepackage{hyperref}

\usepackage[utf8]{inputenc}
\usepackage{xspace}
\usepackage{amsmath}
\usepackage{amsthm}
\usepackage{amsfonts}
\usepackage{amssymb}
\usepackage{mathtools}
\usepackage{textcomp}
\usepackage{stmaryrd}
\usepackage{varwidth}
\usepackage{graphicx}
\usepackage{mathrsfs}
\usepackage{cite}
\usepackage{subcaption}
\usepackage{tikz}
\usetikzlibrary{automata,graphs,positioning,chains,arrows,snakes,decorations.pathmorphing,shapes.geometric,matrix,positioning}
\pgfdeclarelayer{background}
\pgfsetlayers{background,main}
\usepackage{wrapfig}

\newcommand{\cD}{\mathcal{D}}
\newcommand{\cK}{\mathcal{K}}

\newcommand{\bbN}{\mathbb{N}}
\newcommand{\bbR}{\mathbb{R}}

\newcommand{\bbB}{\mathbb{B}}

\renewcommand{\operatorname}[1]{\mathsf{#1}}
\newcommand{\defineas}{\coloneqq \!\!\!\!\!}
\newcommand{\lang}{\ensuremath{\mathsf{MATLANG}}\xspace}
\newcommand{\sumLang}{\ensuremath{\mathsf{sum}\text{-}\mathsf{MATLANG}}\xspace}
\newcommand{\posnat}{\ensuremath{\mathbb{N}_{> 0}}}
\DeclareMathAlphabet{\mymathbb}{U}{BOONDOX-ds}{m}{n}
\DeclareMathOperator{\bigo}{\ensuremath{\mathcal{O}}}
\newcommand{\ksum}{\oplus}
\newcommand{\kprod}{\odot}
\newcommand{\bigksum}{\bigoplus}
\newcommand{\bigkprod}{\bigodot}
\newcommand{\kzero}{\mymathbb{0}}
\newcommand{\kone}{\mymathbb{1}}
\newcommand{\NN}{\mathbb{N}}
\newcommand{\RR}{\mathbb{R}}
\newcommand{\Mvar}{\mathcal{M}}
\newcommand{\Vvar}{\mathcal{V}}
\newcommand{\type}{\texttt{type}}
\DeclareMathOperator{\Sch}{\mathcal{S}}
\newcommand{\argSch}[1]{\ensuremath{\Sch\!\left(#1\right)}}
\newcommand{\I}{\mathcal{I}}
\newcommand{\E}{\mathcal{E}}

\newcommand{\mval}{\mu}

\newcommand{\ones}{\mathbf{1}}
\newcommand{\diag}{\textsf{diag}}
\newcommand{\mA}{\mathbf{A}}
\newcommand{\mB}{\mathbf{B}}
\newcommand{\mC}{\mathbf{C}}
\newcommand{\mV}{\mathbf{V}}
\newcommand{\mU}{\mathbf{U}}
\newcommand{\mQ}{\mathbf{Q}}
\newcommand{\mH}{\mathbf{H}}

\newcommand{\mx}{\mathbf{x}}
\newcommand{\my}{\mathbf{y}}
\newcommand{\mz}{\mathbf{z}}
\newcommand{\mv}{\mathbf{v}}
\newcommand{\mw}{\mathbf{w}}
\newcommand{\mau}{\mathbf{u}}
\newcommand{\mb}{\mathbf{b}}
\newcommand{\ms}{\mathbf{s}}
\newcommand{\mt}{\mathbf{t}}
\newcommand{\sem}[2]{\llbracket #1 \rrbracket(#2)}
\newcommand{\Iden}{\mathbf{I}}

\newcommand{\size}[1]{\ensuremath{\parallel\!#1\!\parallel}}
\newcommand{\card}[1]{\ensuremath{|#1|}}
\newcommand{\tl}{\ensuremath{\mathsf{TL}}}
\newcommand{\fo}{\ensuremath{\mathsf{FO}}\xspace}
\newcommand{\foplus}{\ensuremath{\mathsf{FO}^+}\xspace}
\newcommand{\ddom}{\mathbb{D}}

\newcommand{\seq}[1]{\ensuremath{\overline{#1}}}
\DeclareMathOperator{\arity}{\textit{ar}}
\DeclareMathOperator{\Voc}{\sigma}
\newcommand{\argVoc}[1]{\Voc(#1)}
\newcommand{\db}{\ensuremath{\textit{db}}}
\newcommand{\Vocrel}{\ensuremath{\Voc_{\text{rel}}}}
\newcommand{\dbrel}{\ensuremath{\db_{\text{rel}}}}
\DeclareMathOperator{\atoms}{\textit{at}}
\DeclareMathOperator{\free}{\textit{free}}

\newcommand{\val}{\ensuremath{\nu}}
\newcommand{\fssem}[2]{\llceil{#1}\rrfloor_{#2}}
\newcommand{\restr}[2]{\ensuremath{{#1}|_{#2}}}
\newcommand{\ssem}[2]{\llbracket #1 \rrbracket_{#2}}
\newcommand{\x}{\seq{x}}
\newcommand{\y}{\seq{y}}
\newcommand{\z}{\seq{z}}
\newcommand{\w}{\seq{w}}
\newcommand{\cq}{CQ\xspace}
\newcommand{\fcq}{fc-CQ\xspace}
\newcommand{\qcq}{qh-CQ\xspace}
\newcommand{\cqs}{CQs\xspace}
\newcommand{\fcqs}{fc-CQs\xspace}
\newcommand{\qcqs}{qh-CQs\xspace}
\newcommand{\cqNotIneq}{\ensuremath{\text{CQ}^{\not\leq}}\xspace}
\newcommand{\cqsNotIneq}{\ensuremath{\text{CQs}^{\not\leq}}\xspace}

\newcommand{\fcLang}{\ensuremath{\mathsf{fc}\text{-}\lang}\xspace}
\newcommand{\forLang}{\ensuremath{\mathsf{for}\text{-}\lang}\xspace}
\newcommand{\qhLang}{\ensuremath{\mathsf{qh}\text{-}\lang}\xspace}
\newcommand{\simpleLang}{\ensuremath{\mathsf{simple}\text{-}\lang}\xspace}

\newcommand{\conjLang}{\ensuremath{\mathsf{conj}\text{-}\lang}\xspace}

\newcommand{\e}{\overline{e}}
\DeclareMathOperator{\Rel}{\textit{Rel}}
\DeclareMathOperator{\Mat}{\textit{Mat}}

\DeclareMathOperator{\var}{\textit{var}}

\newcommand{\mR}{\mathbf{R}}
\newcommand{\ans}{\textit{ans}}
\newcommand{\mans}{\mathbf{ans}}
\newcommand{\focon}{\ensuremath{\mathsf{FO}^{\wedge}}\xspace}
\newcommand{\foTwo}{\ensuremath{\mathsf{FO}^{\wedge}_2}\xspace}
\newcommand{\foTwoEq}{\ensuremath{\mathsf{FO}^{\wedge}_{2,=}}\xspace}

\newcommand{\foTwoSimp}{\ensuremath{\mathsf{simple}\text{-}\mathsf{FO}^{\wedge}_2}\xspace}
\newcommand{\foTwoH}{\ensuremath{\mathsf{h}\text{-}\mathsf{FO}^{\wedge}_2}\xspace}
\newcommand{\foTwoEqH}{\ensuremath{\mathsf{h}\text{-}\foTwoEq}\xspace}

\newcommand{\ConstLin}{\ensuremath{\EnumClass{\size{\db}}{1}}}
\newcommand{\Qrel}{Q_{\text{rel}}}
\newcommand{\Qineq}{Q_{\text{ineq}}}
\newcommand{\Split}{\texttt{split}}
\newcommand{\dbQRel}{\db_{\text{rel}}^Q}
\newcommand{\dbRel}{\db_{\text{rel}}}

\newcommand{\AnsEnum}[2]{\texttt{Answer}(#1,#2)}
\newcommand{\ValEnum}[2]{\texttt{Vals}(#1,#2)}
\newcommand{\Eval}[2]{\texttt{Eval}(#1,#2)}
\newcommand{\DynEval}[2]{\texttt{DynEval}(#1,#2)}

\newcommand{\EnumClass}[2]{\texttt{Enum}(#1,#2)}
\newcommand{\DynClass}[3]{\texttt{DynEnum}(#1,#2,#3)}

\newcommand{\Bool}{\texttt{Bool}}

\newcommand{\Insertion}[1]{\texttt{insert}(#1)}
\newcommand{\Deletion}[1]{\texttt{delete}(#1)}
\newcommand{\MultisetInsert}[1]{\texttt{ins}(#1)}
\newcommand{\MultisetDelete}[1]{\texttt{del}(#1)}
\newcommand{\DSQuery}{\texttt{query}}
\newcommand{\Maintainable}{\DynClass{\size{\db}}{1}{1}}

\newcommand{\bools}{\ensuremath{\mathbb{B}}}
\newcommand{\btrue}{\ensuremath{\texttt{t}}}
\newcommand{\bfalse}{\ensuremath{\texttt{f}}}
\DeclareMathOperator{\rr}{\ensuremath{\mathsf{rr}}}

\newcommand{\multileft}{\ensuremath{\{\!\!\{}}
\newcommand{\multiright}{\ensuremath{\}\!\!\}}}
\newcommand{\mset}[1]{\multileft #1 \multiright}
\newcommand{\num}[1]{\# #1}

\DeclareMathOperator{\sig}{sig}

\begin{document}

% If the title is longer than 55 characters, then specify a shorter running title as the optional argument to \title. The running title should be roughyl at most 55 characters:
\title[Enumeration and Updates for Conjunctive Linear Algebra Queries]{Enumeration and Updates for Conjunctive Linear Algebra Queries Through Expressibility}
% \titlecomment{{\lsuper*}OPTIONAL comment concerning the title, \eg,
% if a variant or an extended abstract of the paper has appeared elsewhere.}
% \thanks{thanks, optional.}	%optional

% affiliations are numbered automatically with a, b, c (see below)
% use the optional argument to indicate the affiliation(s) of each author
% omit the argument if there is only one author, or only one affiliation
\author[T.~Mu\~noz]{Thomas Mu\~noz\lmcsorcid{0000-0003-0004-5041}}[a]
\author[C.~Riveros]{Cristian Riveros\lmcsorcid{0000-0003-0832-116X}}[b]
\author[S.~Vansummeren]{Stijn Vansummeren\lmcsorcid{0000-0001-7793-9049}}[a]

% affiliation 1 (automatically numbered a)
\address{UHasselt, Belgium}	%optional
% write emails for all authors having that affiliation
\email{thomas.munozserrano@uhasselt.be, stijn.vansummeren@uhasselt.be}  %optional

% affiliation 2 (automatically numbered b)
\address{PUC Chile, Chile}	%optional
\email{cristian.riveros@uc.cl}  %optional

%% etc.

%% required for running head on odd and even pages, use suitable
%% abbreviations in case of long titles and many authors:

%%%%%%%%%%%%%%%%%%%%%%%%%%%%%%%%%%%%%%%%%%%%%%%%%%%%%%%%%%%%%%%%%%%%%%%%%%%

%% the abstract has to PRECEDE the command \maketitle:
%% be sure not to issue the \maketitle command twice!

\begin{abstract}
\noindent Due to the importance of linear algebra and matrix operations in data analytics, there is significant interest in using relational query optimization and processing techniques for evaluating (sparse) linear algebra programs. In particular, in recent years close connections have been established between linear algebra programs and relational algebra that allow transferring optimization techniques of the latter to the former. In this paper, we ask ourselves which linear algebra programs in MATLANG correspond to the free-connex and q-hierarchical fragments of conjunctive first-order logic. Both fragments have desirable query processing properties: free-connex conjunctive queries support constant-delay enumeration after a linear-time preprocessing phase, and q-hierarchical conjunctive queries further allow constant-time updates. By characterizing the corresponding fragments of MATLANG, we hence identify the fragments of linear algebra programs that one can evaluate with constant-delay enumeration after linear-time preprocessing and with constant-time updates. To derive our results, we improve and generalize previous correspondences between MATLANG and relational algebra evaluated over semiring-annotated relations. In addition, we identify properties on semirings that allow to generalize the complexity bounds for free-connex and q-hierarchical conjunctive queries from Boolean annotations to general semirings.
\end{abstract}

\maketitle

%% start the paper here:
\section{Introduction}\label{section:intro}

Linear algebra forms the backbone of modern data analytics, as most machine learning algorithms are coded as sequences of matrix operations~\cite{DBLP:conf/osdi/AbadiBCCDDDGIIK16,DBLP:conf/cidr/ElgamalLBETRS17,DBLP:journals/pvldb/AndersonSSCZDSW17,DBLP:journals/sigmod/Elgohary0HRR17,DBLP:conf/sigmod/YangHZLJC17}. In practice, linear algebra programs operate over matrices with millions of entries. Therefore, efficient evaluation of linear algebra programs is a relevant challenge for data management systems which has attracted research attention with several proposals in the area~\cite{DBLP:conf/sigmod/HuangB013,DBLP:conf/micro/KanellopoulosVG19,DBLP:journals/pvldb/WangHSHL20,DBLP:journals/sigmod/JankovLYCZJG20,DBLP:journals/sigmod/LuoGGPJ18}.

To optimize and evaluate linear algebra programs, we must first agree on the language in which such programs are expressed. There has been a renewed interest in recent years for designing query languages for specifying linear algebra programs and for understanding their expressive power~\cite{DBLP:journals/tods/ShaikhhaEMEK20,DBLP:conf/icdt/BarceloH0S20,DBLP:journals/tods/BrijderGBW19,DBLP:conf/foiks/BrijderGB20,DBLP:conf/pods/GeertsMRV21}.
One such proposal is \lang~\cite{DBLP:journals/tods/BrijderGBW19}, a formal matrix query language that consists of only the basic linear algebra operations and whose extensions (e.g., \forLang) achieve the expressive power of most linear algebra operations~\cite{DBLP:conf/pods/GeertsMRV21}. Although \lang is a theoretical query language, it includes the core of any linear algebra program and, thus, the optimization and efficient evaluation of \lang could have a crucial impact on today's machine learning systems.

In this work, we study the efficient evaluation of \lang programs over sparse matrices whose entries are taken from a general semiring. We consider \lang evaluation in both the static and dynamic setting. For static evaluation, we want to identify the fragment that one can evaluate by preprocessing the input in linear time to build a data structure for enumerating the output entries with constant-delay. For dynamic evaluation, we assume that matrix entries are updated regularly and we want to maintain the output of a \lang query without recomputing it. For this dynamic setting, we aim to identify the \lang fragment that one can evaluate by taking linear time in the size of the update to refresh the aforementioned data structure so that it supports constant-delay enumeration of the modified output entries. These guarantees for both scenarios have become the holy grail for algorithmic query processing since, arguably, it is the best that one can achieve complexity-wise in terms of the input, the output, and the cost of an update~\cite{DBLP:journals/siglog/BerkholzGS20,DBLP:conf/csl/BaganDG07,DBLP:phd/hal/BraultBaron13,DBLP:conf/sigmod/IdrisUV17,DBLP:conf/pods/0002NOZ20,DBLP:conf/pods/SchweikardtSV18,DBLP:conf/pods/KazanaS13,DBLP:journals/tocl/DurandG07,DBLP:conf/pods/DurandSS14}.

To identify the \lang fragments with these guarantees, our approach is straightforward but effective. Instead of developing evaluation algorithms from scratch, we establish a direct correspondence between linear algebra and relational algebra to take advantage of the query evaluation results for conjunctive queries. Indeed, prior work has precisely characterized which subfragments of conjunctive queries can be evaluated and updated efficiently~\cite{DBLP:conf/csl/BaganDG07,DBLP:conf/pods/BerkholzKS17,DBLP:journals/siglog/BerkholzGS20,DBLP:conf/sigmod/IdrisUV17}. Our main strategy, then, is to link these conjunctive query fragments to corresponding linear algebra fragments. More specifically, our contributions are as follows.
\begin{enumerate}
    \item We start by understanding the deep connection between positive first-order logic (\foplus) over binary relations and \sumLang~\cite{DBLP:conf/pods/GeertsMRV21}, an extension of~\lang. We formalize this connection by introducing schema encodings, which specify how relations simulate matrices and vice-versa, forcing a lossless relationship between both. Using this machinery, we show that \sumLang and positive first-order logic are equally expressive over any relation, matrix, and matrix dimension (including non-rectangular matrices). Moreover, we show that conjunctive queries (CQ) coincide with \sumLang without matrix addition, which we call \conjLang. This result forms the basis for linking both settings and translating the algorithmic results from CQ to subfragments of \conjLang.
        
    \item We propose free-connex \lang (\fcLang) for static evaluation, which is a natural \lang subfragment that we show to be equally expressive as \emph{free-connex CQ}~\cite{DBLP:conf/csl/BaganDG07}, a subfragment of CQ that allows linear time preprocessing and constant-delay enumeration. To obtain our expressiveness result, we show that free-connex CQs over binary relations are equally expressive as the two-variable fragment of conjunctive \foplus, a logical characterization of this class that could be of independent interest.
    
    \item For the dynamic setting we introduce the language \qhLang, a \lang fragment that we show equally expressive to \emph{q-hierarchical CQ}~\cite{DBLP:conf/pods/BerkholzKS17,DBLP:conf/sigmod/IdrisUV17}, a fragment of CQ that allows constant update time and constant-delay enumeration.
    
    \item Both free-connex and q-hierarchical CQ are known to characterize the class of CQs that one can evaluate efficiently on Boolean databases. We are interested, however, in evaluating \lang queries on matrices featuring entries in a \emph{general semiring}. To obtain the complexity bounds for \fcLang and \qhLang on general semirings, therefore, we show that the upper and lower bounds for free-connex and q-hierarchical CQs generalize from Boolean annotations to classes of semirings which includes most semirings used in practice, like the reals. The tight expressiveness connections established in this paper then prove that for such semirings \fcLang and \qhLang can be evaluated with the same guarantees as their CQ counterparts and that they are optimal: one cannot evaluate any other \conjLang query outside this fragment under complexity-theoretic assumptions~\cite{DBLP:conf/pods/BerkholzKS17}.
\end{enumerate}

This article is further organized as follows. Section~\ref{sec:preliminaries} adresses all definitions. Section~\ref{sec:fo-matlang} describes how to go from a relational setting to a linear algebra setting and vice versa. In Section~\ref{sec:sumlang-and-fo} we show \sumLang and positive first-order logic are equally expressive. Section~\ref{sec:fc-cq} defines free-connex CQs and characterizes binary free-connex CQs with the conjunctive two variable fragment of positive first order logic. Afterwards, \fcLang is defined in Section~\ref{sec:free-connex-matlang}. The definition of q-hierarchical CQs is located in Section~\ref{sec:qh-cq} and in that matter binary q-hierarchical CQs are shown to be equally expressive as hierarchical queries of the conjunctive two variable fragment of positive first order logic. Subsequently in Section~\ref{sec:q-hierarchical-matlang}, \qhLang is defined. Later, Section~\ref{sec:enum} defines the enumeration problem and states known upper and lower bounds on the Boolean semiring for free-connex CQs. In Section~\ref{sec:evaluation} the upper and lower bounds are extended to general semirings for free-connex CQs which use inequality atoms. Section~\ref{sec:evaluation-qh} defines the dynamic enumeration problem and known upper and lower bounds on the Boolean semiring for q-hierarchical CQs; and illustrates how to extend these bounds to general semirings and for q-hierarchical CQs which use inequality atoms. We end by presenting some conclusions and future work in Section~\ref{sec:conclusions}.

\subsection{New material.}
Our results have previously been published in the 27th International Conference on Database Theory, {ICDT} 2024. This article is an extended and complete version containing full proofs of all formal statements omitted in the conference version. Moreover, several notions and proofs were revisited to improve the presentation and shorten the article. Specifically, we introduce a \emph{prenex normal form} for $\conjLang$ in Section~\ref{sec:sumlang-and-fo}, which resulted in being especially useful for several proofs, in particular, for the proof that
\sumLang and positive first-order logic are equally expressive.
Furthermore, these changes in the main proof required a new presentation of other results, like the lower bound statements of \fcLang and \qhLang in Section~\ref{sec:evaluation} and Section~\ref{sec:evaluation-qh}, respectively.

\subsection{Related work.}
In addition to the work that has already been cited above, the following work is relevant. Brijder et al.~\cite{DBLP:conf/foiks/BrijderGB20} have shown equivalence between \lang and $\foplus_3$, the 3-variable fragment of positive first order logic. By contrast, we show equivalence between \sumLang and $\foplus$, and study the relationship between the free-connex and q-hierarchical fragments of $\lang$ and $\fo^{\wedge}$, the conjunctive fragment of positive first order logic.

Geerts et al.~\cite{DBLP:conf/pods/GeertsMRV21} previously established a correspondence between \sumLang and \foplus. However, as we illustrate in Section~\ref{subsec:new-correspondence}  their correspondence is (1) restricted to square matrices, (2) asymmetric between the two settings, and (3) encodes matrix instances as databases of more than linear size, making it unsuitable to derive the complexity bounds.

Eldar et al.~\cite{DBLP:conf/icdt/EldarCK24} have recently also generalized complexity bounds for free-connex \cqs from Boolean annotations to general semirings. Nevertheless, this generalization is with respect to \emph{direct access}, not enumeration. In their work the focus is to compute aggregate queries, which is achieved by providing direct access to the answers of a query even if the annotated value (aggregation result) is zero. By contrast, in our setting, zero-annotated values must not be reported during the enumeration of query answers. This distinction in the treatment of zero leads to a substantial difference in the properties that a semiring must have in order to generalize the existing complexity bounds.

There are deep connections known between the treewidth and the number of variables of a conjunctive $\foplus$ formula ($\focon$). For example, Kolaitis and Vardi established the equivalence of boolean queries in $\focon_k$, the $k$-variable fragment of $\focon$, and boolean queries in $\focon$ of treewidth less than $k$. Because they focus on boolean queries (i.e., without free variables), this result does not imply our result that for binary queries free-connex $\focon$ equals $\focon_2$.  Similarly, Geerts and Reutter~\cite{DBLP:conf/iclr/GeertsR22} introduce a tensor logic \tl\xspace over binary relations and show that conjunctive expressions in this language that have treewidth $k$ can be expressed in $\tl_{k+1}$, the $k$-variable fragment of $\tl$. While they do take free variables into account, we show in Appendix~\ref{subsec:finite-var-logics} that there are free-connex conjunctive queries with 2 free variables with treewidth 2 in their formalism---for which their result hence only implies expressibility in $\focon_3$, not $\focon_2$ as we show here.

Several proposals~\cite{DBLP:conf/sigmod/HuangB013,DBLP:conf/micro/KanellopoulosVG19,DBLP:journals/pvldb/WangHSHL20,DBLP:journals/sigmod/JankovLYCZJG20,DBLP:journals/sigmod/LuoGGPJ18} have been made regarding the efficient evaluation of linear algebra programs in the last few years. All these works focused on query optimization without formal guarantees regarding the preprocessing, updates, or enumeration in query evaluation. To the best of our knowledge, this is the first work on finding subfragments of a linear algebra query language (i.e., \lang) with such efficiency guarantees.

\section{Preliminaries}\label{sec:preliminaries}

In this section we recall the main definitions of \lang, a query language on matrices, and first order logic (FO), a query language on relations.

\paragraph{Semirings.}
We evaluate both languages over arbitrary commutative and non-trivial semirings.
A (commutative and non-trivial) \textit{semiring}
$(K, \ksum, \kprod, \kzero, \kone)$ is an algebraic structure where $K$ is a
non-empty set, $\ksum$ and $\kprod$ are binary operations over $K$, and
$\kzero, \kone \in K$ with $\kzero \not = \kone$.
Furthermore, $\ksum$ and $\kprod$ are associative
operations, $\kzero$ and $\kone$ are the identities of $\ksum$ and $\kprod$,
respectively, $\ksum$ and $\kprod$ are commutative operations, $\kprod$
distributes over $\ksum$, and $\kzero$ annihilates $K$ (i.e.
$\kzero \kprod k = k \kprod \kzero = \kzero$).
We use
$\bigksum_L$ and $\bigkprod_L$ to denote the $\ksum$ and $\kprod$ operation over
all elements in $L\subseteq K$, respectively.  Typical examples of semirings are
the reals $(\RR, +, \times, 0,1)$, the natural numbers $(\NN, +, \times, 0,1)$,
and the boolean semiring $\bools = (\{\btrue,\bfalse\}, \vee, \wedge, \bfalse, \btrue)$.

Henceforth, when we say ``semiring'' we mean ``commutative and non-trivial''
semiring. We fix such an arbitrary semiring $K$ throughout the document.  We denote by $\posnat$ the set of non-zero natural numbers.

\paragraph{Matrices and size symbols.}
A $K$\textit{-matrix} (or just matrix) of
dimension $m\times n$ is a $m \times n$ matrix with elements in $K$ as its
entries. 
We write $\mA_{ij}$ to denote the $(i,j)$-entry of~$\mA$. Matrices of
dimension $m\times 1$ are \emph{column vectors} and those of
dimension $1\times n$ are \emph{row vectors}. We also refer to matrices of dimension $1 \times 1$ as \emph{scalars}.

We assume a collection of \emph{size symbols} denoted with greek letters
$\alpha, \beta, \dots$ and assume that the natural number $1$ is a valid size
symbol.  A \emph{type} is a pair $(\alpha,\beta)$ of size symbols. Intuitively,
types represent sets of matrix dimensions.  In particular, we obtain dimensions
from types by replacing size symbols by elements from $\posnat$, where the size
symbol $1$ is always replaced by the natural number $1$.  So, $(\alpha,\beta)$ with $\alpha \not = 1 \not = \beta$
represents the set of dimensions $\{ (m,n) \mid m,n \in \posnat \}$, while
$(\alpha,\alpha)$ represents the dimensions $\{ (m,m) \mid m \in \posnat \}$ of
square matrices; $(\alpha,1)$ represents the dimensions
$\{ (m,1) \mid m \in \posnat\}$ of column vectors; and $(1,1)$ represents the dimension $(1,1)$ of scalars.

\paragraph{Schemas and instances.}
We assume a set $\Mvar = \{\mA,\mB,\mC,\mV, \ldots\}$
of \emph{matrix symbols}, disjoint with the size symbols and denoted by bold uppercase letters. Each matrix symbol $\mA$ has a fixed associated type.
We write $\mA\colon (\alpha,\beta)$ to denote that $\mA$ has type $(\alpha,\beta)$. 

A matrix \emph{schema} $\Sch$ is a finite set  of matrix and size symbols.
We require that the special size symbol $1$ is always in $\Sch$, and that all size symbols occurring in the type of any matrix symbol $\mA \in \Sch$ are also in $\Sch$.
A matrix \emph{instance} $\I$ over a matrix schema $\Sch$ is a function that maps each size symbol $\alpha$ in $\Sch$ to a non-zero natural number $\alpha^\I \in \posnat$, and maps each matrix symbol $\mA\colon (\alpha,\beta)$ in $\Sch$ to a $K$-matrix $\mA^\I$ of dimension $\alpha^\I \times \beta^\I$.
We assume that for the size symbol $1$, we have $1^\I = 1$, for every instance
$\I$. 

\paragraph{Sum-Matlang.} Let $\Sch$ be a matrix schema. 
Before defining the syntax of \sumLang, we assume a set $\Vvar = \{\mau,\mv, \mw, \mx, \dots\}$ of \emph{vector variables} over $\Sch$, which is disjoint with matrix and size symbols in $\Sch$.
Each such variable $\mv$ has a fixed associated type,
which must be a vector type $(\gamma, 1)$ for some size symbol $\gamma \in \Sch$. We also write $\mv\colon (\gamma, 1)$ in that case. 

The syntax of $\sumLang$
expressions~\cite{DBLP:conf/pods/GeertsMRV21} over $\Sch$ is defined by the grammar:
\begin{center}
    \begin{tabular}{lcll@{\qquad}cll}
        $e$ & $\Coloneqq  $ & $\mA\in\Sch$ & (matrix symbol)
        & $|$  & $\mv\in\Vvar$ & (vector variable)\\
        & $|$ & $e^T$ & (transpose)
        & $|$ & $e_1 \cdot e_2$ & (matrix multiplication)\\   
        & $|$ & $e_1 + e_2$ & (matrix addition)
        & $|$ & $e_1\times e_2$ & (scalar multiplication)\\
        & $|$ & $e_1 \kprod e_2$ & (pointwise multiplication) 
        & $|$ & $\Sigma \mv. e$ & (sum-iteration).   
    \end{tabular}
\end{center}
In addition, we require that expressions $e$ are \emph{well-typed}, in the sense that their \emph{type} $\type(e)$ is correctly defined as follows:
\allowdisplaybreaks
\[
\renewcommand{\arraystretch}{1.3}
\begin{array}{rl}
    \type(\mA) \ \defineas \  & (\alpha,\beta) \ \text{ for a matrix symbol } \mA\colon (\alpha,\beta) \\
    \type(\mv) \ \defineas \  & (\gamma,1) \ \text{ for vector variable } \mv\colon (\gamma,1) \\
    \type(e^T) \ \defineas \  & (\beta,\alpha) \ \text{ if  } \type(e)=(\alpha,\beta) \\
    \type(e_1 \cdot e_2) \ \defineas \  & (\alpha,\gamma) \ \text{ if } \type(e_1)=(\alpha,\beta) \text{ and } \type(e_2)=(\beta,\gamma) \\
    \type(e_1 + e_2) \ \defineas \  & (\alpha,\beta) \ \text{ if } \type(e_1)=\type(e_2)=(\alpha,\beta) \\
    \type(e_1\times e_2) \ \defineas \  & (\alpha,\beta) \ \text{ if } 
        \type(e_1)=(1,1) \text{ and }  \type(e_2)=(\alpha,\beta) \\ 
        \type(e_1 \kprod e_2) \ \defineas \  & (\alpha,\beta) \ \text{ if } 
        \type(e_1) = \type(e_2) = (\alpha,\beta) \\
        \type(\Sigma \mv. e) \ \defineas \  & (\alpha,\beta) \ \text{ if } 
        e\colon (\alpha,\beta) \text{ and } \type(\mv) = (\gamma,1).
\end{array}
\]
In what follows, we always consider well-typed expressions and write $e\colon (\alpha,\beta)$ to denote that $e$ is well typed, and its type is~$(\alpha,\beta)$. 

For an expression $e$, we say that a vector variable $\mv$ is \emph{bound} if it is under a sum-iteration $\Sigma \mv$, and \emph{free} otherwise. 
We say that an expression $e$ is a \emph{sentence} if $e$ does not have free vector variables. 
To evaluate expressions that are not sentences, we require the following notion of a \emph{valuation}.
Fix a matrix instance $\I$ over $\Sch$. A \emph{vector valuation} over $\I$ is a
function $\mval$ that maps each vector symbol $\mv\colon (\gamma,1)$ to a column
vector of dimension $\gamma^\I \times 1$. Further, if $\mb$ is a vector of dimension
$\gamma^\I \times 1$, then let $\mval[\mv:= \mb]$ denote the \emph{extended}
vector valuation over $\I$ that coincides with $\mval$, except that
$\mv\colon (\gamma,1)$ is mapped to  $\mb$.

Let $e\colon(\alpha,\beta)$ be a $\sumLang$ expression over $\Sch$. 
When one evaluates $e$ over a matrix instance $\I$ and a matrix valuation $\mval$ over~$\I$, it produces a matrix $\sem{e}{\I, \mval}$ of dimension $\alpha^\I \times \beta^\I$ such that each entry $i,j$ satisfies:
\[
\renewcommand{\arraystretch}{1.3}
\begin{array}{rlrl}
    \sem{\mA}{\I, \mval}_{ij} \ \defineas \   & \mA^\I_{ij} \, \text{ for } \mA\in\Sch \\
    \sem{\mv}{\I, \mval}_{ij} \ \defineas \  &  \mval(\mv)_{ij} \, \text{ for } \mv\in\Vvar \\
    \sem{e^T}{\I,\mval}_{ij} \ \defineas \  &   \sem{e}{\I,\mval}_{ji} \\
    \sem{e_1\cdot e_2}{\I,\mval}_{ij} \ \defineas \  & \multicolumn{3}{l}{\bigoplus_{k} \sem{e_1}{\I,\mval}_{ik} \kprod \sem{e_2}{\I,\mval}_{kj}} \\ 
    \sem{e_1+e_2}{\I,\mval}_{ij} \ \defineas \  & \sem{e_1}{\I,\mval}_{ij} \ksum \sem{e_2}{\I,\mval}_{ij}  \\
    \sem{e_1\times e_2}{\I,\mval}_{ij} \ \defineas \  & \multicolumn{3}{l}{a\kprod \sem{e_2}{\I,\mval}_{ij} \text{ with } \sem{e_1}{\I,\mval}=[a]} \\
    \sem{e_1 \kprod e_2}{\I,\mval}_{ij} \ \defineas \  & \sem{e_1}{\I,\mval}_{ij} \kprod \sem{e_2}{\I,\mval}_{ij} \\
    \sem{ \Sigma \mv. \, e\,}{\I,\mval}_{ij}\ \defineas \  & \bigksum_{k=1}^{\gamma^\I} \sem{e}{\I, \mval[\mv \coloneqq \mb^{\gamma^\I}_k]}_{ij}   
\end{array}
\]
In the last line, it is assumed that $\mv\colon (\gamma,1)$, and  $\mb^{n}_1, \mb^{n}_2, \ldots, \mb^{n}_n$ denote the $n$-dimensional canonical vectors, namely, the vectors $[\kone \, \kzero \, \ldots \, \kzero]^T$, $[\kzero \, \kone \, \ldots \, \kzero]^T$\!, $\ldots$, $[\kzero \,\, \kzero \, \ldots \, \kone]^T$, respectively.

Note that if $e$ is a sentence, then $\llbracket e \rrbracket$ is independent of the vector valuation $\mval$ in the sense that $\sem{e}{\I, \mval} = \sem{e}{\I, \mval'}$ for all vector valuations $\mval$ and $\mval'$. We therefore sometimes also simply write  $\sem{e}{\I}$ instead of $\sem{e}{\I, \mval}$ when $e$ is a sentence.

\begin{exa}\label{example:sum-iteration-1}
    Let $\Sch=\{ \mA \}$ where $\mA\colon (\alpha,\alpha)$. Let $\I$ be an instance over $\Sch$ such that $\alpha^\I=3$ and
    $
    \mA^{\I}=\begin{bmatrix}
        a_{11} & a_{12} & a_{13}\\
        a_{21} & a_{22} & a_{23} \\
        a_{31} & a_{32} & a_{33}
    \end{bmatrix}.
    $
    Let $\mv\in\Vvar$ where $\mv\colon (\alpha, 1)$. The expression $\Sigma \mv. \mA\cdot \mv$ is well-typed and
    \[ \sem{\Sigma \mv. \mA\cdot \mv}{\I,\emptyset} =
        \mA\cdot \mb^3_1 +
        \mA\cdot \mb^3_2 +
        \mA\cdot \mb^3_3 =
        \begin{bmatrix}
            a_{11} \\
            a_{21} \\
            a_{31} 
        \end{bmatrix} +
        \begin{bmatrix}
            a_{12} \\
            a_{22} \\
            a_{32} 
        \end{bmatrix} +
        \begin{bmatrix}
            a_{13} \\
            a_{23} \\
            a_{33} 
        \end{bmatrix}.
    \]
\end{exa}
\begin{exa}\label{example:sum-iteration-2}
    Let $\Sch$ and $\I$ be as in Example~\ref{example:sum-iteration-1}. Let $\mv\in\Vvar$ where $\mv\colon (\gamma, 1)$.
    The expression $\Sigma \mv. \mA$ is well-typed and
    \[ \sem{\Sigma \mv. \mA}{\I,\emptyset} = \underbrace{\mA + \cdots + \mA}_{\gamma^\I \text{ times}}. \]
\end{exa}

\paragraph{Fragments of Sum-Matlang.} We recall here two fragments of $\sumLang$ that will be important for the paper. First, we define $\conjLang$ as the \sumLang fragment that includes all operations except matrix addition ($+$).
Second, we define $\lang$ that is a fragment of $\sumLang$.
Specifically, define the syntax of $\lang$ expressions over $\Sch$ by the
following grammar:
\begin{center}
    \begin{tabular}{lcll}
        $e$ & $\Coloneqq$ & $\mA\in \Sch \, \mid \, e^T \, \mid \, e_1 \cdot e_2 \, \mid \, e_1 + e_2 \, \mid \, e_1\times e_2 \, \mid \, e_1 \kprod e_2 \, \mid \, \ones^\alpha  \, \mid \, \Iden^\alpha$
    \end{tabular}
\end{center}
for every size symbol $\alpha \in \Sch$. 
Here, $\ones^\alpha$ and $\Iden^\alpha$  denote the \emph{ones-vector} and \emph{identity-matrix}, respectively, of type 
$\type(\ones^\alpha) = (\alpha, 1)$ and $\type(\Iden^\alpha) = (\alpha, \alpha)$. Their semantics can be defined by using $\sumLang$~as: 
\[
\begin{array}{rlrl}
\sem{\ones^\alpha}{\I,\mval} \defineas & \sem{\Sigma \mv. \mv}{\I,\mval}, \ \ \ \ \ \ \ \ \ &
\sem{\Iden^\alpha}{\I,\mval} \defineas &  \sem{\Sigma \mv. \mv\cdot\mv^T}{\I,\mval}   .
\end{array}
\]
Note that the original $\lang$ language as introduced in~\cite{DBLP:journals/tods/BrijderGBW19} included an operator for  \emph{vector diagonalization}. This operator can be simulated using the $\Iden^\alpha$-operator, and conversely $\Iden^\alpha$ can be simulated using vector diagonalization.
Furthermore, we have included the pointwise multiplication $\kprod$ in \lang, also known as the \emph{Hadamard product}. This
operation will be essential for our characterization
results. In~\cite{DBLP:journals/tods/BrijderGBW19,DBLP:conf/pods/GeertsMRV21}, the syntax of $\lang$ was more generally parameterized by a family of $n$-ary
functions that could be pointwise applied. Similarly to
\cite{DBLP:conf/foiks/BrijderGB20,DBLP:conf/icdt/Geerts19} we do not include such functions
here, but  leave their detailed study to future
work.

\paragraph{Sum-Matlang queries.}
A $\sumLang$ \emph{query} $\mQ$ over a matrix schema $\Sch$ is an expression of the form $\mH \coloneqq e$
where $e$ is a well-typed $\sumLang$ sentence, $\mH$~is a ``fresh'' matrix symbol that does not occur in $\Sch$, and $\type(\mH) = \type(e)$.
When evaluated on a matrix instance $\I$ over schema $\Sch$,
$\mQ$ returns a matrix instance $\E$ over the extended schema $\Sch \cup\{\mH\}$: $\E$ coincides with $\I$ for every matrix and size symbol in $\Sch$ and additionally maps
$\mH^\E = \sem{e}{\I, \emptyset}$ with $\emptyset$ denoting the empty vector valuation.
We denote the instance resulting from evaluating $\mQ$ by $\sem{\mQ}{\I}$. If $\Sch$ is a matrix schema and $\mQ$ a $\sumLang$ query over $\Sch$ then we use $\argSch{\mQ}$ to denote the extended schema $\Sch \cup \{\mH\}$.

The former notion of query is extended to any fragment of \sumLang in a natural manner, e.g., a \conjLang query is a \sumLang query $\mH \coloneqq e$ where $e$ is a well-typed \conjLang expression.

\paragraph{$K$-relations.}
A \emph{$K$-relation}
over a domain of data values $\ddom$ is a function $f\colon \ddom^a \to K$
such that $f(\seq{d}) \not = \kzero$ for finitely many $\seq{d} \in
\ddom^a$. Here, `$a$' is the \emph{arity} of $R$.
Since we want to compare relational queries with \sumLang queries, we will restrict our attention in what follows to $K$-relations where
the domain $\ddom$ of data values is the set $\posnat$. In this context, we may naturally view a $K$-matrix of
dimensions $n \times m$ as a $K$-relation such that the entry $(i,j)$ of the matrix is
encoded by the $K$-value of the tuple $(i,j)$ in the relation (see also
Section~\ref{sec:fo-matlang}).

\paragraph{Vocabularies and databases.}
We assume an infinite set of \emph{relation symbols} together with an infinite and disjoint set of \emph{constant symbols}. 
Every relation symbol $R$ is associated with a 
number, its \emph{arity}, which we denote by $\arity(R) \in \NN$.
A \emph{vocabulary} $\Voc$ is a finite set of relation and constant symbols. 
A \emph{database} over $\Voc$ is a function $\db$ that maps every constant symbol $c \in \Voc$ to a value $c^\db$ in $\posnat$; and every relation symbol $R \in \Voc$ to a $K$-relation $R^\db$ of arity~$\arity(R)$.

\paragraph{Positive first order logic.} As our relational query language, we will work
with the positive fragment of first order logic (\foplus). In contrast to the
standard setting in database theory, where the only atomic formulas are
relational atoms of the form $R(\seq{x})$, we also allow the ability to compare
variables with constant symbols. To this end, the following definitions are in
order. We assume an infinite set of variables, which we usually denote by
$x,y,z$. We denote tuples of variables by $\seq{x}$, $\seq{y}$, and so
on. A \emph{relational atom} is an expression of the form $R(x_1,\dots, x_k)$ with
$R$ a relation symbol of arity~$k$. A
\emph{comparison atom} is 
of the form $x \leq c$ with $x$ a
variable and $c$ a constant symbol. An \emph{equality atom} is of the form $x = y$ with $x,y$ variables.
A \emph{positive first order logic formula} (\foplus formula) over a vocabulary $\Voc$ is an expression
generated by the following grammar:
\begin{center}
    \begin{tabular}{lcll}
        $\varphi$ & $\Coloneqq$ & $R(\seq{x}) \, \mid \, x \leq c \, \mid \, x = y \, \mid \, \exists y.\,\varphi \, \mid \, \varphi \wedge \varphi \, \mid \, \varphi \vee \varphi$
    \end{tabular}
\end{center}
where $R$ and $c$ range over relations and constants in $\Voc$, respectively. The notions of \emph{free and bound variables} are defined as usual in~FO.  We denote the set of free variables of $\varphi$ by $\free(\varphi)$ and the  multiset of atoms occurring in $\varphi$ by~$\atoms(\varphi)$.

We evaluate formulas over $K$-relations as follows using the well-known semiring semantics~\cite{DBLP:conf/pods/GreenKT07}.
A \emph{valuation} over a set of variables $X$ is a function $\val\colon X \to \posnat$ that assigns a
value in $\posnat$ to each variable in $X$ (recall that $\ddom =\posnat$).  
We denote by $\val\colon X$ that $\val$ is a valuation on $X$ and by $\restr{\val}{Y}$ the
restriction of $\val$ to $X \cap Y$.
As usual, valuations are extended point-wise to tuples of variables, i.e.,
$\val(x_1,\dots, x_n) = \left(\val(x_1),\dots, \val(x_n)\right)$.
Let $\varphi$ be an \foplus formula over vocabulary $\Voc$. When evaluated on a database $\db$ over vocabulary $\Voc$, it defines a mapping $\fssem{\varphi}{\db}$ from valuations over $\free(\varphi)$ to $K$ inductively defined as:
\[
\renewcommand{\arraystretch}{1.3}
\begin{array}{rlrl}
\fssem{R(\seq{x})}{\db}(\val) \ \defineas \  & R^\db(\val(\seq{x})) \\
\fssem{x \leq c}{\db}(\val) \ \defineas \  & 
\begin{cases}
    \kone & \text{if } \val(x) \leq c^\db \\
    \kzero & \text{otherwise}
\end{cases} \\
\fssem{x=y}{\db}(\val)\ \defineas \  & 
\begin{cases}
    \kone &\text{ if } \val(x)=\val(y) \\
    \kzero &\text{ otherwise }
\end{cases} \\
\fssem{\varphi_1 \wedge \varphi_2}{\db}(\val) \ \defineas \  & \fssem{\varphi_1}{\db}(\restr{\val}{\free(\varphi_1)}) \kprod \fssem{\varphi_2}{\db}(\restr{\val}{\free(\varphi_2)}) \\
\fssem{\varphi_1 \vee \varphi_2}{\db}(\val) \ \defineas \  & \fssem{\varphi_1}{\db}(\restr{\val}{\free(\varphi_1)}) \ksum \fssem{\varphi_2}{\db}(\restr{\val}{\free(\varphi_2)}) \\
\fssem{\exists y. \varphi}{\db}(\val) \ \defineas \  &  \bigksum_{\mu\colon \free(\varphi) \text{s.t.} \restr{\mu}{\free(\varphi)\setminus \{y\}} = \val} \fssem{\varphi}{\db}(\mu).
\end{array}
\]
The \emph{support} of the mapping $\fssem{\varphi}{\db}$ is the set of
valuations that it maps to a nonzero annotation.  As the reader may notice, it
is possible that $\fssem{\varphi}{\db}$ has infinite support. This situation is
remeniscent of classical \fo (evaluated on the Boolean semiring) which may also
define infinite relations. To circumvent this problem, we restrict ourselves to
\emph{safe} formulas~\cite{DBLP:books/aw/AbiteboulHV95}, which in classical FO
ensures that output relations are finite. To define safety, we first define the set $\rr(\varphi) \subseteq \free(\varphi)$ of \emph{range restricted} variables of a formula:
\begin{align*}
\rr(R(x_1, \ldots, x_n))& =\{x_1, \ldots, x_n\} & \rr(x\leq c)&=\{x\} & \rr(x=y)&=\emptyset \\
\rr(\varphi_1\wedge\varphi_2) &=\rr(\varphi_1)\cup\rr(\varphi_2) &
\rr(\varphi_1\vee \varphi_2)&=\rr(\varphi_1)\cap\rr(\varphi_2) &
\rr(\exists y.\varphi)&=\rr(\varphi)\setminus \{y\}.
\end{align*}

  An \foplus formula $\varphi$ is \emph{safe}
if it satisfies the following conditions:
\begin{enumerate}
\item every occurrence of an equality atom $x=y$ is a part of a subformula of
  $\varphi$ of the form $\psi \wedge (x=y)$ or $(x=y) \wedge \psi$ where at
  least one of $\{x,y\}$ is in $\rr(\psi)$;
\item $\varphi_1\vee \varphi_2$ is only used when $\varphi_1$ and $\varphi_2$
  have the same set of free variables.
\end{enumerate}

\begin{prop}\label{prop:foplus-safeness}
    Every safe \foplus formula has finite support.
\end{prop}
The proof is remeniscent of that of classical \fo; we therefore  delegate it to Appendix~\ref{sec:app-preliminaries}. For the rest of this paper, we restrict to safe formulas. Hence when we say ``formula'' we mean ``safe formula'', and similarly for queries.

\paragraph{FO queries.}
An $\foplus$ \emph{query} $Q$ over vocabulary $\Voc$ is an expression of the form $H(\x)\gets\varphi$ where
$\varphi$ is an \foplus formula over $\Voc$,
$\seq{x} = (x_1,\dots,x_k)$ is a sequence of (not necessarily distinct) free
variables of $\varphi$, such that every free variable of $\varphi$
occurs in $\seq{x}$, and $H$ is a ``fresh'' relation
symbol not in $\Voc$ with $\arity(H) = k$.
The formula $\varphi$ is called the \emph{body} of $Q$, and $H(\seq{x})$ its~\emph{head}.

When evaluated over a database $\db$ over $\Voc$,  $Q$ returns a database
$\ssem{Q}{\db}$ over the extended vocabulary $\Voc \cup\{H\}$.
This database $\ssem{Q}{\db}$ coincides with $\db$ for every relation and constant symbol in $\Voc$,
and maps the relation symbol $H$ to the $K$-relation of arity $k$ defined as follows. 
For a sequence of domain values $\seq{d} = (d_1,\dots, d_k)$, we write $\seq{d} \models \seq{x}$ if,
for all $i \not = j$ with $x_i = x_j$ we also have $d_i = d_j$.  Clearly, if
$\seq{d} \models \seq{x}$ then the mapping
$\{ x_1 \to d_1,\dots, x_k \mapsto d_k\}$ is well-defined. Denote this mapping
by $\seq{x} \mapsto \seq{d}$ in this case. Then
\[
\ssem{Q}{\db}(H) \coloneqq  \seq{d} \mapsto   
\begin{cases}
    \fssem{\varphi}{\db}(\seq{x} \mapsto \seq{d})  & \text{if } \seq{d} \models \seq{x} \\
    \kzero  & \text{otherwise}
\end{cases}
\]
In what follows, if $Q$ is a query, then  we will often use the notation
$Q(\seq{x})$ to denote that the sequence of the variables in the head of $Q$ is
$\seq{x}$. If $Q$ is a query over $\Voc$ and $H(\x)$ its head, then we write $\argVoc{Q}$ for the extended vocabulary $\Voc \cup \{H\}$.

\paragraph{Formula and query equivalence}
Two \foplus formulas $\varphi_1$ and $\varphi_2$ are said to be
\emph{equivalent}, denoted $\varphi_1 \equiv \varphi_2$ if
$\fssem{\varphi_1}{\db} = \fssem{\varphi_2}{\db}$ for every database.  Note that
because $\fssem{\varphi_1}{\db}$ is a set of mappings, all with domain
$\free(\varphi_1)$, and because the same holds for $\fssem{\varphi_2}{\db}$, we
neccesarily have $\free(\varphi_1) = \free(\varphi_2)$ when
$\varphi_1 \equiv \varphi_2$. Similarly, two \foplus queries $Q_1$ and $Q_2$ are equivalent, denoted $Q_1 \equiv Q_2$ if $\ssem{Q_1}{\db} = \ssem{Q_2}{\db}$ for every database. This implies that they have the same head but the variables occurring in the heads need not be the same.

Note that because we have fixed $\mathcal{K}$ arbitrarily, formula and query equivalence cannot use the concrete interpretation of the operators in  $\mathcal{K}$ operators, and must hence hold for every choice of $\mathcal{K}$.

\paragraph{Fragments of \foplus}
We denote by $\focon$ the fragment of $\foplus$ formulas in which disjunction is disallowed. A query $Q = H(\x) \gets \varphi$ is an $\focon$ query if $\varphi$ is in $\focon$. If  additionally $\varphi$ is in prenex normal form, i.e., 
$Q: H(\seq{x}) \gets \exists \seq{y}. a_1 \wedge \dots \wedge a_n$
where $a_1,\dots, a_n$ are relational,  comparison, or equality atoms, 
then $Q$ is a \emph{conjunctive query} (CQ). Note that, while the classes of conjunctive queries and $\focon$ queries are equally expressive, for our purposes conjunctive queries are hence formally a syntactic fragment of $\focon$ queries.

An $\foplus$ formula is binary if every relational atom occurring in it has arity at most two and it has at most two free variables. Similarly, an $\foplus$
query is \emph{binary} if every relational atom occurring in it (body and head)
has arity at most two. Because in \sumLang both the input and output are matrices,
our correspondences between \sumLang and \foplus will focus on binary queries.

\paragraph{Discussion.}
We have added comparison atoms to \foplus in order to establish its correspondence with \sumLang. To illustrate why we will need comparison atoms, consider the \sumLang expression $\Iden^\alpha$ of type $(\alpha,\alpha)$ that computes
the identity matrix. This can be expressed by means of the following~\cq $Q: I(x,x) \gets x \leq \alpha$. We hence use
comparison atoms to align the dimension of the matrices with the domain size of
relations.

To make the correspondence hold, we note that in \sumLang there is a special size symbol, $1$,
which is always interpreted as the constant $1 \in \NN$.
This size symbol is used in particular to represent column and row vectors,
which have type $(\alpha,1)$ and $(1, \alpha)$ respectively.
We endow \cqs with the same property in the sense that we will assume in what follows that $1$ is a valid constant symbol and that $1^\db = 1$ for every database $\db$.

\section{From matrices to relations and back}\label{sec:fo-matlang}

In this section, we lay the grounds to revisit and generalize the connection between $\sumLang$ and $\foplus$. Towards this goal, we introduce next all the formal machinery necessary to provide translations between the two query languages that work for any matrix schema, are symmetric, and ensure that matrices are encoded as databases of linear size. These translations are specified in Section~\ref{sec:sumlang-and-fo}. We start by discussing why we need to revisit this formal machinery to then continue presenting the formal details. 

\subsection{The need to revisit the characterization of Sum-Matlang}\label{subsec:new-correspondence} In \cite{DBLP:conf/pods/GeertsMRV21}, the authors already propose a way of translating between \sumLang and \foplus. Next, we summarize how the translations that we propose in this paper improve over those of \cite{DBLP:conf/pods/GeertsMRV21}, justifying why we need to revisit this characterization.

In \cite{DBLP:conf/pods/GeertsMRV21}, the translation from \sumLang to \foplus goes as follows. It is assumed that on the \foplus side there is a unary relation $R_\alpha$ for each size symbol $\alpha$ and that, in the database $\db$ simulating the \sumLang instance $\I$ we have $R^\db_\alpha(i)=1$ for all $1 \leq i\leq\alpha^\I$. Note that, consequently, the size of $\db$ is larger than that of $\I$: it take unit space to store $\alpha^\I$ in $\I$ while it takes $\alpha^\I$ space in $\db$.  Therefore, $\size{\db}$ (the size of $\db$) is certainly not linear in the size of $\I$. Since we are interested in evaluating \sumLang and \lang expressions with linear time preprocessing by reducing to the relational case, this encoding of matrix instances as databases is hence unsuitable. We resolve this in our work by adding inequality atoms to $\foplus$, and adopting a representation of matrix instances by databases that is linear in size.

Conversely, in \cite{DBLP:conf/pods/GeertsMRV21}, the translation from \foplus to \sumLang only works when the input relations to the \foplus query represent square matrices of type $(\alpha,\alpha)$, or vectors with type $(\alpha,1)$ or $(1,\alpha)$, for a fixed single size symbol $\alpha$ such that $\alpha^\db=n$ where $\{ d_1,\ldots,d_n \}$ is the \emph{active domain} of $\db$. This assumption implies that if one translates a \sumLang query into \foplus, optimizes the \foplus query, and then translates back, one obtains a \sumLang query whose input schema is different than that of the original query. This phenomemon is illustrated by means of the scenario below. By contrast, we establish a more general correspondence, parametrized by schema encodings, that preserves the original input schema in such a round-trip translation.

Consider for example the schema $\Sch=\{ \mA \}$ with $\mA\colon (\alpha,\beta)$. Let the instance $\I$ over $\Sch$ be such that $\alpha^\I=3$, $\beta^\I=2$, and:
\[
\mA^\I=\begin{bmatrix} 1 & 0 \\ 0 & 0 \\
	0 & 1\end{bmatrix}.
\]
Let $\mQ$ be a \sumLang query over $\Sch$ to be evaluated on $\I$. In \cite{DBLP:conf/pods/GeertsMRV21} the corresponding relational schema is $\{A, R_1^\alpha,R_2^\beta\}$. The instance $\db$ is such that $A^\db(1,1)=1$, $A^\db(3,2)=1$ and $0$ for any other tuple. Also, $(R_1^\alpha)^\db(i)=1$ for all $i\leq \alpha^\I=3$ and $(R_2^\beta)^\db(i)=1$ for all $i\leq \beta^\I=2$. Then, if we translate this relational setting back to a matrix schema and instance, the result is $\Sch'=\{ \mA, \mR_1, \mR_2\}$ with $\mA\colon (\alpha,\alpha), \mR_1\colon(\alpha,1)$ and $\mR_2\colon(\alpha,1)$. The instance $\I'$ over $\Sch'$ is such that $\alpha^\I=3$ (because of the active domain), and:
\[
\mA^{\I'} \ = \ \begin{bmatrix} 1 & 0 & 0\\ 0 & 0 & 0 \\ 0 & 1 & 0 \end{bmatrix}.
\] Additionally, we have $\mQ'$ over $\Sch'$ and that can be evaluated on instance $\I'$. Note that $\ssem{\mQ}{\I}=\ssem{\mQ'}{\I'}$ does not hold in general, even for the expression $\mA$ since $\mA^\I\neq\mA^{\I'}$.

Given the previous discussion, in the following we determine precisely in what sense relations can encode matrices, or matrices can represent relations, and how this correspondence transfers to queries. Then, in Section~\ref{sec:sumlang-and-fo} we show how to generalize the expressibility results in~\cite{DBLP:conf/pods/GeertsMRV21} for any matrix sizes and every encoding between schemas.

\subsection{How we relate objects.} Let $\mA$ be a matrix of dimension $m \times n$.  There exist multiple natural ways to encode $\mA$ as a relation, depending on the dimension $m \times n$.
\begin{itemize}
\item We can always encode $\mA$, whatever the values of $m$ and $n$ (i.e., $\mA$ could be a matrix, a column or row vector, or a scalar), as the
binary $K$-relation $R$ such that (1) $\mA_{i,j} = R(i,j)$ for every
$i \leq m$, $j \leq n$ and (2) $R(i,j) = \kzero$ if $i > m$ or $j > n$.
\item If $\mA$ is a column vector ($n = 1$) then we can also encode it as the
unary $K$-relation $R$ such that $\mA_{i,1} = R(i)$ for every $i \leq m$ and $R(i) = \kzero$ if $i > m$.
\item Similarly, if $\mA$ is a row vector ($m=1$) then we can encode it as the
unary $K$-relation~$R$ with $\mA_{1,j} = R(j)$ for every $j \leq n$ and
$R(i) = \kzero$ if $j > n$.
\item If $\mA$ is a scalar ($m=n=1$), we can encode it as a nullary $K$-relation $R$ with~$\mA_{1,1} = R()$.
\end{itemize}
Note that if $\mA$ is scalar then we could hence encode it by means of a binary relation, a unary relation, or a nullary relation; and if it is a vector we can encode it by a binary or unary relation. In what follows, we write $\mA \simeq R$ to denote that $R$ encodes $\mA$ following any of the alternatives above.

Conversely, given a (nullary, unary, or binary) $K$-relation $R$ we may
interpret this as a matrix of appropriate dimension. Specifically, we say that
relation $R$ is \emph{consistent} with dimension $m \times n$ if there exists a
matrix $\mA$ of dimension $m\times n$ such that $\mA \simeq R$. This is
equivalent to requiring that relation is $\kzero$ on entries outside of
$m \times n$. Note that, given $R$ that is consistent with
$m\times n$ there is exactly one matrix $\mA\colon m\times n$ such that
$\mA \simeq R$.

\subsection{How we relate schemas.} 
A \emph{matrix-to-relational schema encoding} from a matrix schema $\Sch$ to a relational vocabulary $\Voc$ is a function $\Rel\colon \Sch \to \Voc$ that maps every
matrix symbol $\mA$ in $\Sch$ to a unary or binary relation symbol $\Rel(\mA)$ in $\Voc$, and every size symbol $\alpha$ in $\Sch$ to a constant symbol $\Rel(\alpha)$ in $\Voc$. Here,
$\Rel(\mA)$ can be unary only if $\mA$ is of vector type, and nullary only if $\mA$ is of scalar type.  Intuitively, $\Rel$
specifies which relation symbols will be used to store the encodings of which
matrix symbols. In addition, we require that $\Rel(1) = 1$ and that $\Rel$ is a bijection between $\Sch$ and $\sigma$.
This makes sure that we can always invert~$\Rel$. In what follows, we will only specify that $\Rel$ is a matrix-to-relational schema
encoding \emph{on $\Sch$}, leaving the vocabulary $\Voc$ unspecified. In that
case, we write $\Rel(\Sch)$ for the relational vocabulary $\Voc$.

Conversely, we define a \emph{relational-to-matrix schema encoding} from $\Voc$
into $\Sch$ as a function $\Mat\colon \Voc \to \Sch$ that maps every relation
symbol $R$ to a matrix symbol $\Mat(R)$ and every constant symbol $c$ to a size
symbol $\Mat(c)$.  We require that all unary relations are mapped to matrix
symbols of vector type, either row or column, and that all nullary relations are mapped to matrix symbols of scalar type. Furthermore, $\Mat$ must map $1 \mapsto 1$ and be bijective.  Similarly, we denote by
$\Mat(\Voc)$ the matrix schema $\Sch$ mapped by $\Mat$.

Note that the bijection assumption over $\Mat$ imposes some requirements over $\Voc$ to encode it as matrices. For example, $\Mat$ requires the existence of at least one constant symbol in $\Voc$ for encoding matrices dimensions, since every matrix symbol has at least one size symbol in its type, and that size symbol is by definition in~$\Sch$. The bijection between constant and size symbols is necessary in order to have lossless encoding between both~settings.

Given that $\Rel$ and $\Mat$ are bijections between $\Sch$ and $\Voc$, their inverses $\Rel^{-1}$ and $\Mat^{-1}$ are well defined. Furthermore, by definition we have that $\Rel^{-1}$ and $\Mat^{-1}$ are relational-to-matrix and matrix-to-relational schema encodings, respectively.

\subsection{How we relate instances.} 
We start by specifying how to encode matrix instances as database instances.  Fix a matrix-to-relational schema encoding $\Rel$ between $\Sch$ and $\Rel(\Sch)$.
Let $\I$ be a matrix instance over $\Sch$ and $\db$ a database over $\Rel(\Sch)$. We say that \emph{$\db$ is a relational encoding of $\I$ w.r.t. $\Rel$}, denoted by $\I \simeq_{\Rel} \db$, if
\begin{itemize}
    \item $\mA^{\I} \simeq \Rel(\mA)^{\db}$ for every matrix symbol $\mA$ in $\Sch$, and 
    \item $\Rel(\alpha)^{\db} = \alpha^\I$ for every size symbol $\alpha$ in $\Sch$.
\end{itemize}
Note that, given $\I$ and $\Rel$, the relational encoding $\db$ is uniquely defined. As
such, we also denote this database by $\Rel(\I)$.

We now focus on interpreting database instances as matrix instances, which is more subtle. 
Fix a relational-to-matrix schema encoding $\Mat$ from $\Voc$ to $\Mat(\Voc)$. We need to first leverage the consistency requirement from relations to databases. Formally, we say that a database $\db$ over $\Voc$ is \emph{consistent with} $\Mat$ if for every relation symbol $R$ in $\Voc$, $R^\db$ is consistent with
dimension $c^\db \times d^\db$ where $\Mat(R)\colon (\Mat(c), \Mat(d))$.
In other words, a consistent database specifies the value of each dimension, and the relations are themselves consistent with them.

Let $\db$ be a database over $\Voc$, consistent with $\Mat$ and let $\I$ be a matrix instance of $\Mat(\Voc)$.
We say that \emph{$\I$ is a matrix encoding of $\db$ w.r.t. $\Mat$}, denoted $\db \simeq_{\Mat} \I$, if
\begin{itemize}
    \item $\Mat(R)^{\I} \simeq R^{\db}$  for every relation symbol~$R \in \Voc$; and
    \item $c^\db = \Mat(c)^\I$ for every constant symbol $c \in \Voc$.
\end{itemize}
Given $\Mat$ and a consistent database $\db$, the matrix encoding $\I$
is uniquely defined. As such, we also denote this instance by $\Mat(\db)$.

From the previous definitions, one notes an asymmetry between both directions. Although an encoding always holds from matrices to relations,  we require that the relations are consistent with the sizes (i.e., constants) from relations to matrices. Nevertheless, this asymmetry does not impose a problem when we want to go back and forth, as the next result shows. 
\begin{prop}\label{prop:encodings}
    Let $\Rel$ and $\Mat$ be matrix-to-relational and relational-to-matrix schema encodings from $\Sch$ to $\Voc$ and from $\Voc$ to $\Sch$, respectively, such that $\Mat = \Rel^{-1}$. Then
    \begin{itemize}
        \item $\Rel^{-1}(\Rel(\Sch)) = \Sch$ and $\Mat^{-1}(\Mat(\Voc)) = \Voc$;
        \item $\Rel(\I)$ is consistent with $\Rel^{-1}$, for every instance $\I$ over $\Sch$;
        \item $\Rel^{-1}(\Rel(\I)) = \I$, for every instance $\I$ over $\Sch$; and 
        \item $\Mat^{-1}(\Mat(\db)) = \db$, for every  $\db$ consistent with $\Mat$.
    \end{itemize}
\end{prop}

The previous proposition is a direct consequence of the definitions; however, it shows that the consistency requirement and schema encodings provide a lossless encoding between the relational and matrix settings. This fact is crucial to formalize the expressiveness equivalence between $\sumLang$ and $\foplus$, and their subfragments in the following sections.

\section{Sum-matlang and positive FO}\label{sec:sumlang-and-fo}

The previous section has established the  formal basis for comparing $\sumLang$ and $\foplus$. In this section, we give our first expressibility results, proving that $\sumLang$ and $\foplus$ are equally expressive in some precise sense. 
Furthermore, we show that its fragments $\conjLang$ and binary conjunctive queries (CQs) are also equally expressive. These results will be our starting point to find fragments of $\sumLang$ that captures fragments of $\foplus$ with good algorithmic properties, like free-connex and q-hierarchical CQs.

\subsection{From Sum-Matlang to positive FO}

We first aim to simulate every $\sumLang$ query with an \foplus query w.r.t. some matrix-to-relational schema encoding. 
In what follows, fix a matrix schema $\Sch$. Let $\mQ$ be a \sumLang query over $\Sch$, and $\Rel$ be a matrix-to-relational schema encoding on $\argSch{\mQ}$. We say that \foplus query $Q$ \emph{simulates} $\mQ$ w.r.t.\ $\Rel$ if $\Rel(\sem{\mQ}{\I}) = \ssem{Q}{\Rel(\I)}$ for every matrix instance $\I$ over $\Sch$. Note that the definition implies that the output matrix symbol of $\mQ$ must be mapped to the output relation symbol of $Q$ by $\Rel$, since $\Rel$ is a bijection and the condition must hold for every matrix instance. Indeed, it is equivalent to $\sem{\mQ}{\I} = \Rel^{-1}(\ssem{Q}{\Rel(\I)})$, namely, that one can evaluate $\mQ$ by first evaluating $\ssem{Q}{\Rel(\I)}$ and then mapping the results back.

Our main goal will be to prove that one can simulate every $\sumLang$ query in the relational setting by some \foplus query. 
For the sake of presentation, we will take a detour by first showing that we can simulate every $\conjLang$ query by some CQ to then leverage this result to $\sumLang$ and \foplus. The first step towards the proof is to note that \conjLang expressions have a prenex normal form as follows. We say that a \conjLang expression $e\colon (\alpha,\beta)$ is in \emph{prenex normal form} if it has the
form:
\[
e \ \coloneqq \ \Sigma \mx, \my, \mv_1,\ldots,\mv_k. \hspace{1ex}\ms \times \big(\mx\cdot \my^T\big)
\]
such that $\mx\colon(\alpha, 1)$, $\my\colon(\beta,1)$, and $\ms$ is an expression satisfying the following grammar:
\[ 
\ms \coloneqq \mv^T\cdot\mA\cdot \mw \mid \mv^T\cdot \mw
\mid \ms \times \ms
\] 
where $\mA\in\Sch$; $\mv$ ranges over $\{\mx,\my,\mv_1,\ldots,\mv_k\}$; and $\mw$ ranges over $\{\mx,\my,\mv_1,\ldots,\mv_k\}$ and the free vector variables of $e$. 

\begin{lem}\label{lemma:conjlang-prenex-normal-form}
	Every $\conjLang$ expression can be %
	expressed in prenex normal form.
\end{lem}

The proof of Lemma~\ref{lemma:conjlang-prenex-normal-form} is by direct induction on the expression syntax; the interested reader may find it in Appendix~\ref{sec:app-sumlang-and-fo}. In essence, the form structure is analogous to the classic conjunctive query structure: the expression $\mv^T\cdot\mA\cdot \mw$ mimics access to base relations using variables $\mv$ and $\mw$; the expression $\mv^T\cdot \mw$ mimics equality between $\mv$ and $\mw$; and $\{\mv_1,\ldots,\mv_k\}$ are the projected variables. Variables $\mx$ and $\my$ encode the size of the output matrix.

By using the prenex normal form of \conjLang, we can prove the simulation of $\conjLang$ queries by CQs. 
\begin{prop}\label{prop:conjlang-to-cq}
	For every $\conjLang$ query $\mQ$ over $\Sch$ and every matrix-to-relational schema encoding $\Rel$ on $\argSch{\mQ}$, there exists a CQ $Q$ that simulates $\mQ$ w.r.t.\ $\Rel$.
\end{prop}
\begin{proof}
	Let $\mQ = \mH\coloneqq e$ be a \conjLang query such that
	$e\colon(\alpha,\beta)$. By definition of queries, $e$ is a sentence (it does not have free variables). Then by
	Lemma \ref{lemma:conjlang-prenex-normal-form} we may assume that
	$e= \Sigma \mx, \my,
	\mv_1,\ldots,\mv_k. \hspace{1ex}\ms_1\times\cdots\times\ms_n \times \mx\cdot
	\my^T$ where $\ms_i \coloneqq \mv^T\cdot\mA\cdot \mw \mid \mv^T\cdot\mw$ with
	$\mv,\mw\in \{\mx,\my,\mv_1,\ldots,\mv_k\}$.
	
	Let $x,y,v_1,\ldots,v_k$ be \fo variables. In what follows, the variables $v,w\in\{x,y,v_1,\ldots,v_k\}$ are selected respectively for $\mv,\mw\in \{\mx,\my,\mv_1,\ldots,\mv_k\}$: $x$ is used when $\mx$ appears, $y$ is used when $\my$ appears, and so on. Additionally, let us denote $\Rel(\mA)$ simply by means of $A$ (in non-bold math font), for every matrix symbol $\mA$, and let us denote the constant symbol $\Rel(\alpha)$ simply by $\alpha$, for every size symbol $\alpha$ (recall that $\Rel(1) = 1$ by definition).
	
	We now build a CQ $Q$ that simulates $\mQ$ w.r.t. $\Rel$. For each $1\leq i\leq n$:
	\begin{itemize}
		\item If $\ms_i \coloneqq \mv^T\cdot\mA\cdot \mw$ define:
		\begin{itemize}
			\item $a_i\coloneqq A(v,w)$ if $A$ is binary.
			\item $a_i\coloneqq A(v)\wedge w\leq 1$ if $A$ is unary and $\mA\colon(\alpha, 1)$.
			\item $a_i\coloneqq v\leq 1 \wedge A(w)$ if $A$ is unary and $\mA\colon(1, \alpha)$.
			\item $a_i\coloneqq v\leq 1\wedge w\leq 1 \wedge A()$ if $A$ is nullary.
		\end{itemize}
		\item If $\ms_i \coloneqq \mv^T\cdot\mw$ with $\mv\colon(\gamma,1)$ and $\mw\colon(\gamma,1)$ define:
		\begin{itemize}
			\item $a_i\coloneqq v\leq\gamma$ if $\mv=\mw$.
			\item $a_i\coloneqq v\leq\gamma\wedge w\leq\gamma\wedge v=w$ if $\mv\neq\mw$. 
		\end{itemize}
	\end{itemize}
	Finally, define $\varphi\coloneqq \exists v_1,\ldots,v_k. a_1\wedge\cdots\wedge a_n \wedge x\leq\alpha \wedge y\leq\beta$. Note that this formula is in prenex normal form. Then the CQ $Q\colon H(x,y)\gets \varphi$ simulates $\mQ$ if $\Rel(\mH)$ is binary and $\mH\colon(\alpha, \beta)$; the CQ $Q\colon H(x)\gets \exists y.\varphi$ simulates $\mQ$ if $\Rel(\mH)$ is unary and $\mH\colon(\alpha, 1)$; the CQ $Q\colon H(y)\gets \exists x.\varphi$ simulates $\mQ$ if $\Rel(\mH)$ is unary and $\mH\colon(1, \beta)$; and the CQ $Q\colon H()\gets \exists x, y.\varphi$ simulates $\mQ$ if $\Rel(\mH)$ is nullary.
\end{proof}

Finally, we can provide a formal proof for the simulation from \sumLang to \foplus.

\begin{prop}\label{prop:langsum-to-fo}
    For every $\sumLang$ query $\mQ$ over $\Sch$ and every matrix-to-relational schema encoding $\Rel$ on $\argSch{\mQ}$, there exists an \foplus query $Q$ that simulates $\mQ$ w.r.t.\ $\Rel$.
\end{prop}
\begin{proof}
  Let $\mQ = \mH\coloneqq e$ be a \sumLang query. By definition of query, $e$ is
  a sentence with $\type(e) = (\alpha, \beta)$ for some $\alpha$, $\beta$. 
  Note that every \sumLang sentence can be expressed as a sum of $\conjLang$ sentences. Indeed, $\left( e_1 + e_2 \right)^T = e_1^T + e_2^T$; $e_1\kprod \left(e_1 + e_2\right) = e_1\kprod e_2 + e_1 \kprod e_3$; $\left(e_1 + e_2\right) \kprod e_3 = e_1 \kprod e_3 + e_2\kprod e_3$; $e_1\cdot \left(e_2 + e_3\right) = e_1\cdot e_2 + e_1\cdot e_3$; $\left(e_1 + e_2\right) \cdot e_3 = e_1 \cdot e_3 + e_2\cdot e_3$; and $\Sigma \mv.\left( e_1 + e_2\right) = \Sigma \mv.e_1 + \Sigma \mv. e_2$.
  Therefore, we can assume that $e$ is of the form $e_1 + \cdots + e_l$
  such that each $e_i$ is a \conjLang sentence in prenex normal form.

	By Proposition~\ref{prop:conjlang-to-cq}, for each $i=1,\ldots, l$ there exists a CQ $Q_i\colon H(\bar{x}_i)\gets \psi_i$ that simulates the query $\mH\coloneqq e_i$  w.r.t. $\Rel$. 
	Further, since $e = e_1 + \cdots + e_l$ is well-typed, then $e_i \colon (\alpha, \beta)$ for every $i$. This implies that every formula $\psi_i$ has the same number of free variables and we can assume that $\bar{x}_1 = \ldots = \bar{x}_l = \bar{x}$ without loss of generality. Therefore, the \foplus query $Q\colon H(\bar{x})\gets \psi_1 \vee \cdots \vee \psi_l$ simulates $\mQ$ where $\psi_1 \vee \cdots \vee \psi_l$ is a safe formula. 
\end{proof}

\subsection{From positive FO to Sum-Matlang}

We now aim to simulate every \foplus query with a \sumLang query. Contrary to the previous direction, the expressiveness result here is more subtle and requires more discussion and additional notions.

Fix a vocabulary $\Voc$. Let $Q$ be a \foplus query over $\Voc$ and let $\Mat$ be relational-to-matrix schema encoding on $\argVoc{Q}$. We say that a matrix query $\mQ$ \emph{simulates} $Q$ w.r.t.\ $\Mat$ if $\Mat(\ssem{Q}{\db}) = \sem{\mQ}{\Mat(\db)}$ for every database $\db$ consistent with $\Mat$. We note again that this definition implies that the input vocabulary and output relation symbol of $Q$ coincides with the input schema and output matrix symbol of $\mQ$, respectively. Further, it is equivalent that $\ssem{Q}{\db} = \Mat^{-1}(\sem{\mQ}{\Mat(\db)})$.

Before stating how to connect \foplus with $\sumLang$, we need to overcome the  following problem: a \foplus query can use the same variable within different relational atoms, which can be mapped to matrix symbols of different types. For an illustrative example of this problem, consider the query \[ Q\colon H(x,y) \gets \exists z. \left(R(x,y)\wedge S(y,z)\right) \] and a relational-to-matrix schema encoding such that $\Mat$ maps $R$ and $S$ to symbols of type $(\alpha,\beta)$, $H$ to a symbol of type $(\beta,\beta)$, and $c$ and $d$ to $\alpha$ and $\beta$, respectively. For a consistent database $\db$ w.r.t. $\Mat$, we could have that $R$ and $S$ are consistent with $c^\db \times d^\db$, but $H$ is not consistent with $d^\db \times d^\db$ if $d^\db < c^\db$.  Moreover, $\Mat$ bounds variable $y$ with different sizes $c^\db$ and $d^\db$. It is then problematic to simulate $Q$ under $\Mat$ in $\sumLang$ because $\sumLang$ expressions need to be well-typed.

Given the previous discussion, a \emph{well-typedness} definition of a \foplus formula is necessary. Let $\Mat$ be a relational-to-matrix schema encoding on $\Voc$. Given a \foplus formula $\varphi$ over $\Voc$ and a function $\tau$ from $\free(\varphi)$ to size symbols in $\Mat(\sigma)$, define the rule $\Mat\vdash\varphi\colon \tau$ inductively as shown in Figure~\ref{fig:well-typedness}, where $\tau_1\sim \tau_2$ if and only if $\tau_1(x)=\tau_2(x)$ for every $x\in dom(\tau_1)\cap dom(\tau_2)$. We say that $\varphi$ over $\Voc$ is \emph{well-typed} w.r.t. $\Mat$ if there exists such a function $\tau$ such that $\Mat\vdash\varphi\colon \tau$. Note that if $\varphi$ is well-typed, then there is a unique $\tau$ such that $\Mat\vdash\varphi\colon \tau$.

\begin{figure*}[tbp]
    \centering
    $
    \dfrac{\type(\Mat(R))=(\alpha,\beta)}{\Mat\vdash R(x_1,x_2)\colon \{x_1\mapsto\alpha, x_2\mapsto\beta\}} \quad  \quad	\quad 	\dfrac{\type(\Mat(R))=(\alpha,1) \text{ or } (1,\alpha)}{\Mat\vdash R(x)\colon \{x\mapsto\alpha\}}
    $ \\[1.5ex]
    $
    \dfrac{\type(\Mat(R))=(1,1)}{\Mat\vdash R()\colon \{\}}  \quad \quad 	\quad		
    \dfrac{}{\Mat\vdash x\leq c\colon \{x\mapsto\Mat(c)\}} \quad  \quad \quad
    \dfrac{\Mat\vdash \varphi\colon \tau}{\Mat\vdash \exists\y.\varphi\colon \restr{\tau}{\free(\varphi)\setminus\y}}
    $\\[1.5ex]
    $
    \dfrac{\Mat\vdash \varphi_1\colon \tau_1, \Mat\vdash \varphi_2\colon \tau_2\text{ and }\tau_1\sim \tau_2}{\Mat\vdash \varphi_1\wedge\varphi_2\colon \tau_1\cup \tau_2} \quad \quad \quad
    \dfrac{\Mat\vdash \varphi_1\colon \tau_1, \Mat\vdash \varphi_2\colon \tau_2\text{ and }\tau_1\sim \tau_2}{\Mat\vdash \varphi_1\vee\varphi_2\colon \tau_1\cup \tau_2} 
    $ \\[1.5ex]
    $
    \dfrac{\alpha \text{ a size symbol }}{\Mat\vdash x= y \colon \{x \mapsto \alpha, y \mapsto \alpha\}}
    $
    \caption{Well-typedness of \foplus under relational-to-matrix mapping $\Mat$. }
    \label{fig:well-typedness}
\end{figure*}

Now, let $Q\colon H(\x)\gets \varphi$ be a binary $\foplus$ query over $\Voc$ and $\Mat$ be a matrix encoding specification on $\sigma(Q)$. We say that $Q$ is \emph{well-typed} w.r.t. $\Mat$ if $\varphi$ is well-typed w.r.t. $\Mat$ and for $\tau$ such that $\Mat\vdash\varphi\colon \tau$, we have:
\begin{itemize}
    \item if $\x=(x_1,x_2)$, then $\type(\Mat(H))= (\tau(x_1),\tau(x_2))$; and
    \item if $\x=(x)$, then $\type(\Mat(H))$ is either $(\tau(x),1)$ or $(1,\tau(x))$.
\end{itemize}
We write $\Mat\vdash Q\colon \tau$ to indicate that $Q$ is well-typed w.r.t. $\Mat$, and $\tau$ is the unique function testifying to well-typedness of \foplus formula of $Q$. We note that we can show that the query obtained by Proposition~\ref{prop:langsum-to-fo} is always well-typed.

The next proposition connects well-typedness with consistency.

\begin{prop}\label{prop:welltype-consistent}
For binary \foplus query $Q\colon H(\x)\gets \varphi$ over a vocabulary $\Voc$, if $\Mat\vdash Q \colon \tau$ then for any $\db$ consistent with $\Mat$ we have:
    \begin{itemize}
        \item If $\x=(x_1,x_2)$ then $\ssem{Q}{\db}(H)$ is consistent with dimension $\tau(x_1)^{\db}\times \tau(x_2)^{\db}$.
        \item If $\x=(x)$ then $\ssem{Q}{\db}(H)$ is consistent with both dimension $\tau(x)^{\db}\times 1$ and $1 \times \tau(x)^{\db}$.
    \end{itemize}
\end{prop}
\begin{proof}
	We show a more detailed statement. Namely,  that for any \foplus formula $\varphi$ with $\Mat\vdash \varphi \colon \tau$, any database $\db$ consistent with $\Mat$ and any variable $x \in \free(\varphi)$ if $\val$ is a valuation with $\val(x) > \tau(x)^\db$ then $\fssem{\varphi}{\db}(\val) = \kzero$. The proposition follows from this more general statement since $Q$ is of the form $H(\x)\gets \varphi$ with $\x$ a tuple consisting of the free variables of $\varphi$ (possibly with repetition) and since $\type(\Mat(H)) = \tau(\x)$ as $Q$ is well-typed under $\Mat$.

	The proof of the more detailed statement is by induction on $\varphi$. We illustrate the reasoning when $\varphi$ is an atom, or a disjunction. The other cases are similar.

	If $\varphi$ is a relational atom $R(x_1,x_2)$, then by definition of the typing relation, we know that $(\tau(x_1),\tau(x_2)) = (\alpha,\beta)$ where $\type(\Mat(R)) = (\alpha,\beta)$. Hence, for valuation $\val$ with $\val(x_1) > \tau(x_1)^\db$ or $\val(x_2) > \tau(x_2)^\db$ we have $\fssem{R(x_1,x_2)}{\db}(\val) = \kzero$ because $R^\db$ is consistent with dimensions $\alpha^\db \times \beta^\db = \tau(x_1)^\db \times \tau(x_2)^\db$. If $\varphi$ is a comparison atom $x\leq\alpha$, then $\tau(x) = \alpha$ by definition of the typing relation. Hence, for valuation $\val$ with $\val(x) > \tau(x)^\db = \alpha^\db$ we trivially obtain $\fssem{x\leq\alpha}{\db}(\val) = \kzero$.

	When $\varphi$ is a disjunction $\varphi_1 \vee \varphi_2$, let a valuation $\val \colon \free(\varphi)$ such that $\val(x) > \tau(x)^\db$ for some $x \in \free(\varphi)$. Because we consider only safe formulas, both $\varphi_1$ and $\varphi_2$ have the same set of free variables, which equals the free variables of $\varphi$. Then, from  $x\in \free(\varphi_1)$ and $x\in \free(\varphi_2)$, and so by induction hypothesis both $\fssem{\varphi_1}{\db}(\val)=\kzero$ and $\fssem{\varphi_2}{\db}(\val)=\kzero$ hold. Hence $\fssem{\varphi}{\db}(\val)=\kzero$.
\end{proof}

Similar to the proof from \sumLang to \foplus, we first illustrate how to simulate CQs with \conjLang queries.

\begin{prop}\label{prop:cq-to-conjlang}
    For every binary CQ $Q$ over a vocabulary $\Voc$ and every relational-to-matrix schema encoding $\Mat$ on $\Voc(Q)$ such that $Q$ is well typed w.r.t. $\Mat$ there exists a \conjLang query $\mQ$ that simulates $Q$ w.r.t. $\Mat$.
\end{prop}
\begin{proof}
    Let $Q\colon H(\x)\gets \exists v_1,\ldots,v_k.\varphi$ with $\free(\varphi)=\{ x,y \}$ and $\x\subseteq\{x,y \}$ be a CQ, i.e., $\varphi=a_1\wedge\cdots\wedge a_n$ where each $a_i$ is a relational, comparison, or equality atom using variables from the set $\{x,y,v_1,\ldots,v_k\}$. Assume $\Mat\vdash Q \colon \tau$.
    
    For simplicity, let all input relations be binary (we will deal with the other cases later). Let $\mv,\mw\in \{\mx,\my,\mv_1,\ldots,\mv_k\}$ be vector variables of type $(\tau(x), 1),(\tau(y), 1),(\tau(v_1), 1),\ldots,\allowbreak(\tau(v_k), 1)$, respectively. In what follows, we choose $\mv,\mw\in \{\mx,\my,\mv_1,\ldots,\mv_k\}$ respectively according to $v,w\in \{x,y,v_1,\ldots,v_k\}$. For each $i=1,\ldots,n$ define:
    \begin{itemize}
        \item $\ms_i \coloneqq \mv^T\cdot\mA\cdot \mw$ if $a_i$ is equal to $A(v,w)$. Note that $\Mat(A)\colon(\tau(v),\tau(w))$, $\mv\colon (\tau(v),1)$ and $\mw\colon (\tau(w),1)$.
        \item $\ms_i \coloneqq \mv^T\mv$ if $a_i$ is equal to $v\leq\alpha$. Note that $\mv\colon (\tau(v),1)$.
        \item $\ms_i \coloneqq \mv^T\mw$ if $a_i$ is equal to $v=w$. Note that $\mv\colon (\tau(v),1)$ and $\mw\colon (\tau(w),1)$, with $\tau(v)=\tau(w)$ because $Q$ is well-typed.
    \end{itemize}
    Finally, define
    \begin{itemize}
        \item $e= \Sigma \mx, \my, \mv_1,\ldots,\mv_k. \hspace{1ex}\left(\ms_1\times\cdots\times\ms_n\right) \times \left(\mx\cdot \my^T\right)$ if the head of $Q$ is $H(x,y)$.
        \item $e= \Sigma \mx, \mv_1,\ldots,\mv_k. \hspace{1ex}\left(\ms_1\times\cdots\times\ms_n\right) \times \left(\mx\cdot \mx^T\right)$ if the head of $Q$ is $H(x,x)$.
        \item $e= \Sigma \mx, \mv_1,\ldots,\mv_k. \hspace{1ex}\left(\ms_1\times\cdots\times\ms_n\right) \times \mx$ if the head of $Q$ is $H(x)$ and $\mH\colon (\alpha,1)$. Otherwise, if $\mH\colon (1, \alpha)$, define $e= \Sigma \mx, \mv_1,\ldots,\mv_k. \hspace{1ex}\left(\ms_1\times\cdots\times\ms_n\right) \times \mx^T$.
    \end{itemize}
    Then $\mQ = \mH\coloneqq e$ simulates $Q$ w.r.t. $\Mat$ for any choice of $\x\subseteq\{x,y\}$.

    If some input is not binary, define:
    \begin{itemize}
        \item $\ms_i \coloneqq \mv^T\cdot\mA\cdot \mw$ if $a_i=A(v)$ such that $\Mat(A)\colon(\tau(v),1)$ and $\mv\colon (\tau(v),1)$ for some $\mw\colon (1,1)$.
        \item $\ms_i \coloneqq \mw^T\cdot\mA\cdot \mw$ if $a_i=A()$ such that $\Mat(A)\colon(1,1)$, for some $\mw\colon (1,1)$.
    \end{itemize}

    Finally, if there is no $\mv_i\colon (1,1)$ in $e$ above we add a dummy $\mw\colon (1,1)$ vector variable as $e= \Sigma \mx, \my, \mv_1,\ldots,\mv_k,\mw. \hspace{1ex}\ms_1\times\cdots\times\ms_n \times \mx\cdot \my^T$.
\end{proof}

We have now all the formal machinery to state how to simulate every binary \foplus query over relations with a \sumLang query over matrices. 

\begin{prop}\label{prop:fo-to-langsum}
    For every binary \foplus query $Q$ over a vocabulary $\Voc$ and every relational-to-matrix schema encoding $\Mat$ on $\Voc(Q)$ such that $Q$ is well typed w.r.t. $\Mat$ there exists a \sumLang query $\mQ$ that simulates $Q$ w.r.t. $\Mat$.
\end{prop}
\begin{proof}
    First, note that every $\varphi\in\foplus$ can be written in disjunctive normal form, i.e., $\varphi=\varphi_1\vee\cdots\vee\varphi_l$ where each $\varphi_i$ is in prenex normal form and the big disjunction is safe. On one hand, $\exists x. \psi_1\vee \psi_2\equiv \exists x. \psi_1 \vee \exists x. \psi_2$ and $\left(\psi_1\vee\psi_2\right)\wedge \psi_3 \equiv \left(\psi_1\wedge\psi_2\right)\vee\left(\psi_1\wedge\psi_3\right)$. On the other hand, if $\psi = R(\x)\mid x\leq c \mid x = y\mid \psi_1\wedge\psi_2\mid \exists x.\psi$ then it can be written in prenex normal form, because $\psi_1\wedge \exists x.\psi_2\equiv\exists x. \psi_1\wedge\psi_2$ since without loss of generality we can assume that $x$ does not appear in $\varphi_1$.

    Now, let $Q\colon H(\x)\gets \varphi$ be a binary \foplus query such that $\varphi$ is in disjunctive normal form, i.e., $\varphi=\varphi_1\vee\cdots\vee\varphi_l$. By Proposition~\ref{prop:cq-to-conjlang} there exist \conjLang queries $\mQ_i=\mH\coloneqq e_i$ that simulate the CQs $Q\colon H(\x)\gets \varphi_i$ w.r.t. $\Mat$, respectively.

    It is straightforward to see that $\mQ=\mH\coloneqq e_1 + \cdots + e_l$ simulates $Q$ w.r.t. $\Mat$. \qedhere

\end{proof}

\subsection{Equivalence between Sum-Matlang and positive FO}

Taking into account the correspondence between \sumLang and \foplus established by Propositions~\ref{prop:langsum-to-fo} and~\ref{prop:fo-to-langsum}, in what follows we say that matrix query language $\mathcal{L}_M \subseteq \sumLang$ and relational language $\mathcal{L}_R \subseteq \foplus$ are \emph{equivalent} or \emph{equally expressive} if (1) for every matrix query $\mQ \in \mathcal{L}_M$ over a matrix schema $\Sch$ and every matrix-to-relational schema encoding $\Rel$ on $\argSch{\mQ}$ there exists a query $Q \in \mathcal{L}_R$ that simulates $\mQ$ w.r.t. $\Rel$ and is well-typed w.r.t. $\Rel^{-1}$; and (2) for every binary query $Q \in \mathcal{L}_R$ over a vocabulary $\sigma$ and every relational-to-matrix schema encoding $\Mat$ such that $Q$ is well-typed w.r.t. $\Mat$ there exists $\mQ \in \mathcal{L}_M$ that simulates $Q$ w.r.t $\Mat$.

From Propositions~\ref{prop:langsum-to-fo} and~\ref{prop:fo-to-langsum}, and Propositions~\ref{prop:conjlang-to-cq} and~\ref{prop:cq-to-conjlang} one obtains the following characterization of positive FO and CQs, respectively.

\begin{cor}\label{cor:langsum-to-cq}
	\begin{enumerate}
		\item $\sumLang$ and binary $\foplus$ queries are equally expressive.
		\item $\conjLang$ and binary conjunctive queries are equally expressive.
	\end{enumerate}
\end{cor}

While the result between $\conjLang$ and CQs is a consequence of the connection between \sumLang and \foplus, it provides the basis to explore the fragments of \conjLang that correspond to fragments of CQ, like free-connex or q-hierarchical CQ. We determine these fragments in the next sections.

\section{Free-connex conjunctive queries and \foTwo}\label{sec:fc-cq}

We wish to understand the fragment of \lang that allows efficient
enumeration-based evaluation, i.e., enumerating the query result with constant
delay after a preprocessing phase that is linear in the input instance.  Towards
that goal, we will first specialize in this section and the next the
correspondence between \conjLang and CQs given by
Corollary~\ref{cor:langsum-to-cq} to \emph{free-connex}
CQs~\cite{DBLP:conf/csl/BaganDG07,DBLP:phd/hal/BraultBaron13}. Next, we recall the definition of free-connex CQs and state the main result of this section. 

\subsection{Free-connex CQs.}
First, recall that a query $Q$ is a CQ if it is in prenex normal form:
\begin{equation}\label{eqn:cq}
	Q: H(\seq{x}) \gets \exists \seq{y}. a_1 \wedge \dots \wedge a_n \tag{*}
\end{equation}
where $a_1,\dots, a_n$ are relational,  comparison, or equality atoms.
From now on, unless explicitly stated otherwise, we will assume without loss of generality that a CQ does not have equality atoms. 
Indeed, we can always remove equality atoms from CQ by taking the equivalence classes formed by all equalities in the body and replacing each variable in the body and head of the query with a representative from its class.

A \cq
$Q$ is \emph{acyclic} (also sometimes referred to as
$\alpha$-acyclic~\cite{DBLP:journals/jacm/Fagin83}) if it has a \emph{join
  tree}. A join tree for $Q$ is an undirected tree $T$ whose node-set is the set
of all atoms (i.e., relational or comparison atoms) occurring in $Q$, that satisfies the
\emph{connectedness} property. This property requires that for every variable
$x$ occurring (free or bound) in $Q$, all the nodes containing $x$ form a
connected subtree of $T$.  Note that we consider comparison atoms as unary.
Then, a \cq $Q$ like \eqref{eqn:cq} is
  \emph{free-connex}
  if it is acyclic and the query $Q'$ obtained by adding the head atom $H(\x)$
  to the body of $Q$ is also acyclic. 
The requirement that the query $Q'$ obtained by adding the head atom $H(\x)$ to
the body of $Q$ is also acyclic forbids queries like
$H(x,y) \gets \exists z. A(x,z) \wedge B(z,y)$. Indeed: when we adjoin the head
to the body we get $\exists z. A(x,z) \wedge B(z,y) \wedge H(x,y)$, which is
cyclic. We will usually refer to free-connex \cqs simply as \fcqs in what
follows.

Free-connex CQs are a subset of acyclic CQs
that allow efficient enumeration-based query evaluation: in the Boolean semiring
and under data complexity they allow to enumerate the query result
$\ssem{Q}{\db}(H)$ of free-connex \cq $Q\colon H(\seq{x}) \gets \varphi$ with
\emph{constant delay} after a preprocessing phase that is linear in $\db$.  In
fact, under complexity-theoretic assumptions, the class of \cqs that admits
constant delay enumeration after linear time preprocessing is precisely the
class of free-connex~\cqs~\cite{DBLP:conf/csl/BaganDG07}.

In this section, we prove the following correspondence between \fcqs and \foTwo,
the two-variable fragment of $\fo^{\wedge}$. Formally, a formula $\varphi$ in
$\fo^{\wedge}$ is said to be in \foTwo if it does not use equality  and the
set of all variables used in $\varphi$ (free or bound) is of cardinality at most
two. So, $\exists y \exists z. A(x,y) \wedge B(y,z)$ is not in $\foTwo$, but the
equivalent formula $\exists y. \left(A(x,y) \wedge \exists x. B(y,x)\right)$
is. An $\foTwo$ query is a binary query whose body is an $\foTwo$
formula. Recall that a query is binary if every relational atom occurring in its
body and head have arity at most two.
\begin{thm}\label{theo:free-connex-binary-cq-equiv-foTwo}
    Binary free-connex CQs and \foTwo queries are equally expressive.
\end{thm}
We find this a remarkable characterization of the \fcqs on binary relations
that, to the best of our knowledge, it is new. Moreover, we will use this result
in Section~\ref{sec:free-connex-matlang} to identify the fragment of \lang
queries that expressively correspond to \fcqs. Having that fragment in hand, we
later turn to its actual query evaluation on arbitrary (not necessarily Boolean)
semirings in Section~\ref{sec:evaluation}.

The rest of this section is devoted to proving
Theorem~\ref{theo:free-connex-binary-cq-equiv-foTwo}. We do so in three
steps. First, in Section~\ref{sec:query-plans} we define \emph{query plans},
which allow to give an alternate but equivalent definition of \fcqs that we find
more convenient to use. Then, in Sections~\ref{sec:fcq-to-fo2} and ~\ref{sec:fo2-to-fcq} we prove the two directions of Theorem~\ref{theo:free-connex-binary-cq-equiv-foTwo}.

\subsection{Query plans.}
\label{sec:query-plans}

To simplify notation, in what follows we denote the set of all FO variables that occur in an object $X$ by $\var(X)$. In particular, if $X$ is itself a set of variables, then $\var(X) = X$.

A \emph{query plan} (QP for short) is a pair $(T,N)$ with $T$ a
  \emph{generalized join tree} and $N$ a \emph{sibling-closed connex subset of
    $T$.} A generalized join tree (GJT) is a node-labeled directed tree
  $T = (V,E)$ such that
    \begin{itemize}
        \item $T$ is binary: every node has at most two children.
        \item Every leaf is labeled by an atom (relational or inequality).
        \item Every interior node $n$ is labeled by a set of variables and has
          at least one child $c$ such that $\var(n)\subseteq \var(c)$. Such
          child $c$ is called the \textit{guard} of $n$.
        \item Connectedness property: for every variable $x$ the sub-graph
          consisting of the nodes $n$ where $x\in \var(n)$ is connected.
    \end{itemize}
    A \emph{connex subset of $T$} is a set $N \subseteq V$ that includes the
    root of $T$ such that the subgraph of $T$ induced by $N$ is a tree. $N$ is
    \emph{sibling-closed} if for every node $n \in N$ with a sibling $m$ in $T$,
    also $m$ is in $N$. The \emph{frontier} of a connex set $N$ is the subset
    $F\subseteq N$ consisting of those nodes in $N$ that are the leaves in the
    subtree of $T$ induced by $N$.

\begin{exa}
  To illustrate, Figure~\ref{fig:qp-example} shows two query plans. There and in future illustrations we use the convention that we depict a node with id $n$ and label $l$ as $n; l$. %
  The elements of the query plan's connex set are indicated by the gray area. So, in $T_1$, we have $N_1= \{a,b,e\}$ and in $T_2$ we have $N_2 = \{a,b,e,f\}$.  Since all nodes in a connex set must be connected, an unallowed choice of $N_1$ for the tree $T_1$ would be $\{a,c\}$. Furthermore, also $\{a,b\}$ is not allowed, since the set must be sibling-closed.
\end{exa}

\begin{figure}[tbp]
\begin{center}
  \small
 \tikzset{
        expand bubble/.style={
            preaction={draw,line width=1.4pt},
            gray!20,fill,draw,line width=2pt,
        },
    }

    \begin{tikzpicture}
        \node (a1) at (0,0) {$a; \{y\}$} [level distance=0.8cm, sibling distance=3cm, level 2/.style={sibling distance = 2cm}]
        child {
            node (b1) {$b;\{ y, z\}$}
            child {
                node {$c;S(y,z)$}
            }
            child {
                node {$d;T(z,w)$}
            }
        }
        child {
            node (e1) {$e;R(x,y)$}
        };

        \node (a2) at (6,0) {$a;\{y,w\}$} [level distance=0.8cm, sibling distance=3cm, level 2/.style={sibling distance = 2cm}]
        child {
            node (b2) {$b;\{ y,w\}$}
            child {
                node (c2) {$c;R(x,y)$}
            }
            child {
                node (d2) {$d;S(y,w,z)$}
            }
        }
        child {
            node (e2) {$e; \{w\}$}
            child {
                node (f2) {$f;T(u,w)$}
            }
          };

       \node[above of=a1] (lab1) {$(T_1,N_1)$};
       \node[above of=a2] (lab2) {$(T_2,N_2)$};

       \begin{pgfonlayer}{background}
            \path[expand bubble]plot [smooth cycle,tension=1]
            coordinates {(a1.north) (b1.south west) (e1.south east) };
            \path[expand bubble]plot [smooth cycle,tension=1]
            coordinates {(a2.north) (b2.south west) (e2.south west) (f2.south west) (f2.south east) (e2.south east)};
        \end{pgfonlayer}

    \end{tikzpicture}
\end{center}
  \caption{  \label{fig:qp-example}
Two query plans $(T_1, N_1)$ and $(T_2, N_2)$ where the sets $N_1$ and $N_2$ are indicated by the gray area.}
\end{figure}

For ease of presentation, we abbreviate query plans of the form $(T, \{r\})$ simply by $(T,r)$ when the connex set consists only of the single root node $r$. Furthermore, if $T$ is a GJT and $n$ is a node of $T$, then we write $\restr{T}{n}$ for the subtree of $T$ rooted at $n$.

Let $(T,N)$ be a query plan and assume that $\mset{ a_1, \ldots a_n}$ is the
  multiset of atoms occurring as labels in the leaves of $T$. Then the
  \emph{formula associated to $(T,N)$} is the conjunctive
  formula \[\varphi[T,N] \ := \  \exists y_1 \ldots \exists y_k. \, a_1\wedge\ldots\wedge a_n\]
  where $\var(T)\setminus\var(N) = \{y_1,\dots,y_k\}$. In particular,
  $\var(N)=\free(\varphi[T,N])$.  A query plan $(T,N)$ is a \emph{query plan
    for} a \cq $Q$ with body $\varphi$ if $\varphi[T,N]\equiv\varphi$.

\begin{exa}
  Referring to Figure~\ref{fig:qp-example}, we have
  \begin{align*}
    \varphi[T_1,N_1] &= \exists w. \, S(y,z) \wedge T(z,w) \wedge R(x,y), \\
    \varphi[T_2,N_2] &= \exists x \exists z. \, R(x,y) \wedge S(y,w,z) \wedge T(u,w).
  \end{align*}
\end{exa}

The following proposition is a particular case of the results shown in \cite{DBLP:conf/sigmod/IdrisUV17,DBLP:journals/vldb/IdrisUVVL20}.

\begin{prop}\label{prop:free-connex-iff-query-plan}
    A \cq $Q$ is free-connex if, and only if, it has a query plan.
\end{prop}

\subsection{From binary free-connex CQ to the two variable fragment of FO}
\label{sec:fcq-to-fo2}

We next show that it is possible to translate every binary \fcq $Q$ into an
equivalent \foTwo query. Our construction is by induction on a suitable query
plan for $Q$. To enable this induction, we first observe the following properties of query plans.

Two query plans $(T,N)$ and $(T', N')$ are said to be \emph{equivalent} if (1) the multiset of atoms $\atoms(T)$ appearing in $T$ equals the multiset of atoms $\atoms(T')$ appearing in $T'$, and (2) $\var(N) = \var(N')$.\footnote{Note that in condition (1) we view $\atoms(T)$ and $\atoms(T')$ as \emph{multisets}: so $T$ and $T'$ must have the same set of atoms, and each atom must appear the same number of times in both.}
Clearly if two query plans are equivalent then their associated formulas are
also equivalent.

\begin{lem}\label{lemma:refined-query-plan}
    For every QP there exists an equivalent QP such that for any node $n$ with two children $c_1, c_2$ it holds that $\var(c_1)\subseteq \var(n)$ and $\var(c_2)\subseteq \var(n)$. In particular, either $\var(c_1) = \var(n)$ or $\var(c_2) = \var(n)$ holds.
\end{lem}

\begin{proof}
    Let $n$ be any node of $T$ that has two children $c_1$ and $c_2$.
    Let $(T,N)$ be the original query plan.
    We modify $(T,N)$ into an equivalent plan $(T', N')$ that satisfies the lemma by adding, for every node $n$ with two children $c_1$ and $c_2$, two new intermediate nodes $n_1$ and $n_2$ as follows labeled with $\var(n_1)= \var(n) \cap \var(c_1)$ and $\var(n_2)= \var(n) \cap \var(c_2)$.

    \begin{center}
        \begin{tikzpicture}
            \node (r1) at (0,0) {$n$} [level distance=0.7cm]
            child {
                node {$c_1$}
            }
            child {
                node {$c_2$}
            };

            \node (r2) at (3,0) {$n$} [level distance=0.7cm]
            child {
                node {$n_1$}
                child { node {$c_1$} }
            }
            child {
                node {$n_2$}
                child { node {$c_2$} }
            };

            \draw[->, dotted, >=stealth] (r1) -- (r2);
        \end{tikzpicture}
    \end{center}

    We obtain the set $N'$ by adding also $n_i$ to $N$ if $c_i\in N$, $i=1,2$.
    It is readily verified that $T'$ is a GJT; $N'$ is a sibling-closed connex
    subset of $T'$; that $T$ and $T'$ have the same multiset of atoms; and that
    $\var(N) = \var(N')$. Hence $(T,N) \equiv (T', N')$. Moreover, in $T'$
    every node $n$ with two children $n_1$ and $n_2$ is such that
    $\var(n_1)\subseteq \var(n)$ and $\var(n_2)\subseteq \var(n)$. Finally, because $T$ is a GJT, it is the case that either $\var(n) \subseteq \var(c_1)$ or $\var(n) \subseteq \var(c_2)$. As such, either $\var(n) = \var(n_1)$ or $\var(n) = \var(n_2)$.
\end{proof}

Next, we show how the formulas and subformulas encoded by a query plan are related.

\begin{lem}
  \label{lem:connex-subset-extra-projection}
  Let $(T,M)$ and $(T,N)$ be two QPs over the same GJT such that
  $N \subseteq M$. Then $\varphi[T,N] \equiv \exists \z. \varphi[T,M]$ with
  $\z = var(M) \setminus \var(N)$.
\end{lem}
\begin{proof}
  Let $\mset{a_1,\dots, a_k}$ be the multiset of atoms in $T$.  Let $\x$ be the set of all variables occurring in $a_1,\dots, a_k$ that are not mentioned in $\var(N)$. Let $\y$ be the set of all variables occurring in $a_1,\dots,a_k$ that are not mentioned in $\var(M)$. Because $N \subseteq M$ also $\var(N) \subseteq \var(M)$ and hence $\x \supseteq \y$. Moreover, $\x = \y \cup \z$.  Then, by definition,
  \begin{equation*}
    \varphi[T,N] = \exists \x. (a_1 \wedge \dots \wedge a_k) 
                  \equiv \exists \z. \exists \y. (a_1 \wedge \dots \wedge a_k) 
     \equiv \exists \z. \varphi[T,M] \qedhere
  \end{equation*}
\end{proof}

\begin{lem}\label{lemma:formula-of-query-plan-is-conjunction-of-frontier-induced-trees-formulas}
    Let $(T,N)$ be a query plan and let $F=\{ r_1, \ldots, r_l \}$ be the frontier of $N$, i.e, the bottom nodes of $N$. Then $\varphi[T,N]\equiv \varphi[\restr{T}{r_1}, r_1] \wedge \dots \wedge \varphi[\restr{T}{r_l}, r_l]$.
\end{lem}
\begin{proof}
    Let, for every frontier node $r_i$, $A_i$ be the multiset of atoms appearing in $\restr{T}{r_i}$. Then, by definition \[ \varphi[T,N] = \exists \seq u.\ \bigwedge_{a_1 \in A_1} a_1 \wedge \dots \wedge \bigwedge_{a_l \in A_l} a_l\] where $\seq u$ is the sequence of all variables in $T$ except the variables in $N$.
    
    We can write $\seq u$ as union of $l$ sets $\seq{v_1} \cup \dots \cup \seq{v_l}$ where $\seq{v_i}$ consists of all variables in $T_i$ except $\var(N)$.
    Because $T$ is a GJT, it satisfies the connectedness property, and hence we know that any variable that is shared by atoms in $T_i$ and atoms in $T_j$, for $1 \leq i < j \leq l$ must be in $\var(r_i) \cap \var(r_j) \subseteq \var(N)$.
    It follows that the sets $\seq{v_i}$ are disjoint: assume for the purpose of obtaining a contradiction that variable $x$ occur in both $\seq{v_i}$ and $\seq{v_j}$. Then it occurs in $T_i$ and $T_j$, and hence, by the running intersection property of $T$ in $\var(r_i) \cap \var(r_j)$.
    Therefore, $x \in \var(N)$, which gives the desired contradiction. Then we can rewrite $\varphi[T,N]$ as follows
    \begin{align*}
        \varphi[T,N] & = \exists \seq u.\ \bigwedge_{a_1 \in A_1} a_1 \wedge \dots \wedge \bigwedge_{a_l \in A_l} a_l \\    
        & = \exists \seq{v_1} \exists \seq{v_2} \dots \exists \seq{v_l}.\ \bigwedge_{a_1 \in A_1} a_1 \wedge \dots \wedge \bigwedge_{a_l \in A_l} a_l \\    
        & \equiv \left(\exists \seq{v_1}.\ \bigwedge_{a_1 \in A_1} a_1\right) \wedge \dots \wedge \left(\exists \seq{v_l}.\ \bigwedge_{a_l \in A_l} a_l\right) \\    
        & \equiv \varphi[T_1, r_1] \wedge \dots \wedge \varphi[T_l, r_l].
    \end{align*}
    Here, the last equivalence holds because $\seq{v_i}$ is exactly the set of variables in $T_i$ except $\var(r_i)$.
    This can be seen as follows. By definition, $\seq{v_i} = \var(T_i) \setminus \var(N)$.
    Because $\var(N) \supseteq \var(r_i)$, it follows that $\seq{v_i} \subseteq \var(T_i) \setminus \var(r_i)$.
    Now assume, for the purpose of obtaining a contradiction that this inclusion is strict, i.e., that there is some $x \in \var(T_i) \cap \var(N)$ that is not in $\var(r_i)$.
    Then, because it is in $\var(N)$ there is some $r_j \not = r_i$ with $x \in \var(r_j)$.
    Because also $x \in \var(T_i)$, it appears in an atom of $T_i$ and is shared with an atom of $T_j$.
    Because $T$ has the running intersection property, we know that hence $x \in \var(r_i) \cap \var(r_j)$, which is the desired contradiction since then $x \in \var(r_i)$.
\end{proof}

\begin{cor}\label{lemma:child-formula-projection}
    Let $(T,r)$ be a query plan where $r$ has one child $c$. Then $\varphi[T,r] \equiv \exists \y. \varphi[\restr{T}{c},c]$ with $\y = \var(c) \setminus \var(r)$.
\end{cor}
\begin{proof}
  Consider the set $M = \{r,c\}$, which is connex in $T$. By
  Lemma~\ref{lem:connex-subset-extra-projection} we have
  $\varphi[T,r] \equiv \exists \y. \varphi[T,M]$. Moreover, since the frontier of
  $M$ is $\{c\}$ we obtain by
  Lemma~\ref{lemma:formula-of-query-plan-is-conjunction-of-frontier-induced-trees-formulas}
  that $\varphi[T,M] \equiv \varphi[\restr{T}{c},c]$. Hence
  $\varphi[T,r] \equiv \exists \y. \varphi[\restr{T}{c},c]$.
\end{proof}

\begin{cor}\label{lemma:children-formulas-conjunction}
    Let $(T,r)$ be a query plan where $r$ has two children $r_1,r_2$ such that one of the following holds:
    \begin{itemize}
        \item $\var(r_1)=\var(r)$ and $\var(r_2)\subseteq\var(r)$; or
        \item $\var(r_2)=\var(r)$ and $\var(r_1)\subseteq\var(r)$.
    \end{itemize}
    Then $\varphi[T,r]\equiv \varphi[\restr{T}{r_1},r_1]\wedge \varphi[\restr{T}{r_2},r_2]$.
  \end{cor}
\begin{proof}
  Let $N = \{r,r_1,r_2\}$. This set is a connex subset of $T$. Hence, $(T,N)$ is
  also a query plan. Observe that, because $\var(N) = \var(r)$ by the
  assumptions on $\var(r_1)$ and $\var(r_2)$, we have
  $\varphi[T,r] = \varphi[T,N]$. Further observe that the frontier of $N$ is
  $\{r_1, r_2\}$. Then, by  Lemma~\ref{lemma:formula-of-query-plan-is-conjunction-of-frontier-induced-trees-formulas} $\varphi[T,r] = \varphi[T,N] \equiv \varphi[T_1,r_1] \wedge \varphi[T_2,r_2]$ as desired.
\end{proof}

Finally, we note a property of query plans over binary relations.

\begin{lem}\label{lemma:binary-query-plan-two-variables-per-node}
    Let $(T,N)$ be a query plan such that $\varphi[T,N]$ is binary then $|\var(N)| \leq 2$ and any node of $T$ contains at  most two variables.
\end{lem}

\begin{proof}
  By definition $\var(N)=\free(\varphi[T,N])$, then necessarily
  $|\var(N)| \leq 2$ because $\varphi[T,N]$ is binary.  Now, suppose that
  there exists some node $g$ of $T$ that has more than two variables. This
  implies that the child which is the guard of $g$ also has more than two
  variables, and so on. Thus there is an leaf atom that uses more than two
  variables, which is not possible since $\varphi[T,N]$ is binary.
\end{proof}

At this point we are ready to prove that we can translate any binary query plan of the form $(T,r)$ into an equivalent \foTwo formula.

\begin{prop}\label{prop:from-rooted-query-plan-to-foTwo}
    For every QP $(T,r)$ such that $\varphi[T,r]$ is binary, there exists an \foTwo formula $\psi$ equivalent to $\varphi[T,r]$.
\end{prop}
\begin{proof}
    By Lemma~\ref{lemma:refined-query-plan} we may assume without loss of generality that for every node $n$ in $T$ with children $c_1,c_2$ either $\var(c_1)=\var(n)$ and $\var(c_2)\subseteq\var(n)$; or $\var(c_2)=\var(n)$ and $\var(c_1)\subseteq\var(n)$. Moreover, by Lemma~\ref{lemma:binary-query-plan-two-variables-per-node} we know that all nodes $n$ in $T$ have $\card{\var(n)} \leq 2$.
    We build the \foTwo formula $\psi$ equivalent to $\varphi[T,r]$ by induction on the height of $T$.
    
    The base case is where the height of $T$ is zero.  Then $T$ consists of just its root, $r$, which must be labeled by atom $a$ and $\varphi[T,r]=a$. Then it suffices to take $\psi \coloneqq a$. Clearly, $\psi \equiv \varphi[T,r]$ and, because $a$ is of arity at most two by assumption, $\psi\in\foTwo$.
    
    Now, for the inductive step. Assume $T$ has height $k>0$. By inductive hypothesis we have that the statement holds for height $k-1$.
    We distinguish the cases where $r$ has only one child or it has two children.
    \begin{itemize}
        \item First, if $r$ has only one child $c$. By induction hypothesis there exists a formula $\psi_c\in\foTwo$ such that $\psi_c \equiv\varphi[\restr{T}{c},c]$. Note that this implies that $\free(\psi_c)=\free(\varphi[\restr{T}{c},c])$. Take $\psi\coloneqq \exists \var(c)\setminus\var(r).\psi_c$. By Corollary~\ref{lemma:child-formula-projection} we have \[\psi= \exists \var(c)\setminus\var(r).\psi_c \equiv \exists \var(c)\setminus\var(r).\varphi[\restr{T}{c},c]\equiv\varphi[T,r].\] Clearly, $\psi\in\foTwo$, since it is a projection of $\psi_c\in\foTwo$.

        \item Second, assume that $r$ has two children, $r_1$ and $r_2$. We may
          assume w.l.o.g. that $\var(r_1) = \var(r)$ and
          $\var(r_2) \subseteq \var(r)$. By the inductive
          hypothesis there exist $\foTwo$ formulas
          $\psi_1\equiv\varphi[\restr{T}{r_1},r_1]$ and $\psi_2\equiv\varphi[\restr{T}{r_2},r_2]$. We
          next construct an \foTwo formula equivalent to $\psi_1\wedge
          \psi_2$. This suffices since by
          Corollary~\ref{lemma:children-formulas-conjunction}
          $\varphi[\restr{T}{r_1},r_1]\wedge \varphi[\restr{T}{r_2},r_2]\equiv \varphi[T,r]$ and hence
          $\psi_1\wedge \psi_2\equiv \varphi[T,r]$.
        
          Note that $\free(\psi_1)=\var(r_1) = \var(r)$ and $\free(\psi_2)=\var(r_2)\subseteq \var(r)$. Moreover, $\varphi[T,r]$ is binary. Thus, there exists a set of two variables, say $\{ x,y \}$, such that $\var(r)\subseteq \{ x,y \}$. Consequently, $\free(\psi_1)\subseteq\{ x,y \}$, $\free(\psi_2)\subseteq\{ x,y \}$. Because $\psi_1, \psi_2\in\foTwo$ we know that  when $\psi_1$ (or $\psi_2$) has exactly two free variables, $x$ and $y$ are the only variables used in $\psi_1$ (resp. $\psi_2$). When $\psi_1$ (resp. $\psi_2$) has fewer than two free variables it is possible that also  some other variable $z \not \in \{x,y\}$ occurs in $\psi_1$. However, because $\psi_1 \in \foTwo$ it is then always possible to rename bound variables to use only $x,y$. The same holds for $\psi_2$. We conclude that there exist
 formulas $\psi_1', \psi_2'\in\foTwo$ such that $\psi_1\equiv \psi_1'$, $\psi_2\equiv \psi_2'$ and $\psi_1',\psi_2'$ only use the variables $x$ and $y$. Thus $\varphi[\restr{T}{r_1},r_1]\wedge \varphi[\restr{T}{r_2},r_2]\equiv \psi_1'\wedge\psi_2'\in\foTwo$. \qedhere
    \end{itemize}
\end{proof}

Finally, we are able to show that every formula encoded by a binary query plan has an equivalent \foTwo formula.

\begin{prop}\label{prop:from-qp-to-foTwo}
    For every QP $(T,N)$ such that $\varphi[T,N]$ is binary, there exists an \foTwo formula $\psi$ equivalent to $\varphi[T,N]$.
\end{prop}
\begin{proof}
    By Lemma~\ref{lemma:refined-query-plan} we may assume w.l.o.g. that for every node $n$ in $T$ with two children $c_1$ and $c_2$ we have $\var(c_1)=\var(n)$ and $\var(c_2)\subseteq\var(n)$; or $\var(c_1)=\var(n)$ and $\var(c_2)\subseteq\var(n)$.
    Assume that the frontier of $N$ is $F=\{ r_1, \ldots, r_l \}$. By Proposition~\ref{prop:from-rooted-query-plan-to-foTwo},
    there are \foTwo formulas $\psi_1,\ldots,\psi_l$ equivalent to
    $\varphi[\restr{T}{r_1}, r_1], \ldots, \varphi[\restr{T}{r_l}, r_l]$, respectively.  By
    Lemma~\ref{lemma:formula-of-query-plan-is-conjunction-of-frontier-induced-trees-formulas}
    we have
    $\varphi[T,N]\equiv \varphi[\restr{T}{r_1}, r_1] \wedge \dots \wedge \varphi[\restr{T}{r_l}, r_l]
    \equiv \psi_1 \wedge \dots \wedge \psi_l$. To prove the proposition, it
    hence suffices to show that the conjunction $\psi_1\wedge\ldots\wedge\psi_l$
    is expressible in $\foTwo$. 
    For this we reason as follows.
 
    By assumption $\varphi[T,N]$ is binary, thus there exists a set of two
    variables, say $\{ x,y \}$, such that $\var(N)\subseteq\{ x,y \}$. Hence,
    $\var(r_i) \subseteq \var(N) \subseteq \{x,y\}$. Then, because $\psi_i$ is
    equivalent to $\varphi[T_i, r_i]$, also
    $\free(\psi_i)=\var(r_i) \subseteq \{x,y\}$ for $i=1,\ldots,l$. Similar to
    our argumentation in the proof of Proposition~\ref{prop:from-rooted-query-plan-to-foTwo}
    we can then exploit the fact that every $\psi_i \in \foTwo$ to obtain
    formulas $\psi_1',\ldots,\psi_l'\in\foTwo$ such that $\psi_i'\equiv \psi_i$
    and $\psi_i'$ only uses the variables $x$ and $y$, for $i=1,\ldots,l$. The
    formulas $\psi'_i$ are obtained from $\psi_i$ by renaming bound variables, which is only necessary if $\psi_i$ has fewer than two free variables but also mentions some bound variable $z \not \in \{x,y\}$. %
    Thus
    $\varphi[T,N]\equiv\varphi[\restr{T}{r_1}, r_1] \wedge \dots \wedge \varphi[\restr{T}{r_l},
    r_l]\equiv\psi_1'\wedge\ldots\wedge\psi_l'\in\foTwo$
\end{proof}

\subsection{From the two variable fragment of FO to binary free-connex CQ}
\label{sec:fo2-to-fcq}

We end this section by proving the other direction, namely, that every formula in \foTwo is equivalent to some binary free-connex CQ. Similar to the previous subsection, we use binary query plans. 

\begin{prop}\label{prop:from-foTwo-to-qp}
    For every \foTwo formula $\psi$ there exists a binary QP $(T,N)$ such that $\varphi[T,N]$ is equivalent to $\psi$.
\end{prop}
\begin{proof}
  Fix two distinct variables $x$ and $y$. Because $\foTwo$ formulas use at most
  two variables, we may assume without loss of generality that the only variables occurring in $\foTwo$
  formulas are $x$ and $y$.
  We obtain the proposition by proving a more detailed statement: for every
  $\psi\in\foTwo$ there exists a binary QP $(T,N)$ such that $\varphi[T,N] \equiv \psi$  and either:
  \begin{itemize}
  \item $N = \{r\}$ with $r$ the  root node $r$; or
  \item $N$ consists of three nodes, $N = \{r,c_1,c_2\}$ with $r$ the root of
    $T$ and $c_1,c_2$ its children, which are labeled by
    $\var(r) = \emptyset, \var(c_1) = \{x\}$, and $\var(c_2) = \{y\}$,
    respectively. This case only happens when $\psi$ is a cross product
    $\psi = \psi_1 \wedge \psi_2$.
  \end{itemize}
  Note that in the second case it follows from Lemma~\ref{lemma:formula-of-query-plan-is-conjunction-of-frontier-induced-trees-formulas} that $\psi = \psi_1 \wedge \psi_2 \equiv \varphi[T,N] \equiv \varphi[\restr{T}{c_1},c_1] \wedge \varphi[\restr{T}{c_2},c_2]$ since the frontier of $\{r,c_1,c_2\}$ is $\{c_1, c_2\}$. 

  We prove the stronger statement by induction on $\psi$.  

  \begin{itemize}
  \item When $\psi$ is an atom then the result trivially follows by creating GJT
    $T$ with a single node $r$, labeled by the atom, and fixing $N = \{r\}$.
  \item When $\psi = \exists z. \psi'$ with $z \in \{x,y\}$ we may assume
    w.l.o.g. that $z \in \free(\psi')$: if not then $\psi \equiv \psi'$ and the
    result follows directly from the induction hypothesis on $\psi'$. So, assume
    $z\in \free(\psi')$. By induction hypothesis we have a binary QP $(T', N')$
    equivalent to $\psi'$.  We next consider two cases.
  \begin{itemize}
  \item $N'$ is a singleton, $N'=\{r'\}$ with $r'$ the root of $T'$. 
    We create the claimed binary QP $(T,N)$ as follows: let $T$ be obtained by
    adjoining a new root $r$ to $T'$; label $r$ by $\var(r') \setminus \{z\}$;
    and make $r'$ the only child of $r$. Take $N = \{r\}$. Then $(T,N)$ remains
    binary. Equivalence of $\varphi[T,N]$ to $\psi$ follows by induction
    hypothesis and Lemma~\ref{lemma:child-formula-projection}.
  \item Otherwise $N' = \{r', c_1, c_2\}$ with $c_1$ and $c_2$ the children of
    $r'$ in $T'$; $\var(r') = \emptyset$; $\var(c_1) =\{x\}$;
    $\var(c_2) = \{y\}$; and
    $\psi'\equiv \varphi[\restr{T'}{c_1}, c_1] \wedge
    \varphi[\restr{T'}{c_2},c_2]$. Since $z \in \{x,y\}$ we have either
    $\var(c_1) = \{z\}$ or $\var(c_2) = \{z\}$ (but not both). Assume
    w.l.o.g. that $\var(c_1) = \{z\}$; the other case is similar.  We create the
    claimed QP $(T,N)$ as follows: let $T$ be obtained by removing $r'$ from
    $T'$ and replacing it with a new root, $r$, which has children $c_1$ and
    $c_2$ and which is labeled by $\var(c_2)$. Since $\var(c_1)$ and $\var(c_2)$
    were disjoint, this new tree satisfies the guardedness and connectedness
    properties, and remains binary. Take $N = \{r\}$. Then
    \[ \var(N) = \var(c_2) = \var(N') \setminus \var(c_1) = \free(\psi')
      \setminus \{z\} = \free(\psi),\] as desired. To see why
    $\varphi[T,N] \equiv \psi$ first define $M = \{r, c_1, c_2\}$. By
    lemma~\ref{lem:connex-subset-extra-projection} we obtain that
    $\varphi[T,N] \equiv \exists z. \varphi[T,M]$. Furthermore, since the
    frontier of $M$ is $\{c_1,c_2\}$ we know from
    Lemma~\ref{lemma:formula-of-query-plan-is-conjunction-of-frontier-induced-trees-formulas}
    that
    $\varphi[T,M] \equiv \varphi[\restr{T}{c_1},c_1] \wedge
    \varphi[\restr{T}{c_2}, c_2] = \varphi[\restr{T'}{c_1},c_1] \wedge
    \varphi[\restr{T'}{c_2}, c_2] \equiv \psi'$. Hence
    $\varphi[T,N] \equiv \exists z. \psi' = \psi.$

  \end{itemize}    

\item Otherwise $\psi$ is a conjunction, $\psi = \psi_1 \wedge \psi_2$. By induction
  hypothesis, we have binary QPs $(T_1, N_1)$ and $(T_2, N_2)$ equivalent to $\psi_1$
  and $\psi_2$, respectively. We may assume w.l.o.g.\@ that $T_1$ and $T_2$ have
  no nodes in common, as we can always rename nodes otherwise. We may further
  assume w.l.o.g.\@ that
  $\var(T_1) \cap \var(T_2) \subseteq \var(N_1) \cap \var(N_2)$, i.e., that all
  variables that are mentioned in both $T_1$ and $T_2$ are also in
  $\var(N_1) \cap \var(N_2)$. Indeed, for every variable
  $z \in (\var(T_1) \cap \var(T_2)) \setminus (\var(N_1) \cap \var(N_2))$ we can
  simply rename all occurrences of $z$ in $T_1$ to a new unique variable $z'$
  that does not occur anywhere in $T_2$. Assume that we apply this reasoning to
  all such variables $Z$ and let $T'_1$ be the result of applying this renaming
  on $T_1$. Note that $N_1$ is still a connex subset of $T'_1$. Moreover,
  $\varphi[T_1,N_1] \equiv \varphi[T'_1, N_1]$: essentially we have only renamed
  bound variables in the formula $\varphi[T_1,N_1]$ to obtain
  $\varphi[T'_1,N_1]$ . So we can continue our reasoning with $(T'_1,N_1)$
  instead of $(T_1, N_1)$ and the former has no variables in common with $T_2$,
  except those in $\var(N_1) \cap \var(N_2)$.

  By our inductive hypothesis, $N_1$ and $N_2$ are either singletons, or contain
  exactly three nodes. We consider three cases.
  \begin{itemize}
  \item Both $N_1$ and $N_2$ are singletons, say $N_1 = \{r_1\}$ and
    $N_2 = \{r_2\}$ with $r_1,r_2$ the roots of $T_1,T_2$ respectively. In
    particular, because $\varphi[T_i, N_i]$ is equivalent to $\psi_i$ for
    $i=1,2$ we necessarily have $\var(r_i) = \free(\psi_i)$.  Construct the
    claimed binary QP $(T,N)$ by taking  the (disjoint) union of $T_1$ and $T_2$
    and adjoining a new root node $r$ with children $r_1, r_2$ such that
    $\var(r) = \var(r_1) \cap \var(r_2)$.  Since the only variables shared
    between $T_1$ and $T_2$ are in
    $\var(N_1) \cap \var(N_2) = \var(r_1) \cap \var(r_2)$ it follows that $T$
    satisfies the connectedness property, and is hence a GJT. Moreover, let
    $N = \{r,r_1,r_2\}$. Clearly
    \[ \var(N) = \var(r_1) \cup \var(r_2) = \free(\psi_1) \cup \free(\psi_2) =
      \free(\psi) \subseteq \{x,y\}.\] Correctness follows by
    Lemma~\ref{lemma:formula-of-query-plan-is-conjunction-of-frontier-induced-trees-formulas},
    since $r_1$ and $r_2$ are the frontier nodes of $N$:
    \[\varphi[T,N] \equiv \varphi[\restr{T}{r_1},r_1] \wedge
      \varphi[\restr{T}{r_2},r_2] = \varphi[T_1,r_1] \wedge \varphi[T_2,r_2]
      \equiv \psi_1 \wedge \psi_2. \]
  \item Both $N_1$ and $N_2$ contain exactly three nodes, say
    $N_1 = \{r_1, c^1_1, c^1_2\}$ and $N_2 = \{r_2, c^2_1, c^2_2\}$ with $r_i$
    the root node of $T_i$ and $c^i_j$ the two children of $r_i$ in $T_i$, for
    $i,j \in \{1,2\}$. Furthermore, $\var(r^i) = \emptyset$;
    $\var(c^i_1) = \{x\}$ and $\var(c^i_2) = \{y\}$ for $i=1,2$ and $\psi_1$ and
    $\psi_2$ must themselves be conjunctions,
    $\psi_1 = \psi^1_1 \wedge \psi^1_2$ and $\psi_2 = \psi^2_1 \wedge
    \psi^2_2$.  By induction hypothesis, 
    \begin{equation}
      \label{eq:detailed-1}
      \psi_i \equiv \varphi[T_i, N_i] \equiv \varphi[\restr{T_i}{c^i_1}, c^i_1] \wedge \varphi[\restr{T_i}{c^i_2}, c^i_2] 
    \end{equation}
    We construct the claimed QP $(T,N)$ by taking $T$ as follows, and $N = \{r, d_1, d_2\}$. 

\tikzset{
itria/.style={
  draw,dashed,shape border uses incircle,
  isosceles triangle,shape border rotate=90,yshift=-0.80cm, minimum height=0.6cm,
        minimum width=0.4cm,
        shape border rotate=#1,
        isosceles triangle stretches,
        inner sep=0pt},
rtria/.style={
  draw,dashed,shape border uses incircle,
  isosceles triangle,isosceles triangle apex angle=90,
  shape border rotate=-45,yshift=0.2cm,xshift=0.5cm},
ritria/.style={
  draw,dashed,shape border uses incircle,
  isosceles triangle,isosceles triangle apex angle=110,
  shape border rotate=-55,yshift=0.1cm},
letria/.style={
  draw,dashed,shape border uses incircle,
  isosceles triangle,isosceles triangle apex angle=110,
  shape border rotate=235,yshift=0.1cm}
}

 \tikzset{
        expand bubble/.style={
            preaction={draw,line width=1.4pt},
            gray!20,fill,draw,line width=2pt,
        },
    }

    \begin{center}
        \begin{tikzpicture}[sibling distance=0.7cm, level distance =0.4cm]
            \node (r1) {$r; \emptyset$} [sibling distance=2.5cm]
            child {
              node (d1) {$d_1; \{x\}$} [sibling distance=1.25cm,level distance=0.8cm]
              child {
                node {$c^1_1$}
                { node [itria,yshift=-6pt] {$T^1_1$}}
              }
              child {
                node {$c^2_1$}
                { node [itria,yshift=-6pt] {$T^2_1$}}
              }
            }
            child {
              node (d2) {$d_2; \{y\}$} [sibling distance=1.25cm, level distance=0.8cm]
              child {
                node {$c^1_2$}
                { node [itria,yshift=-6pt] {$T^1_2$}}
              }
              child {
                node {$c^2_2$}
                { node [itria,yshift=-6pt] {$T^2_2$}}
                }
              };

              \begin{pgfonlayer}{background}
                \path[expand bubble]plot [smooth cycle,tension=1]
                coordinates {(r1.north) (d1.south west) (d2.south east) };
              \end{pgfonlayer}

        \end{tikzpicture}
      \end{center}
      
      Here $r$, $d_1$ and $d_2$ are new nodes with $\var(r)= \emptyset$;
      $\var(d_1) = \{x\}$, and $\var(d_2) = \{y\}$, and
      $T^i_j = \restr{T_i}{c^i_j}$ for $i = 1,2$. It is straightforward to
      verify that $T$ satisfies the guardedness and connectedness properties. It
      is hence a GJT.  To see why $\varphi[T,N] \equiv \psi$, first define
      $M = N \cup \{c^1_1, c^1_2, c^2_1, c^2_2\}$. Note that
      $\var(M) = \var(N)$. Hence, by
      Lemma~\ref{lem:connex-subset-extra-projection}, we have
      $\varphi[T,N] \equiv \varphi[T,M]$. Then, because the frontier of $M$ is
      $\{c^1_1, c^1_2, c^2_1, c^2_2\}$ we have by
      Lemma~\ref{lemma:formula-of-query-plan-is-conjunction-of-frontier-induced-trees-formulas}
      that $\varphi[T,M] \equiv \bigwedge_{i,j \in \{1,2\}} \varphi[\restr{T}{c^i_j}, c^i_j] \equiv \bigwedge_{i,j \in \{1,2\}} \varphi[\restr{T_i}{c^i_j}, c^i_j] \equiv \psi_1 \wedge \psi_2$,
    where the last equivalence follows from \eqref{eq:detailed-1}.

  \item One of $N_1$ and $N_2$ has exactly three nodes; the other has one node. Assume w.lo.g. that $N_1 = \{r_1\}$ and $N_2 = \{r_2, c_1, c_2\}$. We then construct the claimed binary QP $(T,N)$ as follows. If $\var(r_1) = \{x,y\}$  then we take $T$ as follows, where $r, r'$ are new nodes with $\var(r) = \var(r')  = \var(r_1) = \{x,y\}$ and $N = \{r\}$. Moreover, $T^2_j =  \restr{T_2}{c_j}$ for $j = 1,2$.
    \begin{center}
        \begin{tikzpicture}[sibling distance=0.7cm, level distance =0.4cm]
            \node (r1) {$r$} [sibling distance=2.5cm]
            child {
              node {$r'$} [sibling distance=1.25cm,level distance=0.7cm]
              child {
                node {$r_1$}
                { node [itria,yshift=-6pt] {$T_1$}}
              }
              child {
                node {$c_1$}
                { node [itria,yshift=-6pt] {$T^2_1$}}
              }
            }
            child {
                node {$c_2$}
                { node [itria,yshift=-6pt] {$T^2_2$}}
              };

              \begin{pgfonlayer}{background}
                \path[expand bubble]plot [smooth cycle,tension=1]
                coordinates {(r1.north) (r1.south west) (r1.south east) };
              \end{pgfonlayer}

        \end{tikzpicture}
      \end{center}
When $\var(r_1) \not = \{x,y\}$ then either $\var(r_1) \subseteq \var(c_1)$ or $\var(r_1) \subseteq \var(c_2)$ (or both). If $\var(r_1) \subseteq \var(c_1)$ then we take $T$ as above but set $\var(r) = \emptyset$; $\var(r') = \var(c_1) = \{x\}$; and $N = \{r, r_1, c_2\}$. The case where $\var(r_1) \subseteq \var(c^2_2)$ but $\var(r_1) \not \subseteq \var(c_1)$ is similar. In all cases, equivalence with $\psi$ follows from the same reasoning used earlier. \qedhere
  \end{itemize}
  
\end{itemize}

\end{proof}

\section{The free-connex fragment of matlang queries}\label{sec:free-connex-matlang}

Define $\fcLang$ to be the class of all \lang expressions generated by the grammar:
\[ 
e \ \ \Coloneqq \ \ \mA \ \mid \ \ones^\alpha \ \mid \ \Iden^\alpha \ \mid \ e^T \ \mid \ e_1\times e_2 \ \mid \ e_1\kprod e_2 \ \mid \ e_1 \cdot v_2 \ \mid \ v_1 \cdot e_2 
\]
where $v_1$ and $v_2$ are \fcLang expressions with type $(\alpha, 1)$ or $(1, \alpha)$. In other words, matrix multiplication $e_1 \cdot e_2$ is only allowed when at least one of $e_1$ or $e_2$  has a row or column vector~type.

In this section, we prove the following correspondence.

\begin{thm}\label{theo:fcLang-equiv-foTwo}
    \fcLang and \foTwo are equally expressive.
\end{thm}

Combined with Theorem~\ref{theo:free-connex-binary-cq-equiv-foTwo} we  obtain the correspondence with binary free-connex~CQ:

\begin{cor}\label{cor:fcLang-equiv-free-connex-cq}
    \fcLang and binary free-connex CQs are equally expressive.
  \end{cor}

We prove both directions of Theorem~\ref{theo:fcLang-equiv-foTwo} separately.

\subsection{From \foTwo to \fcLang} \label{sec:fotwo-to-fclang}

We first simulate \foTwo queries by means of \fcLang queries. In order to achieve this, free variables of formulas need to describe row and column indexes of expressions. Moreover, it is crucial to know which free variable will correspond to the row index and which free variable will correspond to the column index. For this reason we assume  the existence of an arbitary \emph{canonical order} $\prec$ over $\fo$ variables, where $x \prec y$ denotes that $x$ strictly precedes $y$ in this order.

Fix a vocabulary $\Voc$. We first define the notion of when a \sumLang
expression simulates a \emph{formula} over $\Voc$ instead of a query.
Let $\psi\in\foplus$ over $\Voc$. Let $\Mat$ be a relational-to-matrix schema encoding on $\Voc$ such that $\Mat\vdash \psi\colon\tau$. We say that \sumLang sentence \emph{$e$  simulates $\psi$ w.r.t. $\Mat$} if the \sumLang query $\Mat(\ans)\coloneqq e$ simulates the query $\ans(\x) \gets \psi$ w.r.t. $\Mat$, where
\begin{enumerate}
    \item $\x$ is a tuple containing exactly the free variables of $\psi$, without repetition and canonically ordered: i.e., if $\x = (x_1,x_2)$ then $x_1 \prec x_2$.
    \item $\ans$ is an arbitrary relation symbol, not occurring in $\varphi$, of the same arity as $\free(\varphi)$ such that $\type(\Mat(\ans))=\tau(\x)$.
\end{enumerate}

So, in the second bullet if $\varphi$ has two free variables then $\type(\Mat(\ans)) = (\tau(x_1), \tau(x_2))$; if $\varphi$ has one free variable then $\type(\Mat(\ans)) = (\tau(x),1)$; and if $\varphi$ has zero free variables then $\type(\Mat(\ans)) = (1,1)$. In particular, due to well-typedness, when $\varphi$ has two free variables $\Mat(\ans)$ is consistent with dimensions $\tau(x_1)^\db \times \tau(x_2)^\db$; with $\tau(x)^\db \times 1$ when $\varphi$ has one free variable; and with $1 \times 1$ when $\varphi$ has no free variable, for every database $\db$ consistent with $\Mat$.
Further, we note that, while relation symbol $\ans$ is fixed arbitrarily, it is clear that if $e$ simulates $\varphi$ w.r.t. one particular choice of $\ans$, then it simulates it w.r.t. all valid choices.

It is important to stress that (1) in the query $\ans(\x) \gets \varphi$ no variable is ever repeated in the head; and (2)  if $\varphi$ hence has a single free variable, and hence computes a unary relation, then the simulating \fcLang expression will always simulate it by means of a \emph{column} vector. We fix this merely to simplify the proof, and could also have fixed it to be a row vector instead.

\begin{prop}\label{prop:fotwo-to-fc-lang}
    Let $\psi\in\foTwo$ over $\Voc$.
    Let $\Mat$ be a relational-to-matrix schema encoding on $\Voc$ such that
    $\Mat\vdash \psi\colon\tau$. Then there exists an expression $e\in \fcLang$
    that simulates $\psi$ w.r.t. $\Mat$.
\end{prop}

\begin{proof}
  Fix two distinct variables $x_1 \prec x_2$. We may assume without loss of generality that the
  only variables occurring in $\psi$ are $x_1,x_2$. We prove the statement by induction on $\psi$. For every relation symbol $R$, we denote $\Mat(R)$ simply by $\mR$. In particular, $\mans = \Mat(\ans)$.

  First, when $\psi=a$ and $a$ is a relational atom we have multiple cases.
  \begin{itemize}
  \item Atom $a$ is a binary atom, and its variables occur in canonical order, i.e. $a = R(x_1,x_2)$. Then take $e := \mR$.
  \item Atom $a$ is a binary atom, but its variables occur in reverse canonical order, i.e., $a = R(x_2,x_1)$. Then take $e := \mR^T$.
    
  \item
    Atom $a$ is a binary atom that mentions the same variable twice, 
    $a = R(y,y)$ for some $y \in \{x_1,x_2\}$. Note that $\mans = \Mat(\ans)$ will be of column vector type $(\tau(y), 1)$ since $\x = (y)$. Also note that, because of well-typedness, $\mR$ must be of type $(\tau(y), \tau(y))$, i.e., $\mR$ is a square matrix. Then construct $e$ as follows. First consider the expression $e' := \Iden^{\tau(y)} \kprod\mR.$ This selects from $\mR$ all the entries on the diagonal, and sets to $\kzero$ all other entries. Then take $e := e'\cdot \ones^{\tau(y)}$, which converts the diagonal entries into a column vector.
    
  \item Atom $a$ is a unary atom of the form $a = R(y)$ with $y \in \{x_1,x_2\}$. We distinguish two subcases.
    \begin{itemize}
    \item $\mR$ has column vector type. In that case, take $e := \mR$.
    \item $\mR$ has row vector type. Because $\mans$ always has
      column vector type by definition, we convert the input row vector into a
      column vector by taking $e := \mR^T$.
    \end{itemize}
  \item Atom $a$ is a nullary atom, $a =R()$. Then $\mR$ necessarily has type $1 \times 1$, as does $\mans$. So, it suffices to take $e := \mR$.
  \end{itemize}
  
  Second, when $\psi$ is comparison of the form $x \leq \alpha$ then the simulating expression $e$ is $\ones^\alpha$. This expression has type $(\tau(x),1)$, which conforms to the desired type of the definition since $\tau(x) = \alpha$.
  
  Third, when $\psi=\exists y.\psi_1$ we may assume w.l.o.g. that
  $y \in \free(\psi_1)$ since otherwise $\psi \equiv \psi_1$ and the result
  follows directly from the induction hypothesis. So, assume
  $y \in \free(\psi_1) \subseteq \{x_1,x_2\}$. There are two cases.
  \begin{itemize}
  \item If $\free(\psi_1)=\{x_1,x_2\}$ then %
    $e_1\colon (\tau(x_1),\tau(x_2))$ simulates $\psi_1$. If $y = x_2$ then  we take $e = e_1\cdot\ones^{\tau(y)}$ to simulate $\psi$ w.r.t. $\Mat$. If $y = x_1$ then we take    $e =e_1^T\cdot\ones^{\tau(y)}$ to simulate $\psi$ w.r.t. $\Mat$.
    Note that in both cases the simulating expression has type
    $(\tau(x),1)$ with $x$ the unique variable in
    $\{x_1,x_2\} \setminus \{y\}$, as desired.
  \item If $\free(\psi_1)=\{ y \}$, then $e_1\colon(\tau(y),1)$ simulates $\psi_1$ and thus $e_1^T \cdot \ones^{\tau(y)}\colon (1,1)$ simulates $\psi$ w.r.t. $\Mat$. 
  \end{itemize}

  The fourth and final case is when $\psi=\psi_1\wedge\psi_2$.
  By inductive hypothesis there exists \fcLang expressions $e_1,e_2$ that simulate $\psi_1$ and $\psi_2$ w.r.t. $\Mat$, respectively. %
  If $\free(\psi_1)=\free(\psi_2)$ then the simulating expression is simply
  $e_1\kprod e_2$. It will be of type $(\tau(x_1),\tau(x_2))$,
  $(\tau(x_1),1)$, $(\tau(x_2),1)$ or $(1,1)$ depending if the free variables
  are $\{ x_1,x_2 \}$, $\{ x_1\}$, $\{ x_2 \}$ or $\emptyset$, respectively. When $\free(\psi_1) \not = \free(\psi_2)$, we make the following case distinction.
  \begin{enumerate}
  \item If $\free(\psi_1)=\{ x_1,x_2 \}$ then $e_1\colon (\tau(x_1),\tau(x_2))$.
    \begin{enumerate}
    \item If $\free(\psi_2)=\{ x_1 \}$ then $e_2\colon (\tau(x_1),1)$ and we take  $e := e_1\kprod \left(e_2\cdot\left(\ones^{\tau(x_2)}\right)^T\right).$
    \item If $\free(\psi_2)=\{ x_2 \}$ then $e_2\colon (\tau(x_2),1)$ and we take $e := e_1\kprod \left(\ones^{\tau(x_1)}\cdot e_2^T\right).$ 
    \item If $\free(\psi_2)=\emptyset$ then $e_2\colon (1,1)$ and the simulating expression is $e_2\times e_1$.
    \end{enumerate}
  \item If $\free(\psi_1)=\{ x_1 \}$ then $e_1\colon (\tau(x_1),1)$.
    \begin{enumerate}
    \item If $\free(\psi_2)=\{ x_1,x_2 \}$ then we reason analogous to case (1)(b).
    \item If $\free(\psi_2)=\{ x_2 \}$ then $e_2\colon (\tau(x_2),1)$ and define $e := \left( e_1\cdot\left(\ones^{\tau(x_2)}\right)^T \right)\kprod \left(\ones^{\tau(x_1)}\cdot e_2^T\right).$
    \item If $\free(\psi_2)=\emptyset$ then $e_2\colon (1,1)$ and the simulating expression is $e_2\times e_1$.
    \end{enumerate}
  \item If $\free(\psi_1)=\{ x_2 \}$ we reason analogous to case (2).
  \item If $\free(\psi_1)=\emptyset$ then $\free(\psi_2) \not = \emptyset$ and we reason analogous to the cases above where $\free(\psi_2) = \emptyset$ but $\free(\psi_1) \not = \emptyset$.\qedhere
  \end{enumerate}
\end{proof}

Finally, we extend Proposition~\ref{prop:fotwo-to-fc-lang} from formulas to queries. 

\begin{cor}
  Let $Q$ be an \foTwo query over $\Voc$ and let $\Mat$ be a relational-to-matrix schema encoding on $\Voc(Q)$ such that
    $\Mat\vdash Q\colon\tau$. Then there exists $\fcLang$ query $\mQ$
    that simulates $Q$ w.r.t. $\Mat$.
\end{cor}
\begin{proof}
    Let $Q\colon H\gets\psi$ be a \foTwo query over $\Voc$ and let $\Mat$ be a relational-to-matrix schema encoding on $\Voc(Q)$ such that $\Mat\vdash \psi\colon\tau$. By Proposition~\ref{prop:fotwo-to-fc-lang} there exists $e\in\fcLang$
    that simulates $\psi$ w.r.t. $\Mat$. Let $\Mat(H) = \mH$. We derive a $\fcLang$ query $\mQ\colon \mH := \e$ that simulates $Q$ as follows.
    \begin{itemize}
        \item When $\x=(x,y)$, if $x \prec y$ then define $\e:=e$, otherwise define $\e:=e^T$.
        \item When $\x=(x,x)$ then $Q$ outputs the encoding of a square matrix, where the value of diagonal entry $(x,x)$ is computed by $\psi(x)$. To simulate this query, we observe that $e\colon (\tau(x),1)$ is a column vector that simulates $\psi$. We first compute the square matrix of type $(\tau(x),\tau(x))$ with the vector $e$ in all its columns as $e\cdot\left(\ones^{\tau(x)}\right)^T.$ Then set all non-diagonal entries of the former expression to zero, by taking $\e\coloneqq\Iden^{\tau(x)} \kprod \left( e\cdot\left(\ones^{\tau(x)}\right)^T \right).$
        \item When $\x= (x)$ we have $e\colon (\tau(x),1)$. If $\mH\colon (\alpha, 1)$ define $\e:=e$ otherwise if $\mH\colon (1, \alpha)$ define $\e:=e^T.$
        \item When $\x=()$ we have $e\colon (1,1)$. Then define $\e:=e$. \qedhere
        \end{itemize}
      \end{proof}

\subsection{From \fcLang to \foTwo}\label{sec:fclang-to-fotwo} To establish the converse direction of Theorem~\ref{theo:fcLang-equiv-foTwo} we first define when an \foplus \emph{formula} simulates a \sumLang sentence instead of a query. 
  Let $e\colon(\alpha,\beta)$ be a \sumLang sentence over $\Sch$ and let $\Rel$ be a matrix-to-relational schema encoding on $\Sch$.  We say that $\foplus$ formula \emph{$\psi$  simulates $e$ w.r.t. $\Rel$} if $\psi$ has exactly two free variables  $x \prec y$ and for every matrix instance $\I$ and all $i,j$ we    \[\fssem{\psi}{\Rel(\I)}(x\mapsto i, y\mapsto j) = 
    \begin{cases}
      \sem{e}{\I}_{i,j} & \text{ if } 1 \leq i \leq \I(\alpha), 1 \leq j \leq \I(\beta) \\
      \kzero & \text{ otherwise }
    \end{cases}
.\] 

Let $\foTwoEq$ denote the two-variable fragment of $\focon$ where  equality atoms are allowed (subject to being safe), as long as only two variables are used in the entire formula. For proving the direction from \fcLang to \foTwo, we first do the proof by using equality, and later show how to remove it. 

\begin{prop}\label{prop:from-fcLang-exp-to-foTwo-formulas}
  Let $e$ be a \fcLang expression over $\Sch$ and let $\Rel$ be a matrix-to-relational schema encoding on $\Sch$. %
  There exists $\psi_e \in\foTwoEq$ that simulates $e$ w.r.t. $\Rel$.%
\end{prop}

\begin{proof}
  Let $e\in\fcLang$ and fix two distinct variables $x \prec y$. We build the  formula $\psi_e\in\foTwoEq$ that simulates $e$ w.r.t. $\Rel$ by induction on $e$. In particular, we build $\psi_e$ such that $\free(\psi_e) = \{x,y\}$.
  For a matrix symbol $\mA$, let us denote $\Rel(A)$ simply as $A$. 

  \begin{enumerate}
  \item If $e=\mA$ then take:
    \begin{itemize}
    \item $\psi_e\coloneqq A(x,y)$ when $\type(\mA)=(\alpha,\beta)$; $\type(\mA)=(1,1)$; $A$ is binary and $\type(\mA)=(\alpha,1)$ or $\type(\mA)=(1,\beta)$.
    \item $\psi_e\coloneqq A(x) \wedge y \leq 1$ if $A$ is unary and $\type(\mA)=(\alpha,1)$.
    \item $\psi_e\coloneqq x\leq 1 \wedge A(y)$ if $A$ is unary and $\type(\mA)=(1,\beta)$.
    \end{itemize}

  \item If $e=\ones^\alpha\colon (\alpha,1)$ then take  $\psi_e\coloneqq x\leq \alpha \wedge y\leq 1$. 

  \item If $e=\Iden^\alpha\colon (\alpha,\alpha)$ then take $\psi_e    \coloneqq x\leq \alpha \wedge y \leq \alpha \wedge x = y$.  This will be the only case where we include an equality atom.
  \item If $e=(e')^T\colon(\beta,\alpha)$ then take $\psi_e \coloneqq \psi_{e'}[x\leftrightarrow y]$ where  $\psi_{e'}[x\leftrightarrow y]$ denotes  the formula obtained from $\psi_{e'}$ by swapping $x$ and $y$. (I.e., simultaneously replacing all occurrences of $x$---free and bound---with $y$ and all occurrences of $y$ with $x$.)
  \item If $e = e_1 \times e_2$ with $e_1 \colon (1,1)$ and $e_2\colon (\alpha, \beta)$ we reason as follows. First observe that because $\psi_{e_1}$ simulates $e_1$ and because $e_1$  always outputs scalar matrices, $\fssem{\psi_{e_1}}{\Rel(\I)}(x \mapsto i, y \mapsto j) = \kzero$ whenever $i\not =1$ or $j \not = 1$. We conclude that therefore, $\exists x,y. \psi_{e_1}$ will always return $\sem{e_1}{\I}(1,1)$ when evaluated on $\Rel(\I)$. Hence, we take $\psi_{e} \coloneqq \left(\exists x,y. \psi_{e_1}\right) \wedge \psi_{e_2}$.
  \item If $e = e_1 \kprod e_2$ then take $\psi_{e} \coloneqq \psi_{e_1} \wedge \psi_{e_2}$.
  \item If $e = e_1 \cdot v_2$ with $e_1\colon (\alpha,\beta)$ and with $v_2\colon (\beta, 1)$ evaluating to a column vector, we reason as follows. First observe that because $\psi_{v_2}$ simulates $v_2$ and because $v_2$  always outputs column vectors, $\fssem{\psi_{v_2}}{\Rel(\I)}(x \mapsto i, y \mapsto j) = \kzero$ whenever $j \not = 1$. Let $\varphi_{v_2} = \psi_{v_2}[x \leftrightarrow y]$ be the $\foTwoEq$ formula obtained from $\psi_{v_2}$ by simultaneously replacing all occurrences (free and bound) of $x$ by $y$, and all occurrences of $y$ by $x$. Then $\fssem{\varphi_{v_2}}{\Rel(\I)}(y \mapsto i,  x \mapsto j) = \kzero$ whenever $j \not = 1$. 
    We conclude that therefore
    \begin{equation}
      \label{eq:prop:from-fcLang-exp-to-foTwo-formulas-1}
      \sem{v_2}{\I}_{i,1} = \fssem{\exists x. \varphi_{v_2}}{\Rel(\I)}(y\mapsto i) \text{ for all } \I \text{ and } i. 
    \end{equation}
    Next observe that $e$  also always outputs a column vector, and that
    \[   \sem{e}{\I}_{i,1} = \sem{e_1\cdot v_2}{\I}_{i,1} = \bigoplus_{k}\, \sem{e_1}{\I}_{i,k} \kprod \sem{v_2}{\I,\mval}_{k,1}
    \]
    Take $\psi' = \exists y. \left(\psi_{e_1}(x,y) \wedge \varphi'_{v_2}(y)\right)$ where  $\varphi'_{v_2} = \exists x. \varphi_{v_2}$. Then $\psi'$ is in $\foTwoEq$. Because $\psi_{e_1}$ simulates $e_1$ and because of \eqref{eq:prop:from-fcLang-exp-to-foTwo-formulas-1} it follows that $\sem{e}{\I}_{i,1} = \fssem{\psi'}{\Rel(\I)}(x \mapsto i)$ for all $\I$ and $i$. Hence, we take $\psi_{e} := \psi' \wedge y \leq 1$ to simulate $e$.

  \item The case $e = v_1 \cdot e_2$ is similar to the previous case. \qedhere
  \end{enumerate}
 \qedhere
\end{proof}

To establish our desired result at the level of queries we require the following
two lemmas that will help to remove the equalities from \foTwoEq. The proofs can be found in Appendix~\ref{sec:app-free-connex-matlang}.

\begin{lem}\label{lem:fotwo-unify}
  Let $\varphi$ be a $\foTwo$ formula (hence, without equality) using only the
  distinct variables $x,y$ such that
  $\free(\varphi) \cap \{x,y\} \not = \emptyset$. Let $\psi$ be the $\foTwoEq$
  formula $\exists x. (\varphi \wedge x=y)$. Note that $\free(\psi) =
  \{y\}$. There exists an $\foTwo$ formula $\psi'$ equivalent to $\psi$.
\end{lem}

\begin{lem}
  \label{lem:fotwoeq-equiv-equality-toplevel}
  For every \foTwoEq formula $\psi$ there exists an equivalent \foTwoEq formula $\psi'$ where equality atoms only occur at the top level: for every subformula $\exists z. \phi'$ of $\psi'$ it holds that $\phi' \in \foTwo$ does not contain equality atoms.
\end{lem}

Finally, we can prove the direction from \fcLang to \foTwo.

\begin{cor}\label{cor:from-fcLang-to-foTwo-binary-encoding}
  Let $\mQ$ be an \fcLang query over $\Sch$ and let $\Rel$ be a matrix-to-relational schema encoding on $\Sch$. There exists an $\foTwo$ query $Q$ that simulates $\mQ$ w.r.t. $\Rel$.

\end{cor}
  
\begin{proof}
  Let $\mH\coloneqq e$ be an \fcLang query and let $H = \Rel(\mH)$.  By
  Proposition~\ref{prop:from-fcLang-exp-to-foTwo-formulas} there exists
  $\psi\in\foTwoEq$ that simulates $e$ w.r.t. $\Rel$. By construction, $\psi$
  has two distinct free variables. Assume that $x, y$ are these two distinct
  variables, and $x \prec y$.
  There are three cases to consider.

  (1) $H$ is a binary relation.  By
  Lemma~\ref{lem:fotwoeq-equiv-equality-toplevel} there exists
  $\psi' \in \foTwoEq$ that is equivalent to $\psi$ and where equality atoms only
  occur at the top level. Because $\psi$ has $\{x,y\}$ as free
  variables, so does $\psi'$. If no equality atom occurs in $\psi'$ then we take $Q$ to be the query $H(x,y) \gets \psi'$. If some equality atom does occur in $\psi'$ then we may observe w.l.o.g. that the equality is $x = y$ since the equalities $x = x$ and $y=y$ are trivial and can always be removed. As such, $\psi' \equiv  \psi'' \wedge x = y$ with $\psi'' \in \foTwo$ and at least one of $x,y$ appearing free in $\psi''$. Assume w.l.o.g. that $x$ occurs free in $\psi''$. Then the query $H(x,x) \gets \exists y. \psi'$ simulates $\mQ$ w.r.t. $\Rel$. Although this is formally a query in $\foTwoEq$, we can we can next apply
  Lemma~\ref{lem:fotwo-unify} to $\psi''$ to obtain an equivalent formula in $\foTwo$ for the body $\exists y. \psi' \equiv \exists y. (\psi'' \wedge x=y)$, yielding the desired result.

  (2) $H$ is a unary relation. Assume that $\mH$ is of column vector type
  $(\alpha, 1)$; the reasoning when $\mH$ has row vector type is similar. Take
  $\varphi = \exists y. \psi$. Because $\psi$ simulates $e$ and because $e$
  always outputs column vectors,
  $\fssem{\psi}{\Rel(\I)}(x \mapsto i, y \mapsto j) = \kzero$ whenever
  $j \not = 1$, for every $\I$. We conclude that therefore
  $\sem{e}{\I}(i,1) = \fssem{\exists y. \psi}{\Rel(\I)}(x\mapsto i)$ for all $i$
  and $\I$. As such, the query $Q'\colon H(x) \gets \varphi$ simulates
  $\mQ$. While $Q'$ is a query in $\foTwoEq$, we can next apply
  Lemma~\ref{lem:fotwoeq-equiv-equality-toplevel} to $\varphi$ to obtain an
  equivalent formula $\varphi' \equiv \varphi$ in which equality only occurs at
  the top level. Note that, since $\varphi$ has only $\{x\}$ as free variables,
  so does $\varphi'$. Hence, any equality that occurs at the top level in
  $\varphi'$ must be of the form $x=x$, which is trivial and can therefore be
  removed. After removing such trivial equalities, take
  $Q\colon H(x) \gets \varphi'$ as the simulating query.

  (3) $H$ is a nullary relation. The reason is entirely similar to the previous case. 
\end{proof}

\section{Q-hierarchical conjunctive queries}\label{sec:qh-cq}

We next turn our attention to  specializing the
correspondence between \conjLang and CQs given by
Corollary~\ref{cor:langsum-to-cq} to \emph{q-hierarchical}
CQs~\cite{DBLP:conf/csl/BaganDG07,DBLP:phd/hal/BraultBaron13}.
Q-hierarchical CQs are relevant since, over the Boolean semiring, they capture  the free-connex CQs that, in addition to supporting constant delay enumeration after linear time preprocessing, have the property that every single-tuple update (insertion or deletion) to the input database can be processed in constant time, after which the enumeration of the updated query result can again proceed with constant delay~\cite{DBLP:conf/pods/BerkholzKS17}.
Formally, q-hierarchical CQs are defined as follows.
Let $Q\colon H(\x) \gets \exists\y. a_1\wedge\ldots\wedge a_n$ be a \cq. For every variable $x$, define $\atoms(x)$ to be the set $\{ a_i \mid x \in \var(a_i)\}$ of
relational atoms from the body of $Q$ that mention $x$. Note that, contrary to acyclic queries, here we make the distinction between relational atoms and inequalities. Then $Q$ is \emph{q-hierarchical} if for any two variables $x,y$ the following hold:
\begin{enumerate}
    \item $\atoms(x)\subseteq \atoms(y)$ or $\atoms(x)\supseteq \atoms(y)$ or $\atoms(x)\cap \atoms(y)=\emptyset$, and 
    \item if $x\in\x$ and $\atoms(x)\subsetneq \atoms(y)$ then $y\in\x$.
\end{enumerate}

For example, $H(x) \gets \exists y. A(x,y) \wedge U(x)$ is
q-hierarchical. By contrast, the variant
$H(x) \gets \exists y. A(x, y) \wedge U(y)$ is not q-hierarchical, as it
violates the second condition. Furthermore,
$H(x,y) \gets  A(x,y) \wedge U(x) \wedge V(y)$ violates the first condition,
and is also not q-hierarchical.  Note that all these examples are
free-connex. We refer to q-hierarchical \cqs simply as~\qcqs.

In this section, we first prove a correspondence between \qcqs and a fragment of \foTwo. Using this, we characterize the fragment of \lang that corresponds to \qcqs in Section~\ref{sec:q-hierarchical-matlang}.
Concretely, we prove the following result in this section.
We say that a $\foTwo$ formula $\varphi$ is \emph{simple} if  every subformula of the form $\varphi_1 \wedge \varphi_2$ in $\varphi$ satisfies $\free(\varphi_1)=\free(\varphi_2)$. We denote by \foTwoSimp the class of all simple \foTwo formulas.
Let $\varphi_1,\ldots,\varphi_k\in\foTwoSimp$ with $k \geq 1$ such that  $\free(\varphi_i) \subseteq \{x,y\}$ for two distinct variables $\{x,y\}$. We say that $\varphi_1\wedge\ldots\wedge\varphi_k$ is a \emph{hierarchical conjunction} if $\{ \{x,y\}, \{x\}, \{y\}\} \not \subseteq \{ \free(\varphi_i) \mid 1 \leq i \leq n\}$. 
For example, $R(x,y) \wedge S(y) \wedge U()$ is a hierarchical conjunction, but $R(x,y) \wedge S(y) \wedge T(x)$ is not. 
We denote by  $\foTwoH$ all hierarchical conjunctions of \foTwoSimp formulas.
Observe that, by definition, $\foTwoSimp \subseteq \foTwoH$. As usual, an \foTwoH query is a query with body $\psi$ such that $\psi\in\foTwoH$.

\begin{thm}\label{theo:q-hierarchical-binary-cq-equiv-foTwoH}
    Q-hierarchical binary \cqs and \foTwoH queries are equally expressive.
\end{thm}

To prove Theorem~\ref{theo:q-hierarchical-binary-cq-equiv-foTwoH} we first define \emph{guarded query plans} in Section~\ref{sec:guarded-qps}. %
 Similar to how query plans allowed us to  prove equivalence between \fcqs and \foTwo in Section~\ref{sec:fc-cq},  we use guarded query plans to prove equivalence between \qcqs and \foTwoH. We show that \foTwoH is included in \qcqs in Section~\ref{sec:qcq-fotwoh}, and the converse in Section~\ref{sec:fotwoh-qcq}.

\subsection{Guarded query plans.} \label{sec:guarded-qps}

A generalized join tree (GJT) $T$ (see Section~\ref{sec:query-plans}) is \emph{guarded} 
if for every node $n$ and every child $c$ of $n$ it holds that $\var(n) \subseteq \var(c)$, i.e., in a guarded GJT every child is a guard of its parent. A query plan (QP) is guarded if its GJT is guarded.\footnote{Guarded GJTs and QPs are called \emph{simple} in \cite{DBLP:journals/vldb/IdrisUVVL20}.} The following proposition is a particular case of the results shown in \cite{DBLP:conf/sigmod/IdrisUV17,DBLP:journals/vldb/IdrisUVVL20}.

\begin{prop}\label{prop:q-hierarchical-iff-guarded-query-plan}
    A \cq $Q$ is q-hierarchical if and only if it has a guarded query plan.
\end{prop}

For later use we observe the following property. Its proof is completely analogous to the proof of Lemma~\ref{lemma:refined-query-plan}. 
\begin{lem}\label{lemma:full-guarded-query-plan}
    For every guarded QP there exists an equivalent guarded QP such that for any node $n$ with two children $c_1, c_2$
    it holds that $var(n)= var(c_1)=var(c_2)$.
\end{lem}

\subsection{From q-hierarchical \cqs to \foTwoH queries}
\label{sec:qcq-fotwoh}

\begin{lem}\label{lemma:from-rooted-query-plans-to-foTwoSimp-formulas}
    For every guarded QP $(T,r)$ such that $\varphi[T,r]$ is binary, there exists a \foTwoSimp formula $\varphi$ equivalent to $\varphi[T,r]$.
\end{lem}
\begin{proof}
    By Lemma~\ref{lemma:full-guarded-query-plan} we may assume w.l.o.g. that 
    $T$ is such that for every node $n$ with two children $c_1,c_2$ it holds
    $\var(c_1)=\var(n)$ and $\var(c_2)=\var(n)$. In the proof of Proposition~\ref{prop:from-rooted-query-plan-to-foTwo} we showed by induction on $T$ that we can construct  an $\foTwo$ formula $\varphi$ equivalent to $\varphi[T,r]$. It is readily verified that when $T$ is guarded this construction  yields a simple formula. \qedhere
\end{proof}

\begin{prop}\label{prop:from-query-plans-to-foTwoH-formulas}
    For every guarded QP $(T,N)$ such that $\varphi[T,N]$ is binary, there exists an \foTwoH formula $\psi$ equivalent to $\varphi[T,N]$.
\end{prop}
\begin{proof}
  By Lemma~\ref{lemma:full-guarded-query-plan} we may assume w.l.o.g. that $T$
  is such that for every node $n$ with two children $c_1,c_2$ we have
  $\var(c_1)=\var(c_2)=\var(n)$.
  Let $F=\{ r_1, \ldots, r_l \}$ be
  the frontier of $N$.  Let $T_1,\dots, T_l$ be the subtrees of $T$ rooted at
  $r_1, \dots, r_l$, respectively.  By
  Lemma~\ref{lemma:from-rooted-query-plans-to-foTwoSimp-formulas}, there are
  \foTwoSimp formulas $\psi_1,\ldots,\psi_l$ equivalent to
  $\varphi[T_1, r_1], \ldots, \varphi[T_l, r_l]$, respectively.  By
Lemma~\ref{lemma:formula-of-query-plan-is-conjunction-of-frontier-induced-trees-formulas}, 
$\varphi[T,N]\equiv \varphi[T_1, r_1] \wedge \dots \wedge \varphi[T_l, r_l] \equiv \psi_1 \wedge \dots \wedge \psi_l$. Because $\varphi[T,N]$ is binary, there exists a set $\{x,y\}$ of two distinct variables such that $\var(N) \subseteq \{x,y\}$. It follows that $\free(\psi_i) = \var(r_i) \subseteq \var(N) \subseteq \{x,y\}$ for every $i$. It remains to show that $\psi_1 \wedge \dots \wedge \psi_l$ is a hierarchical conjunction.  To that end, assume, for the purpose of obtaining a contradiction, that $\{ \{x,y\}, \{x\}, \{y\}\} \subseteq \{ \free(\psi_i) \mid 1 \leq i \leq n\}$. Because $\free(\psi_i) = \var(r_i)$ for every $i$ there must exist nodes $r_{x,y}$, $r_{x}$ and $r_y$ in $F$  that are labeled by $\{x,y\}$, $\{x\}$ and $\{y\}$ respectively. Now note that, because $T$ is guarded and satisfies Lemma~\ref{lemma:full-guarded-query-plan}, all ancestors of $r_{x}$ (resp. $r_{y}$) can only contain $x$ (resp. $y$) in their label, but not $y$ (resp $x$). At the same time, because of the connectedness property, there is an ancestor $a$ of both $r_{x,y}$ and $r_x$ that must contain $x$, implying that $var(a) = \{x\}$. Similarly, there is an ancestor $b$ of both $r_{x,y}$ and $r_y$ that must contain $y$, implying that $\var(b) = \{y\}$. Now note that $a$ and $b$ are both ancestors of $r_{x,y}$, so either $a$ is an ancestor of $b$, or the converse holds.  Assume $a$ is an ancestor of $b$. But then $a$ is also an ancestor of $r_y$ and $a$ has $x$ in its label, which yields the desired contradiction. The case where $b$ is an ancestor of $a$ is similar. \qedhere
\end{proof}

\subsection{From \foTwoH queries to q-hierarchical \cqs}
\label{sec:fotwoh-qcq}

\begin{lem}\label{lemma:from-foTwoSimple-formulas-to-query-plans}
    For every \foTwoSimp formula $\varphi$ there exists a guarded QP $(T,r)$ such that $\varphi[T,r]$ is binary and equivalent to $\varphi$.
\end{lem}

\begin{proof}
    The proof is analogous to the proof of Proposition~\ref{prop:from-foTwo-to-qp}. Fix two distinct variables $x$ and $y$ and formula $\varphi \in \foTwoSimp$. Because $\foTwo$ formulas use at most two variables we may assume w.l.o.g. that the only variables occurring in $\varphi$ are $x$ and $y$. We prove by induction on $\varphi$ that there exists a guarded query plan $(T,r)$ such that $\varphi[T,r]\equiv\varphi$.
    
    When $\varphi$ is an atom (relational or inequality) then the result trivially follow by creating a GJT $T$ with single node $r$, labeled by the atom, and fixing $N = \{r\}$.%

    When $\varphi = \exists z. \varphi'$ with $z \in
    \{x,y\}$ we may assume w.l.o.g. that $z \in
    \free(\varphi')$: if not then $\varphi \equiv
    \varphi'$ and the result follows directly from the induction hypothesis on
    $\varphi'$. So, assume $z\in
    \free(\varphi)$. By induction hypothesis we have a guarded and binary QP
    $(T', r')$ equivalent to $\varphi'$ with $r'$ the root of $T'$. Let
    $T$ be obtained by adjoining a new root $r$ to $T'$; label $r$ by $\var(r')
    \setminus \{z\}$; and make $r'$ the only child of $r$. Take $N =
    \{r\}$. Then
    $(T,N)$ remains binary and guarded. Equivalence of $\varphi[T,N]$ to
    $\psi$ follows by induction hypothesis and
    Lemma~\ref{lemma:child-formula-projection}.

    Otherwise $\varphi$ is a conjunction,
    $\varphi = \varphi_1 \wedge \varphi_2$. By induction hypothesis, we have
    binary and guarded QPs $(T_1, r_1)$ and $(T_2, r_2)$ equivalent to $\psi_1$
    and $\psi_2$, respectively with $r_1$ and $r_2$ the roots of
    $T_1,T_2$. Using the same reasoning as in the proof of
    Proposition~\ref{prop:from-foTwo-to-qp} we may assume w.l.o.g.\@ that $T_1$
    and $T_2$ have no nodes in common, and that moreover
    $\var(T_1) \cap \var(T_2) \subseteq \var(N_1) \cap \var(N_2)$, i.e., that
    all variables that are mentioned in both $T_1$ and $T_2$ are also in
    $\var(N_1) \cap \var(N_2)$. Because $(T_i,r_i)$ is equivalent to $\varphi_i$ we have $\var(r_i) = \free(\varphi_i)$, for $i=1,2$. Then, because $\varphi$ is simple, we necessarily have $\var(r_1) = \free(\varphi_1) = \free(\varphi_2) = \var(r_2)$. As such, $\var(r_1) = \var(r_2)$.
We construct the claimed QP $(T,r)$ by taking the disjoint union of $T_1$ and $T_2$ and adjoining a new root $r$ with children $r_1,r_2$ such that  $\var(r) = \var(r_1) = \var(r_2)$. It is readily verified that $T$ satisfies the connectedness property and is hence a GJT. Moreover, $T$ is guarded. Correctness follows by Corollary~\ref{lemma:children-formulas-conjunction}.\qedhere

\end{proof}

\begin{prop}\label{prop:from-foTwoH-formulas-to-query-plans}
    For every \foTwoH formula $\psi$ there exists a guarded QP $(T,N)$ such that $\varphi[T,N]$ is binary and equivalent to $\psi$.
\end{prop}  
\begin{proof}
  Let $x,y$ be distinct FO variables, let $\psi=\varphi_1\wedge\cdots\wedge\varphi_k\in\foTwoH$ with
  $\varphi_i\in\foTwoSimp$ and $\free(\varphi_i)\subseteq \{x,y\}$ for
  $i=1,\ldots,k$. Consider the set $F = \{ \free(\varphi_i) \mid 1 \leq i \leq k\}$. Note that $F \not = \emptyset$ since it includes the root. Furthermore, because $\psi$ is a hierarchical conjunction, $\{ \{x,y\}, \{x\}, \{y\}\} \not \subseteq F$. Since $F\neq\emptyset$, there are hence $\binom{4}{1}+\binom{4}{2}+\left(\binom{4}{3} - 1\right)= 19$ possibilities for $F$. We construct $(T,N)$ by case analysis on $F$. We only illustrate one case; the other cases are similar.

  Assume $F = \{ \{x,y\}, \{x\}, \emptyset \}$. Define, for each $S \in F$ the formula $\varphi_S$ to be the conjunction of all $\varphi_i$ with $\free(\varphi_i) = S$. Then
  $\psi \equiv \varphi_{x,y}\wedge\varphi_{x}\wedge\varphi_{\emptyset}$. Observe that each $\varphi_S$ is simple. Hence  by
  Proposition~\ref{lemma:from-foTwoSimple-formulas-to-query-plans} there exist
  guarded query plans $(T_{x,y}, r_{x,y}), (T_{x},r_{x})$ and
  $(T_\emptyset,r_\emptyset)$ equivalent to
  $\varphi_{x,y}, \varphi_{x}, \varphi_{\emptyset}$, respectively. In
  particular, $\var(r_{x,y}) = \{x,y\}$, $\var(r_{x}) = \{x\}$ and
  $\var(r_{\emptyset}) = \emptyset$. We may assume w.l.o.g. that the GJTs
  $T_{x,y}, T_x$ and $T_{\emptyset}$ do not have nodes in common. We construct the
  claimed guarded QP $(T,N)$ by constructing $T$ as follows and setting $N = \{r, r', r_{x,y}, r_{x}, r_{\emptyset}\}$. 
\tikzset{ itria/.style={
      draw,dashed,shape border uses incircle, isosceles triangle,shape border
      rotate=90,yshift=-0.80cm, minimum height=0.6cm, minimum width=0.4cm, shape
      border rotate=#1, isosceles triangle stretches, inner sep=0pt},
    rtria/.style={ draw,dashed,shape border uses incircle, isosceles
      triangle,isosceles triangle apex angle=90, shape border
      rotate=-45,yshift=0.2cm,xshift=0.5cm}, ritria/.style={ draw,dashed,shape
      border uses incircle, isosceles triangle,isosceles triangle apex
      angle=110, shape border rotate=-55,yshift=0.1cm}, letria/.style={
      draw,dashed,shape border uses incircle, isosceles triangle,isosceles
      triangle apex angle=110, shape border rotate=235,yshift=0.1cm} }

 \tikzset{
        expand bubble/.style={
            preaction={draw,line width=1.4pt},
            gray!20,fill,draw,line width=2pt,
        },
    }

      \begin{center}
        \begin{tikzpicture}[sibling distance=0.7cm, level distance =0.4cm]
            \node (r1) {$r; \emptyset$} [sibling distance=2.5cm]
            child {
              node (r2) {$r'; \{x\}$} [sibling distance=1.25cm,level distance=0.7cm]
              child {
                node (r3) {$r_{x,y}$}
                  { node [itria,yshift=-13pt] {$T_{x,y}$}}
              }
              child {
                node (r4) {$r_x$}
                { node [itria,yshift=-6pt] {$T_x$}}
              }
            }
            child {
                node (r5) {$r_\emptyset$}
                { node [itria,yshift=-6pt] {$T_\emptyset$}}
            };

              \begin{pgfonlayer}{background}
                \path[expand bubble]plot [smooth cycle,tension=1]
                coordinates {(r1.north) (r2.west) (r3.south west) (r4.south east) (r5.south east)};
              \end{pgfonlayer}

        \end{tikzpicture}
      \end{center}
      Here, $r$ and $r'$ are new nodes with $\var(r) = \emptyset$; $\var(r') = \{x\}$. It is readily verified that $T$ is guarded.  Correctness follows from Lemma~\ref{lemma:formula-of-query-plan-is-conjunction-of-frontier-induced-trees-formulas} since $\{r_{x,y}, r_{x}, r_{\emptyset}\}$ is the frontier of $N$.
\end{proof}

\section{The q-hierarchical fragment of matlang queries}\label{sec:q-hierarchical-matlang}

We next define $\qhLang$, a fragment of $\sumLang$ that we will show to be equally expressive as \foTwoH and \qcqs. Like \foTwoH, \qhLang is a two-layered language where expressions in the top layer can only be built from the lower layer. This lower layer, called \simpleLang, is a fragment of \fcLang defined by
\[ e \ \ \Coloneqq \ \  \mA \ \mid \ \ones^\alpha \ \mid \ \Iden^\alpha \ \mid \ e^T \ \mid \ e_1 \times e_2 \ \mid \ e_1\kprod e_2 \ \mid \ e \cdot \ones^\alpha.\]
Note in particular that while \fcLang allows arbitrary matrix-vector multiplication, this is restricted to multiplication with the ones vector in \simpleLang. Intuitively, \simpleLang  can already define q-hierarchical CQs like $H(x) \gets \exists y. A(x,y) \wedge U(x)$, but it cannot define cross-products like $H(x,y) \gets A(x) \wedge B(y)$, which are also q-hierarchical. For this reason, we enhance \simpleLang with the higher layer. Specifically, let us call expressions of the form  $e \cdot (\ones^{\alpha})^T$ and $\ones^{\alpha} \cdot e$ \emph{expansions} of \simpleLang expression $e$. Note that these are well-typed only if $e$ has column resp. row vector type. These expressions construct a matrix from a column (resp. row) vector by duplicating the column (resp. row) $\alpha$ times.
We then define \qhLang to consist of all expressions $g$ of the form
\begin{align*}
    f & \Coloneqq\ \  e \ \mid e \cdot (\ones^{\alpha})^T \mid \ones^{\alpha} \cdot e \\
    g  &  \Coloneqq \ \ f \ \mid f \kprod f 
\end{align*}
where $e$ ranges over $\simpleLang$ expressions, and $f$ over $\simpleLang$ expressions and expansions thereof. 

In this section we prove:
\begin{thm}\label{theo:qhLang-equiv-foTwoH}
    \qhLang queries and \foTwoH queries are equally expressive.
\end{thm}

Combined with Theorem~\ref{theo:q-hierarchical-binary-cq-equiv-foTwoH} we hence obtain.

\begin{cor}\label{theo:qhLang-equiv-q-hierarchical-cq}
    \qhLang and q-hierarchical binary CQs are equally expressive.
\end{cor}

The rest of this section is devoted to proving Theorem~\ref{theo:qhLang-equiv-foTwoH}. We show that \qcqs are at least as expressive as \qhLang in Section~\ref{sec:fotwoh-to-qhlang} and the converse in Section~\ref{sec:qhlang-to-fotwoh}.

\subsection{From \foTwoH formulas to \qhLang expressions}
\label{sec:fotwoh-to-qhlang}

We will first simulate  \foTwoH \emph{formulas} by means of a \qhLang expression according to the definition in Section~\ref{sec:fotwo-to-fclang}.

\begin{prop}\label{prop:from-foTwoSimp-to-simpleLang}
    Let $\varphi\in\foTwoSimp$ be a formula over $\Voc$. Let $\Mat$ be a relational-to-matrix schema encoding on $\Voc$ such that $\Mat\vdash \varphi\colon\tau$. Then there exists a \simpleLang expression $e$ that simulates $\varphi$ w.r.t. $\Mat$. 
\end{prop}
\begin{proof}
    We assume $\var(\varphi) \subseteq \{ x,y \}$ with $x\prec y$. The proof is analogous to the proof of Proposition~\ref{prop:fotwo-to-fc-lang}, with the difference that when $\varphi=\varphi_1\wedge\varphi_2$ the simulating expression is always $e_1\kprod e_2$ because $\free(\varphi_1)=\free(\varphi_2)$ since $\varphi\in\foTwoSimp$. This has the consequence that the simulating expression is always a \simpleLang expression.
\end{proof}

\begin{prop}\label{prop:from-foTwoH-to-qhLang}
    Let $\psi\in\foTwoH$ be a formula over $\Voc$. Let $\Mat$ be a relational-to-matrix schema encoding on $\Voc$ such that $\Mat\vdash \psi\colon\tau$. Then there exists a \qhLang expression $e$ that simulates $\psi$ w.r.t. $\Mat$.
\end{prop}
\begin{proof}
    We assume $\var(\psi) \subseteq \{ x,y \}$ with $x\prec y$. By definition, $\psi=\varphi_1\wedge\cdots\wedge\varphi_k$ with $\varphi_i\in\foTwoSimp$ for every $i$. Consider the set $F = \{ \free(\varphi_i) \mid 1 \leq i \leq k\}$. Note that $F \not = \emptyset$. Furthermore, because $\psi$ is a hierarchical conjunction, $\{ \{x,y\}, \{x\}, \{y\}\} \not \subseteq F$. Since $F\neq\emptyset$, there are hence $\binom{4}{1}+\binom{4}{2}+\left(\binom{4}{3} - 1\right)= 19$ possibilities for $F$. We construct $e$ by case analysis on $F$. We only illustrate one case; the other cases are similar.

    Assume $F = \{ \{x,y\}, \{x\}, \emptyset \}$. Define, for each $S \in F$ the formula $\varphi_S$ to be the conjunction of all $\varphi_i$ with $\free(\varphi_i) = S$. Then $\psi \equiv \varphi_{x,y}\wedge\varphi_{x}\wedge\varphi_{\emptyset}$. Observe that each $\varphi_S$ is simple. Hence  by Proposition~\ref{prop:from-foTwoH-to-qhLang} there exist \simpleLang expressions $e_{x,y}, e_x$, and $e_{\emptyset}$ that simulate $\varphi_{x,y}, \varphi_{x}$, and $\varphi_{\emptyset}$, respectively.  Then the \qhLang expression simulating $\varphi$ is \[e\coloneqq \left(e_{\varphi_{\emptyset}} \times e_{\varphi_{\{x,y\}}}\right) \kprod \left(e_{\varphi_{\{x\}}}\cdot \left(\ones^{\tau(y)}\right)^T\right). \qedhere\]

\end{proof}

To also establish the correspondence at the level of queries we first observe the following straightforward fact. For a \sumLang expression $e\colon(\alpha,1)$ define $\diag(e)$ to return a matrix of type $(\alpha,\alpha)$ where the entries of $e$ are stored in the diagonal and where all non-diagonal entries are $\kzero$. 

\begin{lem}\label{lem:qh-lang-closed-transpose-diag}
    \qhLang is closed under transpose and diagonalisation: for any expression $g \in \qhLang$ also $g^T$ is expressible in $\qhLang$ and, moreover, if $g\colon (\alpha, 1)$ has column vector type, then $\diag(g)$ is expressible in $\qhLang$.
\end{lem}
\begin{proof}
    Closure under transpose follows from the fact that \simpleLang is closed under transpose by definition, and from the following equivalences:
    \begin{align*}
        (f_1 \kprod f_2)^T & = f_1^T \kprod f_2^T  & \left(e \cdot (\ones^{\alpha})^T\right)^T & = \ones^{\alpha} \cdot e^T & \left(\ones^{\alpha} \cdot e\right)^T & = e^T \cdot (\ones^{\alpha})^T 
    \end{align*}
    Closure under diagonalisation is straightforward. Assume that $g\colon (\alpha,1)$. Note that expansions never have column vector type. As such, $g$ is a \simpleLang expression, or a Hadamard product of \simpleLang expressions. Note that in the latter case, this is itself also a \simpleLang expression since \simpleLang is closed under Hadamard products. Then $\diag(g)$ can be define in \qhLang by first computing the column expansion of $g$ and subsequently setting all non-diagonal entries of this expansion to zero by taking \[\diag(g) = \Iden^{\tau(x)} \kprod \left(g\cdot \left(\ones^{\tau(x)}\right)^T\right). \qedhere\]
\end{proof}

\begin{cor}
    Let $Q$ be an \foTwoH query over $\Voc$ and let $\Mat$ be a relational-to-matrix schema encoding on $\Voc(Q)$ such that $\Mat\vdash Q\colon\tau$. Then there exists $\qhLang$ query $\mQ$ that simulates $Q$ w.r.t. $\Mat$.
\end{cor}
\begin{proof}
    Assume $Q:H(\x)\gets \psi$. By Proposition~\ref{prop:from-foTwoH-to-qhLang} there exists a \qhLang expression $e$ that simulates $\psi$ w.r.t. $\Mat$.
    Let $\Mat(H)=\mH$. We build the \qhLang query $\mQ\coloneqq \mH=\e$ as follows, using Lemma~\ref{lem:qh-lang-closed-transpose-diag}:
    \begin{itemize}
    \item When $\x=(x,y)$, if $x \prec y$ then define $\e\coloneqq e$, otherwise define $\e\coloneqq e^T$. 
    \item When $\x=(x,x)$ we note that $\free(\psi) = \{x\}$ and therefore $e\colon (\tau(x), 1)$ outputs the encoding of a square matrix, where the value of diagonal entry $(x,x)$ is computed by $\psi(x)$. All non-diagonal entries of this matrix are $\kzero$. Hence we take $\e := \diag(e)$.
    \item When $\x= (x)$  we have $e\colon (\tau(x),1)$. If $\mH\colon (\tau(x), 1)$ define $\e\coloneqq e$ otherwise if $\mH\colon (1, \tau(x))$ define $\e\coloneqq e^T$. 
    \item When $\x=()$ we have $e\colon (1,1)$ and define $\e\coloneqq e$. \qedhere
    \end{itemize}
\end{proof}

\subsection{From \qhLang expressions to \foTwoH formulas}\label{sec:qhlang-to-fotwoh}

To establish the converse direction of Theorem~\ref{theo:qhLang-equiv-foTwoH} we first show that how to  simulate  \qhLang expressions by means of \foTwoH  formulas that may also use equality, according to the definition in Section~\ref{sec:fclang-to-fotwo}. Define $\foTwoEqH$ to consist of all $\foTwoH$ formulas as well as all formulas of the form $\varphi \wedge x=y$ or $x=y \wedge \varphi$ with $\varphi \in \foTwoH$ and $\free(\varphi) \cap \{x,y\} \not = \emptyset$. Note that in the latter case we assume that $x,y$ are the only two distinct variables that can be used in $\varphi$, so that the result again uses only two variables.

Define the \emph{signature} $\sig(\varphi)$ of formula $\varphi \in \foTwoEqH$ to be the set $\{ \free(\varphi_i) \mid 1 \leq i \leq n\}$ where $\varphi_1,\dots, \varphi_n$ are the non-equality conjuncts of $\varphi$. For example, the signature of $\varphi = x \leq c \wedge y \leq d \wedge x = y$ is $\{\{x\}, \{y\}\}$.

The following lemma shows that projections on $\foTwoEqH$ formula are expressible as $\foTwoH$ formulas (hence, not using equality), provided that the original formulas signature is of a certain shape.

\begin{lem}\label{lem:foTwoH-eq-proj-in-foTwoH}
	Let $\varphi$ be a $\foTwoEqH$ formula such that $\sig(\varphi)$ is a subset of either $\{\emptyset,\{x\},\{y\}\}$ or $\{\emptyset, \{x,y\}\}$. There exists a $\foTwoH$ formula equivalent to $\exists x. \varphi$ (resp. $\exists y. \varphi$) whose signature is a subset of $\{\emptyset,\{x\},\{y\}\}$.
\end{lem}
\begin{proof}
	We first make the following claim: if $\psi \in \foTwoSimp$ and $\free(\psi) \cap \{x,y\} \not = \emptyset$, then there exists a $\foTwoSimp$ formula equivalent to $\exists x. (\varphi \wedge x=y)$ (resp. $\exists y. (\varphi \wedge x=y)$). The proof of this claim is identical to the proof of Lemma~\ref{lem:fotwo-unify}, observing that the construction defined there, when applied to simple formulas, yields a simple formula. Because the resulting formula has at most 1 free variable, its signature (viewed as a conjunction with only one conjunct) is trivially a subset of $\{\emptyset,\{x\},\{y\}\}$.
	
	Next, assume that $\varphi = \varphi_1 \wedge \dots \wedge \varphi_k \wedge x=y$ where the $\varphi_i$ are $\foTwoSimp$ formulas. (The reasoning when $\varphi$ does not contain the equality atom $x=y$ is similar.) We make a case analysis on $\sig(\varphi)$.
	\begin{itemize}
		\item $\sig(\varphi)$ is a subset of $\{\emptyset,\{x\},\{y\}\}$. Since $\varphi$ has at least one conjunct, $\sig(\varphi)$ is non-empty. Moreover, the case where $\sig(\varphi) = \{\emptyset\}$ cannot occur, since otherwise $\varphi$ would not be safe.  There are hence $2^3 -2 = 6$ possibilities to consider for the signature. We only illustrate the reasoning when $\sig(\varphi) = \{\emptyset,\{x\},\{y\}\}$, the other cases are similar. Assume that $\sig(\varphi) = \{\emptyset,\{x\},\{y\}\}$. For each $S \in \{\emptyset,\{x\},\{y\}\}$ let $\psi_S$ be the conjunction of all the formulas $\varphi_i$ with $\free(\varphi_i) = S$. Note that each $\psi_S$ is in $\foTwoSimp$, as it is a conjunction of simple formulas with the same set of free variables,  and that $\varphi \equiv \psi_\emptyset \wedge \psi_{x} \wedge \psi_y \wedge x = y$. Then $\exists x. \varphi \equiv \psi_\emptyset \wedge \exists x. (\psi_{x} \wedge x = y) \wedge \psi_y$. By our earlier claim, $\exists x. (\psi_{x} \wedge x = y)$ is expressible as a \foTwoSimp formula $\psi'_x$. Because $\exists x. (\psi_{x} \wedge x = y)$ has free variables $\{y\}$, so does $\psi'_x$. Hence,
		$\exists x. \varphi \equiv \psi_\emptyset \wedge \psi'_x \wedge \psi_y$, where the right-hand side is a hierarchical conjunction of \foTwoSimp formulas and is hence in $\foTwoH$. Also, observe that the signature of the right-hand side is a subset of $\{\emptyset, \{y\}\}$. This establishes the desired result.
		
		\item $\sig(\varphi)$ is a subset of $\{\emptyset,\{x,y\}\}$. The reasoning is entirely similar. \qedhere
	\end{itemize}
\end{proof}

First we show the reduction from $\simpleLang$ expressions to $\foTwoEqH$ formulas. Here, the signature of the resulting formula is important to later remove the equalities.

\begin{prop}\label{prop:from-simpleLang-to-foTwoSimp}
    Let $e$ be a $\simpleLang$ expression over $\Sch$ and let $\Rel$ be a matrix-to-relational schema encoding on $\Sch$.  Let $x,y$ be two variables such that $x \prec y$. There exists  $\varphi_e \in \foTwoEqH$ simulating $e$ such that $\sig(\varphi_e)$ is a subset of either $\{ \emptyset, \{x\}, \{y\} \}$ or $\{\emptyset, \{x,y\} \}$. 
\end{prop}
\begin{proof}
    The proof is by induction and analogous to the proof of Proposition~\ref{prop:from-fcLang-exp-to-foTwo-formulas}. We only illustrate two interesting cases.
    \begin{itemize}
    \item If $e=\Iden^\alpha\colon (\alpha,\alpha)$ then take $\psi_e    \coloneqq x\leq \alpha \wedge y \leq \alpha \wedge x = y$.  Note that $x\leq \alpha \wedge y \leq \alpha$ is a hierarchical conjunction of simple $\foTwo$ formulas. Therefore, $\psi_e \in \foTwoEqH$, as desired.
    \item If $e = e_1 \times e_2$ with $e_1 \colon (1,1)$ and $e_2\colon (\alpha, \beta)$ we reason as follows. Let $\psi_{e_1}$ and $\psi_{e_2}$ be the $\foTwoEqH$ formulas simulating $e_1$ and $e_2$ obtained by induction.

    First observe that because $\psi_{e_1}$ simulates $e_1$ and because $e_1$  always outputs scalar matrices, $\fssem{\psi_{e_1}}{\Rel(\I)}(x \mapsto i, y \mapsto j) = \kzero$ whenever $i\not =1$ or $j \not = 1$. We conclude that therefore, $\exists x,y. \psi_{e_1}$ will always return $\sem{e_1}{\I}(1,1)$ when evaluated on $\Rel(\I)$. Hence, we take $\psi_{e} \coloneqq \left(\exists x,y. \psi_{e_1}\right) \wedge \psi_{e_2}$. Note that $\left(\exists x,y. \psi_{e_1}\right)$ is expressible in $\foTwoH$ by applying Lemma~\ref{lem:foTwoH-eq-proj-in-foTwoH} twice, i.e., it is expressible as a hierarchical conjunction of simple $\foTwo$ formulas. Each of these formulas must have the empty set of free variables. Therefore, their conjunction with $\psi_{e_1}$ is also a hierarchical conjunction, and hence in $\foTwoEqH$. The signature of the resulting conjunction is the union of $\{\emptyset\}$ with the signature of $\psi_{e_1}$, and therefore also a subset of either $\{ \emptyset, \{x\}, \{y\} \}$ or $\{\emptyset, \{x,y\} \}$. \qedhere
    \end{itemize}
\end{proof}

Next, we show how to translate any $\qhLang$ expression to a $\foTwoEqH$ formula.

\begin{prop}\label{prop:from-qhLang-to-foTwoH}
    Let $g$ be a $\qhLang$ expression over $\Sch$ and let $\Rel$ be a matrix-to-relational schema encoding on $\Sch$. There exists  $\psi\in\foTwoEqH$ that simulates $e$ w.r.t. $\Rel$.%
\end{prop}
\begin{proof}
    Let $x,y$ be two variables such that $x \prec y$. We may assume w.l.o.g. that the formulas $\varphi$ returned by Proposition~\ref{prop:from-simpleLang-to-foTwoSimp} have $\free(\varphi) = \{x,y\}$. We reason by case analysis. Let $e_1,e_2$ range over  $\simpleLang$ expressions.
    \begin{enumerate}
        \item $g \in \simpleLang$. Then the claim follows from Proposition~\ref{prop:from-simpleLang-to-foTwoSimp}.

        \item $g=e_1\kprod \left( e_2 \cdot \left(\ones^\beta\right)^T\right)$ with $e_1\colon (\alpha,\beta)$ and $e_2\colon (\alpha, 1)$. By Proposition ~\ref{prop:from-simpleLang-to-foTwoSimp}, there exists $\varphi_1$ and $\varphi_2$ simulating $e_1$ and $e_2$, respectively. Observe that for all matrix instances $\I$ and all valid indices $i,j$ we have $\sem{g}{\I}_{i,j} = \sem{e_1}{\I}_{i,j} \kprod \sem{e_2}{\I}_{i,1}$. Because $\varphi_2$ simulates $e_2$ and because $e_2$ always outputs a column vector, $\fssem{\varphi_2}{\Rel(\I)}(x \mapsto i, y \mapsto j) = \kzero$ whenever $j \not = 1$, for all $\I$.  We conclude that therefore $\sem{e_2}{\I}_{i,1} = \fssem{\exists y. \varphi_2}{\Rel(\I)}(x \mapsto i)$. Hence it suffices take $\psi = \varphi_1(x,y) \wedge \exists y. \varphi_2(x,y)$. By Lemma~\ref{lem:foTwoH-eq-proj-in-foTwoH} $\exists y. \varphi_2(x,y)$ is expressible in \foTwoH, i.e., it is expressible as a hiearchical conjunction of simple $\foTwo$ formulas, which necessarily all have free variables that are a subset of $\{x\}$. Extending this conjunction with $\varphi_1$, whose signature is a subset of $\{ \emptyset, \{x\}, \{y\}\}$ or $\{\emptyset, \{x,y\}\}$ by Proposition~\ref{prop:from-simpleLang-to-foTwoSimp}, remains hierarchical. We conclude that $\psi$ is expressible in $\foTwoEqH$.

        \item $e=e_1\kprod \left( \ones^\alpha \cdot e_2 \right)$ with $e_1\colon (\alpha,\beta)$ and $e_2\colon (1, \beta)$. The reasoning is similar as in the previous case, but taking $\varphi = \varphi_1(x,y) \wedge \exists x. \varphi_2(x,y)$. 

        \item $e=\left( e_1 \cdot \left(\ones^\beta\right)^T\right)\kprod \left( \ones^\alpha \cdot e_2 \right)$ with $e_1\colon (\alpha,1)$ and $e_2\colon (1, \beta)$. The reasoning is again similar, taking $\varphi = \left(\exists y. \varphi_1(x,y)\right) \wedge \left( \exists x. \varphi_2(x,y) \right)$.

        \item The only other possibilities are commutative variants of cases already considered. \qedhere
    \end{enumerate}
\end{proof}

The next lemmas will aid to remove the equalities of $\foTwoEq$ formulas in the final result. The reader can find their proofs in Appendix~\ref{sec:app-q-hierarchical-matlang}.

\begin{lem}\label{lemma:foTwoH-unify}
    Let $\varphi$ be a $\foTwoH$ formula (hence, without equality) using only the distinct variables $x,y$ such that $\free(\varphi) = \{x,y\}$. Let $\psi$ be the $\foTwoEq$ formula $\exists x. (\varphi \wedge x=y)$. Note that $\free(\psi) = \{y\}$. There exists an $\foTwoH$ formula $\psi'$ equivalent to $\psi$.
\end{lem}

\begin{lem}\label{lem:foTwoH-force-equal}
    If $\varphi \in \foTwoH$ has $\free(\varphi) = \{x,y\}$ and $z \in \{x,y\}$ then there exists $\varphi' \in \foTwoH$ with $\free(\varphi') = \{z\}$ such that for all $\val \colon \{z\}$ and all $\db$, 
    \begin{align*}
        \fssem{\varphi'}{\db}(\val) = \fssem{\varphi \wedge x = y}{\db}(x \mapsto \val(z), y \mapsto \val(z)).
    \end{align*}
\end{lem}

We are ready to show how to translate any $\qhLang$ query to a $\foTwoH$ query.

\begin{cor}\label{cor:from-qhLang-to-foTwo-binary-encoding}
    Let $\mQ$ be an \qhLang query over $\Sch$ and let $\Rel$ be a matrix-to-relational schema encoding on $\Sch$. There exist \foTwoH query $Q$  that simulates $\mQ$ w.r.t. $\Mat$.
\end{cor}

\begin{proof}
    Let $\mH\coloneqq e$ be a \qhLang query and let $H = \Rel(\mH)$. We consider three cases.
    \begin{enumerate}
        \item $H$ is a binary relation. By Proposition~\ref{prop:from-qhLang-to-foTwoH} there exists $\psi\in\foTwoEqH$ that simulates $e$ w.r.t. $\Rel$. By construction, $\psi$ has two distinct free variables. Assume that $x, y$ are these two distinct variables, and $x \prec y$. If $\psi \in \foTwoH$ then we take $Q$ to be the query $H(x,y) \gets \psi$. Otherwise, $\psi =  \psi' \wedge x = y$ with $\psi' \in \foTwoH$ and at least one of $x,y$ appearing free in $\psi'$. If $\free(\psi') = \{x\}$ then we take $H(x,x) \gets \psi'$ as simulating query. If $\free(\psi') = \{y\}$ then we take $H(y,y) \gets \psi'$ as simulating query. Otherwise $\free(\psi') = \{x,y\}$, and we take $H(x,x) \gets \psi''$ as simulating query, where $\psi''$ is the result of applying Lemma~\ref{lem:foTwoH-force-equal} to $\psi'$ and $z = x$.
      
        \item $H$ is a unary relation. Assume that $\mH$ is of column vector type $(\alpha, 1)$; the reasoning when $\mH$ has row vector type is similar. Because of well-typedness, also $e \colon (\alpha,1)$. Now observe that  expansions return matrices that are not column vectors (or, if they do, the result can be written without using an expansion). As such, $e \in \simpleLang$. By Proposition~\ref{prop:from-simpleLang-to-foTwoSimp} there exists $\psi\in\foTwoEqH$ that simulates $e$ w.r.t. $\Rel$ such that $\sig(\psi)$ is a subset of either $\{\emptyset, \{x\},\{y\}\}$ or of $\{\emptyset, \{x,y\}\}$. By construction, $\psi$ has two distinct free variables. Assume that $x, y$ are these two distinct variables, and $x \prec y$. Take $\varphi = \exists y. \psi$. Because $\psi$ simulates $e$ and because $e$ always outputs column vectors, $\fssem{\psi}{\Rel(\I)}(x \mapsto i, y \mapsto j) = \kzero$ whenever $j \not = 1$, for every $\I$. We conclude that therefore $\sem{e}{\I}(i,1) = \fssem{\exists y. \psi}{\Rel(\I)}(x\mapsto i)$ for all $i$ and $\I$. As such, the query $Q'\colon H(x) \gets \varphi$ simulates $\mQ$. While $Q'$ is a query in $\foTwoEqH$, we can next apply Lemma~\ref{lem:foTwoH-eq-proj-in-foTwoH} to $\varphi$ to obtain an equivalent formula $\varphi' \equiv \varphi$ in $\foTwoH$.
        
        \item $H$ is a nullary relation. The reason is entirely similar to the previous case. \qedhere
    \end{enumerate}
\end{proof}

\section{Preliminaries for enumeration algorithms and query evaluation}\label{sec:enum}

In the previous sections we have presented a precise connection between
fragments of $\foplus$ and fragments of $\sumLang$. In the sequel, we move to
translate the algorithmic properties of the free-connex and q-hierarchical
conjunctive fragments of $\foplus$ to properties for the corresponding fragments
of $\sumLang$. Towards this goal, this section introduces all the extra
background and definitions regarding enumeration algorithms. The properties themselves are obtained in
Sections~\ref{sec:evaluation}~and~\ref{sec:evaluation-qh}.

\paragraph{Enumeration problems}
An \emph{enumeration problem} $P$ is a collection of pairs $(I, O)$ where $I$ is an \emph{input} and $O$ is a finite set of answers for $I$. There is a functional dependency from $I$ to $O$: whenever $(I, O)$ and $(I',O')$ are in $P$ and $I = I'$ then also $O = O'$. For this reason, we also denote $O$ simply by $P(I)$. An \emph{enumeration algorithm} for an enumeration problem $P$ is an algorithm $A$ that works in two phases: an \emph{enumeration phase} and a \emph{preprocessing phase}. During preprocessing, $A$ is given input $I$ and may compute certain data structures, but does not produce any output. During the enumeration phase, $A$ may use the data structures constructed during preprocessing to print the elements of $P(I)$ one by one, without repetitions.  The \emph{preprocessing time} is the running time of the preprocessing phase. The \emph{delay} is an upper bound on the time between printing any two answers in the enumeration phase; the time from the beginning of the enumeration phase until printing the first answer; and the time after printing the last answer and the time the algorithm terminates.

\paragraph{Query evaluation by enumeration}
In our setting, we consider query evaluation as an enumeration problem. Specifically, define the \emph{answer} of a query $Q$ on database $\db$ to be the set of all tuples outputed by $Q$ on $\db$, together with their (non-zero) annotation, 
\[ 
\AnsEnum{Q}{\db} \ \coloneqq \ \lbrace (\seq{d}, \ssem{Q}{\db}(H)(\seq{d})) \mid \ssem{Q}{\db}(H)(\seq{d})\neq\kzero\rbrace. 
\]
Here, $H$ denotes the relation symbol in the head of $Q$.

We then associate to each \foplus query $Q$ over vocabulary $\Voc$ and semiring $\cK$ the following enumeration problem
$\Eval{Q,\Voc}{\cK}$:
\begin{center}
  \framebox{
      \begin{tabular}{rl}
          \textbf{Problem:} & $\Eval{Q,\Voc}{\cK}$ \\
          \textbf{Input:} &  A $\cK$-database $\db$ over $\Voc$\\
          \textbf{Output:} & Enumerate $\AnsEnum{Q}{\db}$.
      \end{tabular}
  }
\end{center}
Formally this is the collection of all pairs $(I,O)$, where
$I = \db$ with $\db$ a $\cK$-database over $\Voc$; and
$O = \AnsEnum{Q}{\db}$. Hence, $\Eval{Q,\Voc}{\cK}$ is the evaluation problem of $Q$ on $\cK$-databases.

Similarly, we associate to each \sumLang query $\mQ = \mH := e$ over schema $\Sch$ and semiring $\cK$ the enumeration problem 
$\Eval{\mQ,\Sch}{\cK}$:
\begin{center}
  \framebox{
      \begin{tabular}{rl}
          \textbf{Problem:} & $\Eval{\mQ,\Sch}{\cK}$ \\
          \textbf{Input:} &  A $\cK$-instance $\I$ over $\Sch$\\
          \textbf{Output:} & Enumerate $\AnsEnum{\mQ}{\I}$.
      \end{tabular}
  }
\end{center}
This is, the collection of all pairs $(I,O)$ where
\begin{itemize}
\item $I = \I$ with $\I$ a $\cK$-matrix instance over $\Sch$, sparsely represented by listing for each matrix symbol $\mA\colon (\alpha,\beta)$ the set of entries $\{(i,j,k) \mid i \leq \I(\alpha), j \leq \I(\beta), \mA^\I_{i,j} = k \not = \kzero\}$; and
\item $O = \AnsEnum{\mQ}{\I}$ where, assuming $\mH\colon (\alpha,\beta)$,
\[ \AnsEnum{\mQ}{\I} \coloneqq \lbrace (i,j, \sem{\mQ}{\I}(\mH)_{i,j}) \mid
  \sem{\mQ}{\I}(\mH)_{i,j} \neq\kzero, i \leq \I(\alpha), j \leq \I(\beta)
  \rbrace . \] 
\end{itemize}
Hence, $\Eval{\mQ,\Voc}{\cK}$ is the \emph{sparse} evaluation problem of $\mQ$
on matrix instances over $\Sch$, where only the non-zero output entries must be
computed.

\paragraph{Complexity classes} For two functions $f$ and $g$ from the natural
numbers to the positive reals we define $\EnumClass{f}{g}$ to be the class of
enumeration problems for which there exists an enumeration algorithm $A$ such
that for every input $I$ it holds that the preprocessing time is
$\bigo\left( f(\size{I}) \right)$ and the delay is
$\bigo\left( g(\size{I}) \right)$, where $\size{I}$ denotes the size of $I$. For
example, $\EnumClass{\size{\db}}{1}$ hence refers to the class of query
evaluation enumeration problems that have linear time preprocessing and constant
delay (in data complexity). Note that, since the inputs in an evaluation problem
$\Eval{Q,\Voc}{\cK}$ consists only of the database and not the query, we hence
measure complexity in \emph{data complexity} as is standard in the
literature~\cite{DBLP:conf/csl/BaganDG07,DBLP:journals/vldb/IdrisUVVL20,DBLP:journals/siglog/BerkholzGS20,DBLP:conf/pods/BerkholzKS17}.

Throughout, $\size{\db}$ denotes the size of the input $\cK$-database $\db$,
measured as the length of a reasonable encoding of $\db$, i.e.,
$\size{\db} := \sum_{R \in \Voc} \size{R^\db} + \sum_{c \in \Voc} 1$ where
$\size{R^\db} := (\arity(R) + 1) \cdot \card{R^\db}$ with $\card{R^\db}$ the
number of tuples $\seq{d}$ with $R^\db(\seq{d}) \not =
\kzero_{\cK}$. Intuitively, we represent each tuple as well as its $\cK$-value
and assume that encoding a domain value or $\cK$-value takes unit space. In
addition, each constant symbol is assigned a domain value, which also takes unit
space per constant symbol.

Similarly, if $\I$ is a matrix instance over matrix schema $\Sch$ then $\size{\I}$ denotes the size of a reasonable sparse encoding of $I$, i.e., $\size{\I} := \sum_{\mA \in \Sch} \size{\mA^\I} + \sum_{\alpha \in \Voc} 1$ where $\size{\mA^\I}$ denotes the number of entries $(i,j)$ with $(\mA^\I)_{i,j} \not = \kzero$, for $\mA\colon (\alpha,\beta)$ and $1 \leq i \leq \alpha^\I$ and $1 \leq j \leq \beta^\I$.

\paragraph{Model of computation} As has become standard in the
literature~\cite{DBLP:conf/csl/BaganDG07,DBLP:journals/vldb/IdrisUVVL20,DBLP:journals/siglog/BerkholzGS20,DBLP:conf/pods/BerkholzKS17}, we consider algorithms on Random
Access Machines (RAM) with uniform cost measure~\cite{DBLP:books/aw/AhoHU74}. Whenever we work with $\cK$-databases for an arbitrary semiring
$\cK$, we assume that each semiring operation can be executed in constant
time. %

\paragraph{Complexity Hypotheses} %
We will use the following algorithmic problems and associated complexity hypotheses for obtaining
conditional lower bounds.
\begin{enumerate}
  \item The \emph{Sparse Boolean Matrix Multiplication problem}: given $A$ and $B$ as a list of nonzero entries, compute the nonzero entries of the matrix product $AB$. The \emph{Sparse Boolean Matrix Multiplication conjecture} states that this problem cannot be solved in time $\bigo(M)$, where $M$ is the size (number of non-zero entries) of the input plus output.
  \item The \emph{Triangle Detection problem}: given a graph represented using adjacency lists, decide whether it contains a triangle. The \emph{Triangle Detection conjecture} states that this problem cannot be solved in time $\bigo(M)$, where $M$ is the size of the input plus output.
  \item \emph{$(k,k+1)$-Hyperclique}: Given a hypergraph where every hyperedge consists of exactly $k \geq 3$ vertices, decide if it contains a hyperclique of size $k+1$. A hyperclique is a set of vertices such that each pair of vertices in the set is contained in a hyperedge. The \emph{$(k,k+1)$-Hyperclique conjecture} states that this problem cannot be solved in time $\bigo(M)$, where $M$ is the size of the input plus output.
\end{enumerate}
Note that the three hypotheses are standard in computer science and have been used to prove conditional lower bounds for the enumeration of CQs. See~\cite{DBLP:journals/siglog/BerkholzGS20} for a discussion. 

\paragraph{Known bounds for free-connex CQs}\label{subsec:fc-known-results}
In what follows, let \cqNotIneq denote the class of CQs where no inequality atoms occur in the body. The body hence mentions only relational atoms.
 Also, let $\bools=(\{\btrue,\bfalse\}, \vee, \wedge, \bfalse, \btrue)$ be
the boolean semiring.

Bagan, Durand and Grandjean~\cite{DBLP:conf/csl/BaganDG07} (see also \cite{DBLP:journals/siglog/BerkholzGS20}) proved that, under certain
complexity-theoretic assumptions, the class of free-connex conjunctive queries
characterizes the class of conjunctive queries that, in data complexity, can be
enumerated with linear time preprocessing $\bigo(\size{\db})$ and with constant
delay on boolean input databases.  We next recall their~results.
\begin{thmC}[\cite{DBLP:conf/csl/BaganDG07}]
\label{theo:cqNotIneq-free-connex-cde}
  Let $Q$ be a \cqNotIneq over vocabulary $\Voc$.
  \begin{itemize}
    \item If $Q$ is free-connex then  $\Eval{Q,\Voc}{\bools}\in\EnumClass{\size{\db}}{1}$.  
    \item If $Q$ is a query without self joins and
      $\Eval{Q, \Voc}{\bools}\in \EnumClass{\size{\db}}{1}$, then $Q$ is free-connex
      unless either the Sparse Boolean Matrix Multiplication, the Triangle
      Detection, or the $(k,k+1)$-Hyperclique conjecture is false.
    \end{itemize}
\end{thmC}

\section{Efficient evaluation of free-connex queries}\label{sec:evaluation}

In this section, we focus on the evaluation problem for $\fcLang$, namely, the free-connex queries in $\sumLang$.
Our aim is to transfer the known algorithmic properties of evaluating free-connex CQs to properties of evaluating $\fcLang$.

To apply the known algorithm for \fcqs to \fcLang, we must first
solve two problems: (1) the efficient evaluation algorithm for \fcqs is
only established for evaluation over Boolean semiring, not arbitrary semirings $\mathcal{K}$,
and (2) the algorithm is for CQ without inequalities (comparison atoms). We
will therefore generalize the known evaluation algorithms to
other semirings (e.g., $\bbR$), and to queries that have also inequalities (Section~\ref{subsec:fc-upperbound}). Similarly, we do the same for the lower bounds known for \fcqs (Section~\ref{subsec:fc-lowerbound}).  %

\subsection{Upper bounds for free-connex queries} \label{subsec:fc-upperbound}

A semiring $(K, \ksum, \kprod, \kzero, \kone)$ is \emph{zero-divisor free} if, for all $a,b\in K$,
$a\kprod b=\kzero$ implies $a=\kzero$ or $b=\kzero$.  A zero-divisor
free semiring is called a \emph{semi-integral
domain}~\cite{golan2013semirings}. Note that semirings used in practice,
like $\bbB$,~$\bbN$,~and~$\bbR$, are semi-integral domains. We will establish:

\begin{thm}\label{theo:free-connex-upper-bound-body}
    Let $\cK$ be a semi-integral domain. For every free-connex query $Q$ over~$\Voc$,
    $\Eval{Q,\Voc}{\cK}$ can be evaluated with linear-time preprocessing and constant-delay.
    In particular,  $\Eval{\mQ,\Sch}{\cK}$ can also be evaluated with linear-time preprocessing
    and constant-delay for every \fcLang $\mQ$ over $\Sch$.
\end{thm}

The semi-integral condition is necessary to ensure that zero outputs can only be produced by some zero input entries. For instance, consider a semiring $(K, \ksum, \kprod, \kzero, \kone)$ such that there exist $a,b\in K$ where $a\neq\kzero$, $b\neq\kzero$ and $a\kprod b=\kzero$.
Consider $Q:H(x,y)\gets R(x)\wedge S(y)$ and a database  $\db$ over the previous semiring such that $R(1)=a;R(2)=a;S(1)=b$ and $S(2)=b$.
Then, the output of $\Eval{Q,\Voc}{\cK}$ with input $\db$ is empty, although the body can be instantiated in four different ways, all of them producing $\kzero$ values.

The upper bounds for the similar yet different setting presented in~\cite{DBLP:conf/icdt/EldarCK24} are with respect to direct-access. In this context, Theorem~\ref{theo:free-connex-upper-bound-body} straightforwardly yields linear time preprocessing and linear time direct-access, in contrast to the loglinear time preprocessing and logarithmic time direct-access achieved in~\cite{DBLP:conf/icdt/EldarCK24}. However, in this work, tuples with zero-annotated values will not be part of the query result.

To illustrate the utility of Theorem~\ref{theo:free-connex-upper-bound-body}, consider the \fcLang query $\mA \kprod (\mU\cdot\mV^T)$ where $\mU,\mV$ are  column vectors. When this query is evaluated in a bottom-up fashion, the subexpression $(\mU\cdot\mV^T)$ will generate partial results of size $\size{\mU}\size{\mV}$, causing the entire evaluation to be of complexity $\Omega(\size{\mA} + \size{\mU}\size{\mV})$. By contrast, Theorem~\ref{theo:free-connex-upper-bound-body} tells us that we may evaluate the query in time $\bigo(\size{\mA} + \size{\mU} + \size{\mV})$.

We derive the enumeration algorithm for $\Eval{Q,\Voc}{\cK}$ by extending the algorithm of~\cite{DBLP:conf/csl/BaganDG07} to any semi-integral domain $\cK$ and taking care of the inequalities in $Q$. Then, we derive the enumeration algorithm for $\Eval{\mQ,\Sch}{\cK}$ by  reducing to $\Eval{Q,\Voc}{\cK}$, using~Corollary~\ref{cor:fcLang-equiv-free-connex-cq}.

We now prove Theorem~\ref{theo:free-connex-upper-bound-body} in two steps. We first transfer the upper bound of \cqsNotIneq over $\bools$-databases to \cqsNotIneq over $\cK$-databases. We then generalize the latter to \cqs over $\cK$-databases. 

\paragraph{Remark} In what follows, we will describe enumeration algorithms for $\Eval{Q,\Voc}{\cK}$. Formally, given an input database $\db$, these algorithms  hence have to enumerate the set $\AnsEnum{Q}{\db}$ of output tuples with their non-zero semiring annotation. In the proofs however, we will often find it convenient to instead enumerate the set $\ValEnum{Q}{\db}$ of valuations,
\[ \ValEnum{Q}{\db} \coloneqq \lbrace (\val, \fssem{Q}{\db}(\val)) \mid \val \colon \seq{x}, \fssem{Q}{\db}(\val) \neq \kzero\rbrace . \]
Note that an algorithm that enumerates $\ValEnum{Q}{\db}$ can also be used to enumerate $\AnsEnum{Q}{\db}$ by turning each valuation into an output tuple, and vice versa an algorithm that enumerates $\AnsEnum{Q}{\db}$ can also be used to enumerate $\ValEnum{Q}{\db}$: by definition every output tuple $\seq{d}\models \x$ such that $\ssem{Q}{\db}(H)(\seq{d})\neq\kzero$ induces a valuation $\val\colon\x$ such that $\fssem{Q}{\db}(\val) \neq \kzero$.  Note in particular that if either one of them can be enumerated with constant delay, then so can the other.  We use this insight repeatedly in the proofs.

\paragraph{From $\bools$-databases to $\cK$-databases over a semi-integral domain}

We start our proof of Theorem~\ref{theo:free-connex-upper-bound-body} by extending the algorithmic results of free connex CQ from the boolean semiring to any semi-integral domain. 

\begin{prop}\label{prop:upper-bound-notIneq-for-semi-integral}
    If $Q$ is a free-connex \cqNotIneq over $\Voc$ and $\cK$ a semi-integral domain then $\Eval{Q,\Voc}{\cK}\in\ConstLin$.
\end{prop}
\begin{proof}
    Because $Q$ is free-connex, it has a query plan $(T,N)$ by Proposition~\ref{prop:free-connex-iff-query-plan}. We describe different evaluation algorithms depending on the shape of $Q$ and $(T,N)$.  Let $\db$ be a $\cK$-database over $\Voc$ and assume the semi-integral domain $\cK=(K, \ksum, \kprod, \kzero, \kone)$.
  
    \begin{enumerate}
        \item When $Q$ is full, i.e, no variable is quantified, we reduce to the Boolean case as follows. Assume that $Q = H(\z) \gets R_1(\z_1)\wedge\cdots\wedge R_n(\z_n)$. In $\bigo(\size{\db})$ time construct $\Bool(\db)$, the $\bools$-database obtained from $\db$ defined as:
        \[
            R^{\Bool(\db)}\colon \seq{d}\mapsto
            \begin{cases}
                \btrue \text{ if }R^{\db}(\seq{d})\neq\kzero \\
                \bfalse \text{ otherwise, }
            \end{cases}
        \]
        for every $R\in\Voc$. Furthermore, preprocess $\db$ by creating lookup tables such that for every relation $R$ and tuple $\seq{a}$ of the correct arity we can retrieve the annotation $R^\db(\seq{a})$ in time $\bigo(\card{\seq{a}})$ time. Note that since $\Voc$ is fixed, $\card{\seq{a}}$ is constant. This retrieval is hence $\bigo(1)$ in data complexity. It is well-known that such lookup tables can be created in $\bigo(\size{\db})$ time in the RAM model.
    
        Finally, invoke the algorithm for $\Eval{Q,\Voc}{\bools}$ on $\Bool(\db)$. By Theorem~\ref{theo:cqNotIneq-free-connex-cde}, this algorithm has  $\bigo(\size{\Bool(\db)}) = \bigo(\size{\db})$ preprocessing time, after which we may enumerate $\AnsEnum{Q}{\Bool(\db)}$ with $\bigo(1)$ delay. We use the latter to enumerate $\AnsEnum{Q}{\db}$ with $\bigo(1)$ delay: whenever an element $(\seq{a}, \btrue)$ of $\AnsEnum{Q}{\Bool(\db)}$ is enumerated retrieve $k_1 := R_1(\restr{\seq{a}}{\z_1}), \dots, k_n := R_n(\restr{\seq{a}}{\z_n})$ by means of the previously constructed lookup tables. Here, $\restr{\seq{a}}{\z_n}$ denotes the projection of $\seq{a}$ according to the variables in $\z_i$. Then output $(\seq{a}, k_1 \kprod \dots \kprod k_n)$. Because the query is fixed, so is the number of lookups we need to do, and this hence takes $\bigo(1)$ additional delay in data complexity. In summary the delay between printing one output and the next is hence $\bigo(1)$ while the total processing time---comprising the construction of $\Bool(\db)$, the lookup tables and the preprocessing phase of $\Eval{Q,\Voc}{\Bool}$--- is $\bigo(\size{\db})$.
        
        We claim that the set of pairs $(\seq{a}, k_1 \kprod \dots \kprod k_n)$ hence printed is exactly $\AnsEnum{Q}{\db}$:
        \begin{itemize}
            \item If we print $(\seq{a}, k_1 \kprod \dots \kprod k_n)$ then $(\seq{a}, \btrue)$ was in $\AnsEnum{Q}{\Bool(\db)}$, which can only happen if $R_i^{\Bool(\db)}(\restr{\seq{a}}{\z_i}) = \btrue$ for every $1 \leq i \leq n$. By definition of $\Bool(\db)$ this means that $k_i = R_i^{\db}(\restr{\seq{a}}{\z_i}) \not = \kzero$. Hence, $\ssem{Q}{\db}(H)(\seq{a}) = k_1 \kprod \dots \kprod k_n$ since $Q$ is full. Because $\cK$ is a semi-integral domain, the latter product is non-zero and hence $(\seq{a}, k_1 \kprod \dots \kprod k_n) \in \AnsEnum{Q}{\db}$.
            \item If $(\seq{a}, k) \in \AnsEnum{Q}{\db}$ then by definition $k \not = \kzero$ and, because $Q$ is full,  $k = k_1 \kprod \dots \kprod k_n$ where $k_i = R_i^{\db}(\restr{\seq{a}}{\z_i})$.  As such $k_i \not = \kzero$ for $1 \leq i \leq n$. Hence, $R_i^{\Bool(\db)}(\restr{\seq{a}}{\z_i}) = \btrue$ for every $i$ and thus, $(\seq{a}, \btrue) \in \AnsEnum{Q}{\Bool(\db)}$. Therefore, $(\seq{a},  k_1 \kprod \dots \kprod k_n)$ will be enumerated by our algorithm above.
        \end{itemize}
    
        \item When the query plan $(T,N)$ is such that $N$ consists of a single node $N = \{r\}$ we next show that may compute the entire result of $Q$ in
        $\bigo(\size{\db})$ time. This implies that $\Eval{Q,\Voc}{\cK} \in \EnumClass{\size{\db}}{1}$: compute $\AnsEnum{Q}{\db}$ and use this data structure (e.g., represented as a linked list) to enumerate its elements, which trivially supports constant delay. Essentially, the argument that we use here is the same as the standard argument in the theory of acyclic conjunctive queries that Boolean acyclic queries on Boolean databases can be evaluated in linear time by a series of projection operations and semijoin operations~\cite{DBLP:conf/vldb/Yannakakis81,DBLP:books/aw/AbiteboulHV95}. We repeat the argument in the $\cK$-setting here for completeness only. For notational convenience, if $\psi$ is an $\focon$ formula then we denote by $\AnsEnum{\psi}{\db}$ the set $\AnsEnum{P_\psi}{\db}$ where $P_\psi: H(\x) \gets \psi$ with $\x$ a repetition-free list of the free variables of $\psi$. The algorithm exploits the fact that, since $(T,N)$ is a query plan for $Q$ we have the body of $Q$ is equivalent to $\varphi[T,N] = \varphi[T,r]$. Hence, it suffices to show that $\AnsEnum{\varphi[T,r]}{\db}$ may be computed in linear time.\footnote{The only difference between $Q$ and $\varphi[T,r]$ is that $Q$ may repeat some variables in the head; the result of $Q$ can hence be obtained by computing $\varphi[T,r]$ and repeating values in each result tuple as desired.} 
    
        We argue inductively on the height of $T$ that $\AnsEnum{\varphi[T,N]}{\db}$ may be computed in linear time. By Lemma~\ref{lemma:refined-query-plan}, we may assume without loss of generality that for every node $n$ with children $c_1,c_2$ in $T$ either $\var(c_1)=\var(n)$ and $\var(c_2)\subseteq\var(n)$; or $\var(c_2)=\var(n)$ and $\var(c_1)\subseteq\var(n)$.
            \begin{enumerate}
                \item When $T$ has height 1 then the root $r$ is a leaf and hence an atom, say $R(\x)$. As such, $\AnsEnum{\varphi[T,r]}{\db}$ can trivially be computed in $\bigo(\size{\db})$ time by scanning~$R^\db$.
                
                \item When $T$ has height greater than 1 we identify two cases.
                \begin{itemize}
                    \item Root $r$ has one child $c$. By Lemma~\ref{lemma:child-formula-projection} we have $\varphi[T,r] \equiv \exists \y. \varphi[T,c]$ with $\y = \var(c) \setminus \var(r)$. Here, $\varphi[T,c]$ equals $\varphi[T_c, c]$ with $T_c$ the subtree of $T$ rooted at $c$. As such, we may compute $\AnsEnum{\varphi[T,r]}{\db}$ by first computing $A := \AnsEnum{\varphi[T_c,c]}{\db}$ and then projecting on the variables in $\var(r)$.  This is in linear time by induction hypothesis and the fact that projections on $\cK$-relations can be done in linear time in the RAM model, even on general $\cK$-databases.\footnote{Create an empty lookup table. Iterate over the tuples $(\seq{a},k)$ of $A$. For every such tuple $(\seq{a},k)$, look up $\restr{\seq{a}}{\var(r)}$ in the lookup table. If this is present with $\cK$-value $l$, then set $l := l \ksum k$; otherwise add $(\restr{\seq{a}}{\var(r)},k)$ to the table. After the iteration on $A$, enumerate the entries $(\seq{b},k)$ in the lookup table and add each entry to the output when $k \not =\kzero$. This is exactly the projected result.}
        
                    \item Root $r$ has two children $c_1$ and $c_2$. We may assume w.l.o.g. that $\var(r) = \var(c_1)$ and $\var(c_2) \subseteq \var(r) = \var(c_1)$. By Lemma~\ref{lemma:children-formulas-conjunction} we have $\varphi[T,r] \equiv \varphi[T,c_1] \wedge \varphi[T,c_2]$. Note that because $\var(c_1) \subseteq \var(c_2)$ and $\free(\varphi[T,c_i]) = \var(c_i)$, the formula $\varphi[T,r]$ is hence expressing a semijoin between $\varphi[T,c_1]$ and $\varphi[T,c_2]$. Hence, we may compute $\AnsEnum{\varphi[T,r]}$ by first computing $A_1 := \AnsEnum{\varphi[T_1,c_1]}{\db}$ and $A_2 := \AnsEnum{\varphi[T_2,c_2]}{\db}$ (where $T_1$ and $T_2$ are the subtrees of $T$ rooted at $c_1$, resp. $c_2$) and then computing the semijoin between $A_1$ and $A_2$. This is in linear time by induction hypothesis and the fact that semijoins on $\cK$-relations can be done in linear time in the RAM model, even on general $\cK$-databases.\footnote{Create a lookup table on $A_2$, allowing to retrieve the $\cK$-annotation given a $\var(c_2)$-tuple. Then iterate over the entries $(\seq{a},k)$ of $A_1$. For every such tuple $(\seq{a},k)$, look up $\restr{\seq{a}}{\var(c_2)}$ in the lookup table. If this is present with $\cK$-value $l$, then output $(\seq{a}, k \kprod l)$ provided $k \kprod l \not = \kzero$.}
                \end{itemize}
            \end{enumerate}

        \item When $Q$ is not full and $N$ has more than one node we reduce to the previous two cases as follows. By Lemma~\ref{lemma:formula-of-query-plan-is-conjunction-of-frontier-induced-trees-formulas} we have that $\varphi[T,N]\equiv \varphi[T_1, n_1] \wedge \dots \wedge \varphi[T_l,n_l]$ where $F=\{ n_1, \ldots, n_l \}$ is the frontier of $N$, i.e, the nodes without children in $N$, and $T_1,\dots, T_{l}$ are the subtrees of $T$ rooted at $n_1,\dots, n_l$ respectively.  Hence, we may compute $\AnsEnum{\varphi[T,N]}{\db}$ by computing $A_i := \AnsEnum{\varphi[T_{i}, n_i]}{\db}$ for $1 \leq i \leq n$ and then taking the join of the resulting $\cK$-relations. By item (2) above, each $\AnsEnum{\varphi[T_{i}, n_i]}{\db}$ can be computed in linear time, and is hence of linear size.  The preprocessing and enumeration of the final join  can  be done as in item (1) above, since this is a full~join. \qedhere
    \end{enumerate}
\end{proof}

\paragraph{From \cqNotIneq to \cq}
An inequality atom $x \leq c$ is \emph{covered} in a \cq $Q$ if there is a relational atom in $Q$ that mentions $x$; it is \emph{non-covered} otherwise. So, in the query $H(y,z) \gets R(x,y) \wedge y \leq c \wedge z \leq d$ the inequality $y \leq c$ is covered but $z \leq d$ is not.
For a \cq $Q$ define the \emph{split} of $Q$ to be the pair $\Split(Q)=(\Qrel, \Qineq)$ where:
\begin{align*}
    \Qrel(\x)& \gets\exists \y. \bigwedge_{R(\z)\in\atoms(\varphi)}R(\z) & 
    \Qineq(\seq{u})& \gets\exists \seq{v}. \bigwedge_{w\leq c\in\atoms(\varphi), \text{non-covered}} w\leq c
\end{align*}
with $\x$ and $\y$ the set of all free resp. bound variables in $Q$ that occur in a relational atom in $Q$, and $\seq{u}$ and $\seq{v}$ the set of free resp. bound variables in $Q$ that occur in a non-covered inequality in $Q$.  
In what follows, we call $\Qrel$ the relational part of $Q$.

We note that if $Q$ does not have any non-covered inequality then $\Qineq$ is in principle the empty query. For convenience, we then set it equal to $\Qineq() \gets \exists x. x \leq 1$ in that case, which is equivalent to $\btrue$.

The next proposition relates free-connexness of a CQ with inequalities to the free-connexness of its relational part.
\begin{prop}\label{prop:free-connex-equivalence}
    Let $Q$ be a \cq and assume $\Split(Q)=(\Qrel, \Qineq)$.
    $Q$ is free-connex if, and only if, $\Qrel$ is free-connex.
\end{prop}
\begin{proof}
    This is essentially because unary atoms (like inequality atoms) can never affect the (a)cyclicity of a \cq: if a \cq is acyclic it remains so if we add or remove unary atoms from its body. \qedhere
\end{proof}

The following property relates the evaluation of a \cq (with inequalities) to the evaluation of its relational part.
    Let $Q$ be a \cq over vocabulary $\Voc$. To each relational atom $\alpha = R(\seq{x})$ we associate the query
    \[Q[\alpha]: H(\x) \gets R(\seq{x}) \wedge \iota_1 \wedge \dots \wedge \iota_\ell\]
    where $\iota_1,\dots, \iota_\ell$ are all inequality atoms in $Q$ covered by $\alpha$. We say that a database $\db$ is \emph{atomically consistent} w.r.t $Q$ if for every relation symbol $R \in \Voc$ of arity $n$,  every atom $\alpha = R(\seq{x})$ in $Q$ over relation symbol $R$, and every $n$-tuple $\seq{a}$ we have
    \[ \text{if } R^\db(\seq{a}) \not = \kzero \text{ and } \seq{a} \models \x \qquad \text{then} \qquad  \ssem{Q[\alpha]}{\db}(H)(\seq{a}) \not = \kzero .\]

Intuitively, if $\db$ is atomically consistent, then every tuple in a relation $R$ that satisfies the atom $\alpha = R(\seq{x})$\footnote{Note that $\x$ may repeat variables, so an atom like $S(x,y,x)$ requires the first and third components of $\seq{a}$ to be equal.} also satisfies all inequalities covered by $\alpha$. Note that we may always make a given $\db$ atomically consistent with $Q$ in $\bigo(\size{\db})$ time: simply scan each relation $R$ of $\db$ and for each tuple $\seq{a}$ loop over the $R$-atoms $\alpha = R(\x)$ in $Q$. If $\seq{a} \models \x$ then check that the constraints imposed by $Q[\alpha]$ are satisfied. Since $Q[\alpha]$ is a full query, this can be done in constant time per tuple in data complexity. We call this process \emph{atomically reducing $\db$}, that is, of making $\db$ atomically-consistent with $Q$ .

The following claims are straighforward to verify.

\begin{clm}\label{claim:equal-after-reduction}
    Let $Q$ be a \cq over vocabulary $\Voc$, let $\db$ be a database and let $\db'$ be its atomic reduction. Then $\AnsEnum{Q}{\db} = \AnsEnum{Q}{\db'}$. 
\end{clm}

\begin{clm}\label{claim:cq-is-cart-prod-on-consistent-db}
    Let $Q(\x)$ be a \cq with $\Split(Q)=(\Qrel(\x_1), \Qineq(\x_2))$. If $\db$ is atomically consistent with $Q$ then $\fssem{Q}{\db}(\val) =\fssem{\Qrel}{\db}(\restr{\val}{\x_1}) \kprod \fssem{\Qineq}{\db}(\restr{\val}{\x_2})$ for every valuation $\val\colon\x$.
\end{clm}

\begin{lem}\label{lem:ineq-part-in-cdolin}
    Let $Q$ be a \cq over vocabulary $\Voc$ with $\Split(Q)=(\Qrel, \Qineq)$ and let $\cK$ be a semiring (not necessarily a semi-integral domain). Then $\Eval{\Qineq,\Voc}{\cK} \in \ConstLin$.
\end{lem}
\begin{proof}
    Let $x_1,\dots,x_n$ be the free variables of $\Qineq$ and $y_1,\dots,y_m$ its quantified variables. Let $\num{z}$ denote the number of inequalities mentioning variable $z$ (free or quantified). Assume w.l.o.g. that $\Qineq$ is of the form
    \[ H(x_1,\dots,x_n) \gets \exists y_1,\dots, y_m. \bigwedge_{1 \leq i \leq n} \bigwedge_{j \leq \num{x_i}} (x_i \leq c^i_j) \wedge \bigwedge_{1 \leq p\leq m} \bigwedge_{q \leq \num{y_j}} (y_p \leq d^p_q). \] 

    Let $\db$ be a $\cK$-database over $\Voc$.  It is straightforward to verify that $((a_1,\dots,a_n), k) \in \AnsEnum{\Qineq}{\db}$ if, and only if,
    \begin{itemize}
        \item For every $1 \leq i \leq n$ we have that $a_i \leq c_i$ where $c_i := \min_{1 \leq j \leq \num{x_i}} \db(c^i_j)$; 
        \item $k \not = \kzero$; and
        \item $k = (\underbrace{\kone \kprod \dots \kprod \kone}_{\num{y_1} \text{ times}}) \ksum \dots \ksum (\underbrace{\kone \kprod \dots \kprod \kone}_{\num{y_m} \text{ times}}) = \underbrace{\kone \ksum \dots \ksum \kone}_{m \text{ times}}$
    \end{itemize}
    In particular, the annotation $k$ is the same for each output tuple. Because $m$ depends only on the query, we may compute $k$ in constant time. Furthermore, we may compute the numbers $c_i \in \bbN$ in $\bigo(\size{\db})$ time. Computing $c_i$ and $k$ constitutes the preprocessing step of our enumeration algorithm. Having these numbers computed, it is then straightforward to enumerate the tuples: indicate end of enumeration immediately if $k = \kzero$; otherwise enumerate all tuples $((a_1,\dots,a_n), k) \in \AnsEnum{\Qineq}{\db}$ with constant delay by means of a nested loop.
\end{proof}

\begin{prop}\label{prop:relational-enum-implies-withIneq-enum}
  Let $Q$ be a \cq over $\Voc$ with $\Split(Q)=(\Qrel, \Qineq)$. If it holds that
  $\Eval{\Qrel,\Voc}{\cK}\in\ConstLin$ and $\cK$ is a semi-integral domain then
  $\Eval{Q,\Voc}{\cK}\in\ConstLin$.
\end{prop}
\begin{proof}

  Let $\db$ be the input database.  First make $\db$ atomically consistent with $Q$, which takes time $\bigo(\size{\db})$. Let $\db'$ be the resulting database. Note that, by Claim~\ref{claim:equal-after-reduction}, we have $\ValEnum{Q}{\db} = \ValEnum{Q}{\db'}$. Because $\size{\db'} = \bigo(\size{\db})$ it hence suffices to show that we may enumerate $\ValEnum{Q}{\db'}$ with constant delay after linear time preprocessing.

  Thereto, we reason as follows. Because $\Eval{\Qrel,\Voc}{\cK}\in\ConstLin$ we may compute a data structure in $\bigo(\size{\db'})$ time
  that allows us to enumerate $\ValEnum{\Qrel}{\db'}$ with constant delay. Furthermore, by Lemma~\ref{lem:ineq-part-in-cdolin} we may compute in $\bigo(\size{\db'})$ time another data structure that allows us to enumerate $\ValEnum{\Qineq}{\db'}$ with constant delay. Computing both data structures constitutes our preprocessing step. By Claim~\ref{claim:cq-is-cart-prod-on-consistent-db} we can then use both data structures to enumerate $\ValEnum{Q}{\db'}$ with constant delay as follows:
  \begin{itemize}
    \item For each $(\val_1, k_1) \in \ValEnum{\Qrel}{\dbQRel}$
    \begin{itemize}
        \item For each $(\val_2,k_2) \in \ValEnum{\Qineq}{\db}$, output $(\val_1 \cup \val_2, k_1 \kprod k_2)$.
    \end{itemize}
  \end{itemize}
  Note that $k_1 \kprod k_2 \not = \kzero$ as required since $k_1 \not = \kzero$, $k_2 \not = \kzero$ and $\cK$ is a semi-integral domain.
\end{proof}

We can finally prove Theorem~\ref{theo:free-connex-upper-bound-body}.

\begin{proof}[Proof of Theorem~\ref{theo:free-connex-upper-bound-body}]
    If $Q=(\Qrel,\Qineq)$ is free-connex then by Proposition~\ref{prop:free-connex-equivalence} we have that $\Qrel$ is free-connex. By Proposition~\ref{prop:upper-bound-notIneq-for-semi-integral} and because $\cK$ is a semi-integral domain, $\Eval{\Qrel,\Voc}{\cK}\in\ConstLin$. And finally due to Proposition~\ref{prop:relational-enum-implies-withIneq-enum}, the latter implies $\Eval{Q,\Voc}{\cK}\in\ConstLin$.

    Now we prove the second part of the result, i.e., achieve the same processing complexity for the matrix evaluation problem $\Eval{\mQ,\Sch}{\cK}$. Fix a matrix-to-relational schema encoding $\Rel$ on $\Sch(\mQ)$ such that the head atom of $Q$ is binary. The former requirement arises purely due to $\AnsEnum{\mQ}{\I}$ being defined as $\lbrace (i,j, \sem{\mQ}{\I}(\mH)_{i,j}) \mid \sem{\mQ}{\I}(\mH)_{i,j} \neq\kzero, i \leq \I(\alpha), j \leq \I(\beta) \rbrace$ where $\mQ = \mH := e$ and $\mH\colon (\alpha,\beta)$.

	By Corollary~\ref{cor:fcLang-equiv-free-connex-cq} there exists a free-connex CQ $Q$ over vocabulary $\Voc(Q) = \Rel(\Sch(\mQ))$ that simulates $\mQ$ under $\Rel$. As such, we may reduce $\Eval{\mQ,\Sch}{\cK}$ to $\Eval{Q,\Voc}{\cK}$: given a matrix instance $\I$ over $\Sch$, compute $Rel(\I)$ and call the enumeration algorithm for $\Eval{Q,\Voc}{\cK}$. By definition of simulation, the set $\AnsEnum{Q}{\Rel(\I)}$ is exactly the same as $\AnsEnum{\mQ}{\I}$. Hence, because $\Eval{Q,\Voc}{\cK} \in \ConstLin$ and because $\size{\Rel(\I)} = \bigo(\size{\I})$ it follows that $\Eval{\mQ,\Sch}{\cK} \in \EnumClass{\size{\I}}{1}$.
\end{proof}

\subsection{Lower bounds for free-connex queries} \label{subsec:fc-lowerbound}
Next, we show how to extend the lower bounds of~\cite{DBLP:conf/csl/BaganDG07}
and~\cite{DBLP:journals/siglog/BerkholzGS20} to our setting. As in those works,
our lower bounds are for \cq \emph{without self-joins}. Further, we need some additional
restrictions for inequalities. Recall that an inequality $x \leq c$ in $Q$ is
\emph{covered}, if there exists a relational atom in $Q$ that mentions~$x$.  We
say that $Q$ is \emph{constant-disjoint} if (i) for all covered inequalities
$x \leq c$ we have $c \not =1$, and (ii) for all pairs $(x \leq c, y \leq d)$ in
$Q$ of covered inequality $x \leq c$ and non-covered inequality $y \leq d$ if
$c = d$ then $y \not \in \free(Q)$. In other words, if a constant symbol other
than~$1$ occurs in both a covered and non-covered inequality in $Q$, then it
occurs with a bound variable in the non-covered inequality.

A semiring $\cK=(K, \ksum, \kprod, \kzero, \kone)$
is \emph{zero-sum free} \cite{hebisch1996semirings,hebisch1998semirings} if for all $a,b\in K$ it holds that $a\ksum b=\kzero$ implies $a=b=\kzero$.
The \emph{subsemiring of $\cK$ generated} by $\kzero$ and $\kone$ is the semiring $\cK' = (K', \ksum, \kprod, \kzero, \kone)$ where $K'$ is the smallest set closed that (i) contains $\kzero$ and $\kone$ and (ii) is closed under $\ksum$ and $\kprod$.

We will extend the lower bound from~\cite{DBLP:conf/csl/BaganDG07,DBLP:journals/siglog/BerkholzGS20} as follows.

\begin{thm}\label{theo:lower-bound-free-connex-body}
    Let $Q$ be a \cq over $\Voc$ without self-joins and constant-disjoint. Let 
    $\cK$ be a semiring such that the subsemiring generated by $\kzero_{\cK}$ and
    $\kone_{\cK}$ is zero-sum free. If $\Eval{Q,\Voc}{\cK}$ can be evaluated with linear-time preprocessing and constant-delay, then $Q$
    is free-connex, unless either the Sparse Boolean Matrix Multiplication, the
    Triangle Detection, or the $(k,k+1)$-Hyperclique conjecture is false.
\end{thm}
It is important to note that most semirings used in practice, like $\bbB$, $\bbN$, and $\bbR$, are such that the subsemiring generated by $\kzero_{\cK}$ and $\kone_{\cK}$ is zero-sum free. In contrast, lower bounds in \cite{DBLP:conf/icdt/EldarCK24} require the semiring to be \textit{idempotent}, meaning that $a\ksum a = a$ for all $a\in K$.

To see why the constant-disjointness condition is necessary, consider the CQ:
\[
Q: H(x,z) \gets \exists y. R(x,y) \wedge S(y,z) \wedge y \leq 1.
\] This query violates condition (i) of constant-disjointness, and is not free-connex. Nevertheless, its evaluation is in $\ConstLin$: because $y$ can take only a single value, we can evaluate $Q$ by doing preprocessing and enumeration for $Q'\colon H(x,y,z) \gets R(x,y) \wedge S(y,z) \wedge y \leq 1$ instead, which is free-connex, hence in \ConstLin. Whenever a tuple $(x,y,z)$ is enumerated for $Q'$ we yield the tuple $(x,z)$ for the enumeration of $Q$. This is also constant-delay and without duplicates due to the inequality $y \leq 1$.

A similar argument holds for condition (ii) of
constant-disjointness. Consider
\[ P\colon H(x,z,u) \gets \exists y. R(x,y) \wedge S(y,z) \wedge y \leq c \wedge
  u \leq c,\] which violates condition (ii) of constant-disjointess. Again, this
query is not free-connex but its evaluation is in \ConstLin. To see why,
consider the query
$P'\colon H(x,y,z) \gets R(x,y) \wedge S(y,z) \wedge y \leq c$, which is
free-connex and therefore in \ConstLin. To evaluate $P$ we simply do all the
(linear-time) preprocessing for $P'$ and use $P'$ enumeration procedure to
enumerate $P$ with constant delay.  This works as follows. During the
enumeration of $P$ maintain a lookup table $L$ mapping $(x,z)$ tuples to the
smallest natural number $u$ such that all tuples $(x,z,u)$ that have already
been output for $P$ satisfy $u \leq L(x,z)$. If $L(x,z) = 0$ then no entry
occurs in the lookup table. To enumerate $P$ we invoke $P'$ enumeration
procedure. When a tuple $(x,y,z)$ is enumerated by $P'$, we output
$(x,z,L(x,z)+1)$ for the enumeration of $P$. Additionally, we update
$L(x,z) := L(x,z)+1$. Once $P'$'s enumeration is exhausted, we iterate over the entries of the lookup table, and for each such entry $(x,z)$ output all remaining tuples $(x,z, u)$ with $L(x,z) < u' \leq c$. This last step is not necessarily constant-delay (there may be an unbounded number of entries with $L(x,z) = c$, yielding no output),  but this can be fixed by removing $(x,z)$ from $L$ during the enumeration of $P'$ whenever $L(x,z)$ is updated to become $c$. The constraint that $y \leq c \wedge u \leq c$ ensures that the enumeration is still correct.

Theorem~\ref{theo:lower-bound-free-connex-body} allows to also derive lower bounds for $\conjLang$.
Unfortunately, given the asymmetry between the relational and matrix settings,
the lower bounds do not immediately transfer from the relational to the matrix
setting. Specifically, we need a syntactical restriction for \conjLang that
implies the constant-disjointedness restriction in the translation
of~Theorem~\ref{theo:free-connex-binary-cq-equiv-foTwo}. In order to achieve
this, the prenex normal form of \conjLang expressions (defined in
Section~\ref{sec:sumlang-and-fo}) proves useful. Let
$e = \Sigma \seq{\mv}. \hspace{1ex}\ms_1\times\cdots\times\ms_n
\times \mx\cdot \my^T$ be a \conjLang sentence in prenex normal form. A vector
variable $\mv \in \seq{\mv}$ is \emph{covered} in $e$ if (i) there exists a scalar
subexpression $\ms_i$ of the form $\mv^T\cdot\mA\cdot \mau$ or
$\mau^T\cdot\mA\cdot \mv$; or (ii) if there exists scalar subexpression $\ms_i$
doing a vector multiplication $\mv^T\mau$ or $\mau^T\mv$ with $\mau$ covered in $e$. (Note that the latter notion is recursive.) The sentence $e$ is \emph{constant-disjoint} if for every pair
$\mv,\mw \in \seq{\mv}$ of quantified vector variables of the same type,
$\mv\colon(\gamma,1)$ and $\mw\colon(\gamma,1)$, if $\mv$ is covered and $\mw$
is not then $\mw\notin\{\mx,\my\}$. A \conjLang sentence $e\colon(\alpha,\beta)$ is \emph{constant-disjoint} if its conversion into prenex normal form is constant-disjoint.

The former is straightforwardly extended to a \conjLang query $\mQ$. The following lower bound is now attainable.

\begin{cor}\label{cor:lower-bound-free-connex-matrix}
Let $\mQ$ be a \conjLang query over $\Sch$ such that $\mQ$ does not repeat
matrix symbols and $\mQ$ is constant-disjoint. Let $\cK$ be a semiring such
that the subsemiring generated by $\kzero_{\cK}$ and $\kone_{\cK}$
is zero-sum free.  If $\Eval{\mQ,\Sch}{\cK}$ can be evaluated with linear-time
preprocessing and constant-delay, then $\mQ$ is equivalent to a \fcLang query,
unless either the Sparse Boolean Matrix Multiplication, the Triangle
Detection, or the $(k,k+1)$-Hyperclique conjecture is false.
\end{cor}

The remainder of this section is devoted to proving Theorem~\ref{theo:lower-bound-free-connex-body} and Corollary~\ref{cor:lower-bound-free-connex-matrix}.

\paragraph{Lower bound for free-connex CQs}
We begin by proving Theorem~\ref{theo:lower-bound-free-connex-body}, which is done in two steps. First we show that constant-disjointness is a sufficient condition for us to be able to reduce the evaluation problem of \cqsNotIneq over $\bools$-databases to the evaluation problem of \cqs over $\bools$-databases. Second, we show that the evaluation problem of \cqs over $\bools$-databases can be reduced to the evaluation problem of \cqs over $\cK$-databases, assuming that the subsemiring generated by $\kzero_{\cK}$ and $\kone_{\cK}$ is zero-sum free.

\begin{prop}\label{prop:ineq-enum-implies-relational-enum}
    Let $Q$ be a constant-disjoint \cq over  $\Voc$ with $\Split(Q)=(\Qrel, \Qineq)$. If $Q$ is
    constant-disjoint and $\Eval{Q,\Voc}{\bools}\in\ConstLin$ then
    $\Eval{\Qrel,\Voc}{\bools}\in\ConstLin$.
\end{prop}
\begin{proof}
    We reduce $\Eval{\Qrel,\Vocrel}{\bools}$ to $\Eval{Q,\Voc}{\bools}$. Denote the free variables of $Q$, $\Qrel$ and $\Qineq$ by $\x$, $\x_1$ and $\x_2$, respectively. Note that $\x_1$ and $\x_2$ are disjoint and $\var(\x) = \var(\x_1) \cup \var(\x_2)$. Consider an arbitrary database $\db_{\text{rel}}$ over $\Voc$, input to $\Eval{\Qrel,\Voc}{\bools}$. Construct the database $\db$, input to $\Eval{Q,\Voc}{\bools}$ as follows. 
  
    Recall that all data values occurring in tuples in $\dbrel$ are non-zero natural numbers. Scan $\dbrel$ and compute the largest natural number $M$ occurring in $\dbrel$.  Then construct the database $\db$ by setting
    \begin{itemize}
        \item $R^{\db} = R^{\dbrel}$ for each relation symbol $R \in \Voc$;
        \item $db(c) := M$ for all constant symbols $c \in \Voc$ that occur in a covered inequality in $Q$. Note that $c \not = 1$ because $Q$ is constant-disjoint, and hence we are allowed to map $c \mapsto M$;
        \item $db(c) := 1$ for all other constant symbols $c \in \Voc$.
    \end{itemize}
     The computation of $\db$ is clearly in $\bigo(\size{\dbrel})$ and $\size{\db} = \size{\dbrel}$.
  
    Note that, because $\Qrel$ consists of relational atoms only, its evaluation is hence independent of the values assigned to the constant symbols by
    $\dbrel$. Therefore, $\AnsEnum{\Qrel}{\dbrel} = \AnsEnum{\Qrel}{\db}$. To obtain the proposition, it hence suffices to show that $\AnsEnum{\Qrel}{\db}$ may be enumerated with constant delay after further linear time preprocessing. We do so by showing that $\ValEnum{\Qrel,\db}$ may be enumerated with constant delay after linear time preprocessing as follows. Observe that $\db$ is atomically consistent with $Q$. This is by construction since for every covered inequality $x \leq c$ we have $\db(c)$ equal to $M$, with $M$ larger than any value in $\db$. Hence, by Claim~\ref{claim:cq-is-cart-prod-on-consistent-db} we have
    \[ \ValEnum{Q}{\db} = \{ (\val_1 \cup \val_2, \btrue) \mid (\val_1, \btrue) \in \ValEnum{\Qrel}{\db}, (\val_2, \btrue) \in \ValEnum{\Qineq}{\db} \}. \] As such, $\ValEnum{\Qrel}{\db} = \{ (\restr{\val}{\x_1}, \btrue) \mid (\val, \btrue) \in \ValEnum{Q}{\db} \}$ and we may hence enumerate $\ValEnum{\Qrel}{\db}$ by enumerating the elements of $\ValEnum{Q}{\db}$ and for each such element output $(\restr{\val}{\x_1}, \btrue)$. Note that the resulting enumeration is necessarily duplicate-free as required: by definition of $\db$ the set $\ValEnum{\Qineq}{\db}$ consists of a single
    element: the pair $(\x_2 \mapsto 1, \btrue)$. To see why this is neccesarily the case observe that, by definition, all free variables $y \in \x_2$ of $\Qineq$ must  occur in an non-covered inequality $y \leq c$. Because $Q$ is constant-disjoint this constant $c$ is such that there is no covered inequality $z \leq d$ in $Q$ with $c = d$. (If there were, $y$ would be non-free.) This means that we have set $c^\db = 1$. Therefore, for every free variable $y$ of $\Qineq$, we neccesarily have that $\val(y) = 1$ in a resulting valuation. As such, all valuations in $\ValEnum{Q}{\db}$ are constant on the variables in $\x_2$ and duplicates can hence not occur when projecting on $\x_1$. The proposition then follows because we may enumerate $\ValEnum{Q}{\db}$ with $\bigo(1)$ delay after linear time preprocessing by assumption. \qedhere
\end{proof}
  
\begin{prop}\label{prop:enumeration-of-bool-using-zsf-semiring}
    Let $Q$ be a \cq over $\Voc$ and $\cK$ a semiring such that the subsemiring
    generated by $\kzero_{\cK}$ and $\kone_{\cK}$ is non-trivial and zero-sum free.  If
    $\Eval{Q,\Voc}{\cK}\in\ConstLin$ then $\Eval{Q,\Voc}{\bools}\in\ConstLin$.
\end{prop}
  
\begin{proof}
    Let $\cK=(K, \ksum, \kprod, \kzero_{\cK}, \kone_{\cK})$ be a a semiring such that the subsemiring generated by $\kzero_{\cK}$ and $\kone_{\cK}$ is non-trivial and zero-sum free.  We reduce $\Eval{Q,\Voc}{\bools}$ to $\Eval{Q,\Voc}{\cK}$ as follows. Let $\db_\bools$ be a $\bools$-database over
    $\Voc$, input to $\Eval{Q,\Voc}{\bools}$. Construct the $\cK$-database $\db_\cK$, input to $\Eval{Q,\Voc}{\cK}$ by setting
    \begin{itemize}
        \item for every relation symbol $R\in\Voc$
        \[
            \fssem{R}{\db_\cK}\colon \seq{d}\mapsto
            \begin{cases}
                \kone_{\cK} \text{ if }R^{\db_\bools}(\seq{d})=\btrue \\
                \kzero_{\cK} \text{ otherwise. }
            \end{cases}
        \]

        \item $\db_\cK(c) := \db_\bools(c)$ for every constant symbol $c$.
    \end{itemize}
    Note that this takes time $\bigo(\size{\db_\bools})$.
  
    By assumption, $\Eval{Q,\Voc}{\cK}\in\ConstLin$ thus the set $\ValEnum{Q}{\db_\cK}$ can be enumerated with constant delay after a linear time preprocessing.  Now, it suffices to do the following: for every outputed tuple $(\val, \fssem{Q}{\db_\cK}(\val))$ we output $(\val, \btrue)$.  This forms a constant delay enumeration for the set $\ValEnum{Q}{\db_\bools}$, since we claim that for every valuation $\val$ we have $(\val,\btrue)\in\ValEnum{Q}{\db_\bools}$ if and only if $(\val,k)\in\ValEnum{Q}{\db_\cK}$ for some $\kzero_{\cK}\neq k\in K$. We next  argue why this claim holds.
  
    Let $\varphi=\exists \y. R_1(\z_1)\wedge\cdots\wedge R_n(\z_m)\wedge w_1\leq c_1\wedge\cdots\wedge\w_m\leq c_m$ be the body of $Q(\x)$ and assume $\val\colon\x$.
  
    First, if $(\val, \fssem{Q}{\db_\cK}(\val))\in\ValEnum{Q}{\db_\cK}$ then in particular $\fssem{Q}{\db_\cK}(\val)\neq\kzero_{\cK}$. Because
    \begin{align*}
        \fssem{Q}{\db_\cK}(\val) &= \bigksum_{\mu\colon \var(\varphi) \text{ s.t.} \restr{\mu}{\x} = \val} \fssem{\varphi}{\db_\cK}(\mu) \\
        &= \bigksum_{\mu\colon \var(\varphi) \text{ s.t.} \restr{\mu}{\x} = \val} R_1^{\db_\cK}(\restr{\mu}{\z_1})\kprod\cdots\kprod R_n^{\db_\cK}(\restr{\mu}{\z_n})\kprod \\
        &\qquad\qquad\qquad\qquad\qquad\fssem{w_1\leq c_1}{\db_\cK}(\restr{\mu}{w_1})\kprod\cdots\kprod \fssem{w_m\leq c_m}{\db_\cK}(\restr{\mu}{w_m}) 
    \end{align*}
    we know in particular that there exists some $\mu$ such that
    \[
        R_1^{\db_\cK}(\restr{\mu}{\z_1}) \kprod\cdots\kprod R_n^{\db_\cK}(\restr{\mu}{\z_n})\kprod \fssem{w_1\leq c_1}{\db_\cK}(\restr{\mu}{w_1})\kprod\cdots\kprod \fssem{w_m\leq c_m}{\db_\cK}(\restr{\mu}{w_m}) \neq\kzero_{\cK}.
    \]
    This is because in any semiring, a sum of terms can be non-$\kzero$ only if one term is non-$\kzero$. Moreover, a product is non-$\kzero$ only if all of
    its factors are non-$\kzero$. By construction, $R_i^{\db_\cK}(\restr{\mu}{\z_i}) \not = \kzero$ for all $i=1,\ldots,n$ and $\fssem{w_j\leq c_j}{\db_\cK}(\restr{\mu}{w}) \not = \kzero_{\cK}$ for all $j=1,\ldots,m$ only if $R_i^{\db_\bools}(\restr{\mu}{\z_i})=\btrue$ and $\fssem{w_j\leq c_j}{\db_\bools}(\restr{\mu}{w})=\btrue$ for all $i$ and $j$. Thus $(\val, \btrue)\in\ValEnum{Q}{\db_\bools}$.
  
    Second, if $(\val, \btrue)\in\ValEnum{Q}{\db_\bools}$ then in particular $\fssem{Q}{\db_\bools}(\val)=\btrue$. Because
    \begin{align*}
      \fssem{Q}{\db_\bools}(\val) &= \bigvee_{\mu\colon \var(\varphi) \text{ s.t.} \restr{\mu}{\x} = \val} R_1^{\db_\bools}(\restr{\mu}{\z_1})\wedge\cdots\wedge R_n^{\db_\bools}(\restr{\mu}{\z_n}) \\
      &\qquad\qquad\qquad\qquad\wedge \fssem{w_1\leq c_1}{\db_\bools}(\restr{\mu}{w_1})\wedge\cdots\wedge \fssem{w_m\leq c_m}{\db_\bools}(\restr{\mu}{w_m}) 
    \end{align*}
    this implies there is some valuation $\mu$ such that
    \[
        R_1^{\db_\bools}(\restr{\mu}{\z_1})\wedge \cdots \wedge R_n^{\db_\bools}(\restr{\mu}{\z_n})\wedge\fssem{w_1\leq c_1}{\db_\bools}(\restr{\mu}{w_1})\wedge\cdots\wedge \fssem{w_m\leq c_m}{\db_\bools}(\restr{\mu}{w_m})=\btrue.
    \]
    In particular, $R_i^{\db_\bools}(\restr{\mu}{\z_i})=\btrue$ for all $i=1,\ldots,n$ and $\fssem{w_j\leq c_j}{\db_\bools}(\restr{\mu}{w_j})=\btrue$ for $j=1,\ldots,m$. Then, by construction of $\db_\cK$ we have $R_i^{\db_\cK}(\restr{\mu}{\z_i})=\kone_{\cK}$ for all $i=1,\ldots,n$ and $\fssem{w_j\leq c_j}{\db_\cK}(\restr{\mu}{w_j})=\kone_{\cK}$ for $j=1,\ldots,m$. Hence, 
    \[
        R_1^{\db_\cK}(\restr{\mu}{\z_1}) \kprod\cdots\kprod R_n^{\db_\cK}(\restr{\mu}{\z_n})\kprod \fssem{w_1\leq c_1}{\db_\cK}(\restr{\mu}{w_1})\kprod\cdots\kprod \fssem{w_m\leq c_m}{\db_\cK}(\restr{\mu}{w_m}) = \kone_{\cK}.
    \]
    Thus, since the subsemiring generated by $\kzero_{\cK}$ and $\kone_{\cK}$ is non-trivial and zero-sum free,
    \begin{align*}
        \fssem{Q}{\db_\cK}(\val) &= \bigksum_{\mu\colon \var(\varphi) \text{ s.t.} \restr{\mu}{\x} = \val} \fssem{\varphi}{\db_\cK}(\mu) \not = \kzero_{\cK} \qedhere
    \end{align*}
  \end{proof}
  
Recall that by convention we only consider commutative and non-trivial semirings throughout the paper. As such when we consider an arbitrary semiring $\cK$, then we assume that this semiring is commutative and non-trivial. For such semirings, it is straightforward to check that the sub-semiring generated by $\kzero$ and $\kone$ is also commutative and non-trivial. (We note that we have only explicitly add the ``non-trivial'' requirement in the previous proposition because the proof uses this property.) Given this discussion, we can now prove Theorem~\ref{theo:lower-bound-free-connex-body}.

\begin{proof}[Proof of Theorem~\ref{theo:lower-bound-free-connex-body}]
    Let $\Split(Q) = (\Qrel,\Qineq)$ and assume $\Eval{Q,\Voc}{\cK}\in\ConstLin$. Because the subsemiring of $\cK$ generated by $\kzero_{\cK}$ and $\kone_{\cK}$ is non-trivial and zero-sum free, $\Eval{Q,\Voc}{\bools}\in\ConstLin$ holds by Proposition~\ref{prop:enumeration-of-bool-using-zsf-semiring}.  This implies $\Eval{\Qrel,\Voc}{\bools}\in\ConstLin$, due to Proposition~\ref{prop:ineq-enum-implies-relational-enum}. Theorem~\ref{theo:cqNotIneq-free-connex-cde} yields that $\Qrel$ is free-connex, unless either the Sparse Boolean Matrix Multiplication, the Triangle Detection, or the $(k,k+1)$-Hyperclique conjecture is false. Finally, since $\Qrel$ is free-connex if and only if $Q$ is free-connex (Proposition~\ref{prop:free-connex-equivalence}), this immediately implies that $Q$ is free-connex, unless one of these hypothesis is false.
\end{proof}

\paragraph{Lower bound for \fcLang queries}

We now proceed to prove Corollary~\ref{cor:lower-bound-free-connex-matrix}. %
A necessary ingredient to prove the lower bound is to have \emph{enumeration-suited} matrix-to-relational encodings.
    Let $\mQ\coloneqq \mH = e$ be a \conjLang query. A matrix-to-relational encoding $\Rel$ is \emph{enumeration-suited} for $\mQ$\ if
    \begin{itemize}
        \item $\Rel(\mA)$ is a nullary relation if $\mA$ is of scalar type and a unary relation if $\mA$ is of vector type (but not scalar), for every $\mA\in\argSch{\mQ}\setminus \mH$.
        \item $\Rel(\mH)$ is a binary relation.
    \end{itemize}
   
  One can easily see that for every $\mQ$ there exists some matrix-to-relational encoding $\Rel$ that is enumeration-suited for $\mQ$ (i.e., choose the natural encoding that maps scalars to nullary relations, vectors to unary relations, and matrices to binary relations). Therefore, to prove Corollary~\ref{cor:lower-bound-free-connex-matrix} we may always choose a $\Rel$ that is enumeration-suited, and apply the following proposition to move from the matrix setting to the relational setting.

\begin{prop}\label{prop:simulating-cq-is-constant-disjoint}
    Let $\mQ= \mH \coloneqq e$ be a $\conjLang$ query and let $\Rel$ be an enumeration-suited matrix-to-relational encoding for $\mQ$. If $\mQ$ is constant-disjoint then there exists a constant-disjoint CQ that simulates $\mQ$ w.r.t. $\Rel$. 
\end{prop}
\begin{proof}
Assume that $e$, when converted into prenex normal form, is of the form $e=\Sigma \mx, \my, \mv_1, \dots, \mv_k. \hspace{1ex}\ms_1\times\cdots\times\ms_n \times \mx\cdot \my^T$ and that this prenex normal form is constant-disjoint. Let $\seq{\mv} = \mx, \my, \mv_1, \dots, \mv_k$.

The proof proceeds in two steps. (i) We first  show that because $\Rel$ is enumeration-suited, the  \cq $Q$ \emph{with equality atoms} simulating $\mQ$ that is obtained through the translation given in the proof Proposition~\ref{prop:conjlang-to-cq},  is constant-disjoint. (ii) Then we show that eliminating equality atoms from $Q$ yields a CQ $Q'$ that is still constant disjoint. 

Formally for point (i) to make sense, we say that a \cq with equality atoms is constant-disjoint if satisfies the same conditions as for normal CQs, but with the modification that now an inequality atom $z\leq c$ is covered if variable $z$ is covered: it appears in a relational atom or it appears in an equality atom $z=w$ and $w$ is covered. Note that this definition is recursive. For example, in the formula $\exists u,v,w. A(x,u)\wedge u=v\wedge v=w$ we have that $u,v,w$ are covered.

We first prove (i). The $Q$ CQ simulating $\mQ$ obtained by Proposition~\ref{prop:conjlang-to-cq} is of the form  $Q\colon H(x,y)\gets \exists v_1,\ldots,v_k. a_1\wedge\cdots\wedge a_n \wedge x\leq\alpha \wedge y\leq\beta$ where each $a_i$ is determined as follows, for $1 \leq i \leq n$. Let non-bold symbol $A$ denote $\Rel(\mA)$, and let non-bold $w$ be the FO variable selected respectively for $\mw\in \{\mx,\my,\mv_1,\dots,\mv_k\}$.

\begin{itemize}
\item If $\ms_i = \mv^T\cdot\mA\cdot \mw$ then:
  \begin{itemize}
  \item $a_i\coloneqq A(v,w)$ if $A$ is binary. Note that since $\Rel$ is enumeration suited, this case can only happen if  $\mA\colon(\alpha, \beta)$ with $\alpha\neq 1$ and $\beta\neq 1$. Due to well-typedness, $\mv\colon(\alpha, 1)$ and $\mw\colon(\beta,1)$.
  \item $a_i\coloneqq A(v)\wedge w\leq 1$ if $A$ is
    unary and $\mA\colon (\alpha,1)$.  Note that since $\Rel$ is enumeration suited, necessarily $\alpha \not =1$. For later use we remark that since $\ms_i$ is well-typed, $\mv\colon (\alpha, 1)$ and $\mw\colon (1,1)$ in this case.
  \item $a_i\coloneqq v\leq 1 \wedge A(w)$ if $A$ is unary and $\mA\colon (1,\alpha)$. Since $\Rel$ is enumeration-suited, necessarily $\alpha \not = 1$. For later use we remark that since $\ms_i$ is well-typed, $\mv\colon (1,1)$ and $\mw\colon (\alpha, 1)$. 
			\item $a_i\coloneqq v\leq 1\wedge w\leq 1 \wedge A()$ if $A$ is nullary and $\mA\colon (1,1)$. Note that necessarily $\mv\colon (1,1)$ and $\mw\colon(1,1)$ in this case.
		\end{itemize}
		\item If $\ms_i = \mv^T\cdot\mw$ with $\mv\colon(\gamma,1)$ and $\mw\colon(\gamma,1)$ then:
		\begin{itemize}
			\item $a_i\coloneqq v\leq\gamma$ if $\mv=\mw$.
			\item $a_i\coloneqq v\leq\gamma\wedge w\leq\gamma\wedge v=w$ if $\mv\neq\mw$. 
		\end{itemize}
	\end{itemize}

    Recall that a \cq $Q$ is constant-disjoint if (1) for all covered inequalities $z \leq c$ we have $c \not =1$, and (2) for all pairs $(v \leq c, y \leq d)$ in $Q$ of covered inequality $w \leq c$ and non-covered inequality $w \leq d$ if $c = d$ then $w \not \in \free(Q)$.

  	We start by showing that in $Q$ condition (1) always holds. Consider an inequality $z_{1} \leq 1$ in $Q$. This necessarily appears in some subformula $a_{i_1}$ with $\ms_{i_1} =\mv_{i_1}^T\cdot\mA\cdot \mw_{i_1}$ and $\mz_{1} \in \{\mv_{i_1}, \mw_{i_1}\}$. 
    By inspection of the construction above we observe that necessarily $\mz_1\colon (1,1)$. We further observe that $z_1$ does not appear in a relational atom in the subformula $a_{i_1}$. Hence, if $z$ is covered in $Q$, it must be because there are other subformulas $a_{i_2},\dots, a_{i_\ell}$ that contain equality atoms $z_{j-1} = z_j$ for $2 \leq j \leq \ell$, as well as a subformula $a_{i_{\ell+1}}$ where $z_j$ occurs in a relational atom. The corresponding subexpressions $\ms_{i_j}$ of $e$ with $1 \leq j \leq \ell$ are necessarily of the form $\mv_{i_j}^T\cdot \mw_{i_j}$ with $\mz_{i_j} \in \{\mv_{i_j}, \mw_{i_j}\}$ . Moreover, $\ms_\ell$ must be of the form $\mv_{\ell+1}^T\cdot\mB\cdot \mw_{\ell+1}$ with $\mz_{i_\ell}  \in \{\mv_{i_\ell}, \mw_{i_\ell}\}$. Because each $\ms_{i_j}$ is well-typed and all $\mv_{i_j}, \mw_{i_j}$ are vector variables, it follows that each $\mz_{i_j}$ has type $(1,1)$ for $1\leq j \leq \ell$. This yields a contradiction: it is readily verified that  when the construction generates a relational atom containing a variable $z_{\ell}$, that variable cannot have type $(1,1)$.

    We prove that if $\mQ$ is constant-disjoint, then property (2) holds in $Q$. We also do this by contradiction. Suppose it does not hold, i.e., suppose there is a pair $(v \leq c, w \leq c)$ in $Q$ of covered inequality $v \leq c$ and non-covered inequality $w \leq c$ and assume that $w$ is free in $Q$. Then $w\in\{x,y\}$ and since $v \leq c$ is covered it is the case that there is some $a_i\in\{A(v,z),A(z,v),A(v)\wedge z\leq 1, z\leq 1 \wedge A(v)\}$. Because $\Rel$ is enumeration-suited, we have that $\ms_i\in\{\mv^T\cdot\mA\cdot \mau,\mau^T\cdot\mA\cdot \mv\}$ with $\mv\colon(\gamma,1)$ with $\gamma\neq 1$. Hence we have a pair $(\mv,\mw)$ where $\mv\colon(\gamma,1)$ is covered, $\mw\colon(\gamma,1)$ is not covered and $\mw\in\{\mx,\my\}$, which means that $\mQ$ is not constant-disjoint, a contradiction.

For step (ii), let $Q'$ be the equivalent CQ after unification and removal of equality atoms on $Q$. First note that since $Q$ is well-typed w.r.t. $\Rel^{-1}$ also $Q'$ is well-typed w.r.t $\Rel^{-1}$ since  we only unify variables, and since all  unified variables  must have had the same type in $Q$ by definition of the \foplus type system.  
Further observe that because our definition of constant-disjointness in $Q$ takes equality into account, and because $Q'$ is obtained by unifying variables that must be equal, also $Q'$ must necessarily be constant-disjoint.
\end{proof}

We are ready to prove Corollary~\ref{cor:lower-bound-free-connex-matrix}.

\begin{proof}[Proof of Corollary~\ref{cor:lower-bound-free-connex-matrix}]
    Let $\Rel$ be a matrix-to-relational encoding scheme over $\Sch$ that is enumeration-suited for $\mQ$. Let $\Mat = \Rel^{-1}$ be its inverse relational-to-matrix encoding.
  
    We apply Proposition~\ref{prop:simulating-cq-is-constant-disjoint} on $\mQ$ to obtain binary \cq $Q$ over $\Voc = \Rel(\Sch)$ that simulates $\mQ$ w.r.t. $\Rel$. We know that $Q$ is binary, self-join free and constant-disjoint. We next modify $Q$ into a query $Q'$ over $\Voc$ such that the following properties hold:
    \begin{enumerate}[(P1)]
        \item $Q'$ continues to simulate $\mQ$ w.r.t. $\Rel$;
        \item $Q'$ continues to be constant-disjoint;
        \item $\Eval{Q',\Voc}{\cK} \in \ConstLin$.
    \end{enumerate}
    This suffices to prove the proposition. Indeed, by Theorem~\ref{theo:lower-bound-free-connex-body} $Q'$ is necessarily free-connex unless unless either the Sparse Boolean Matrix Multiplication, the Triangle Detection, or the $(k,k+1)$-Hyperclique conjecture is false. If $Q'$ is free-connex, then by Corollary~\ref{cor:fcLang-equiv-free-connex-cq} there exists a \fcLang query $\mQ'$ that simulates $Q'$ w.r.t $\Mat = \Rel^{-1}$. Consequently, $\mQ$ and $\mQ'$ are then equivalent.

    Intuitively, the modification of $Q$ into $Q'$ is necessary to obtain property (P3): the fact that $\Eval{\mQ,\Sch}{\cK} \in \EnumClass{\size{\I}}{1}$ only implies that we may evaluate $Q$ with linear time preprocessing and constant delay \emph{on databases that encode some matrix instance}. To apply  Theorem~\ref{theo:lower-bound-free-connex-body}, by contrast we need to show that we can evaluate $Q$ on \emph{arbitrary databases}. The crux will be that we obtain $Q'$ by adding extra inequalities to $Q$, such that evaluating $Q'$ on an arbitrary database yields the same result as evaluating it (and $Q$, $\mQ$) on some matrix instance.

    The definition of $Q'$ is as follows. Assume $Q$ is of the form $Q: H(\y) \gets \exists \z. \psi$ with $\psi$ quantifier free. For every relational atom $\alpha = R(\x)$ of $\psi$ we define a formula $\varphi_\alpha$ as follows. This formula is a conjunction of inequalities:
    \begin{itemize}
        \item If $\alpha = R(x_1,x_2)$ with $R$ binary and $\Mat(R)\colon (\beta,\gamma)$ then $\varphi_\alpha := x_1 \leq \Rel(\beta) \wedge x_2 \leq \Rel(\gamma)$. Note that, by definition of $\Rel$, $\Rel(\alpha) \not = 1 \not = \Rel(\beta)$.
        \item If $\alpha = R(x_1)$ with $R$ unary then by definition of $\Rel$ either $\Mat(R)\colon (\beta,1)$ or $\Mat(R)\colon (1,\beta)$ for some size symbol $\beta \not = 1$. Then take $\varphi_\alpha := x_1 \leq \Rel(\beta)$. Note that by definition of matrix-to-relational encoding schemes, $\Rel(\beta) \not =1$ since $\beta \not = 1$.
        \item If $\alpha= R()$ with $R$ nullary then by definition of $\Rel$ we have $\Mat(R)\colon (1,1)$ and we take $\varphi_\alpha$ the empty formula (which is equivalent to true).
    \end{itemize}
    Then we define $Q'$ to be the CQ
    \[ Q'\colon H(\y) \gets \exists z. (\psi \wedge \bigwedge_{\alpha \text{ rel. atom in } \psi} \varphi_\alpha).\]

    Note that $Q'$ continues to simulate $\mQ$ over $\Rel$ (property P1). Indeed, for any matrix instance $\I$ over $\Sch$ we know that $\AnsEnum{\mQ}{\I} = \AnsEnum{Q}{\Rel(\I)}$. By definition, $\Rel(\I)$ is consistent with $\Rel^{-1} = \Mat$, and therefore the tuples in $\Rel(\I)$ vacuously satisfy all the extra inequalities that we have added to $Q$ to obtain $Q'$. As such, $\AnsEnum{Q}{\Rel(\I)} = \AnsEnum{Q'}{\Rel(\I)}$. Thus, $Q'$ continues to simulate $\mQ$.

    Also note that $Q'$ is constant-disjoint (property P2). Indeed:
    \begin{itemize}
        \item We have only added covered inequalities to $Q$ to obtain $Q'$. Any such covered inequality that we have added was of the form $x \leq c$ with $c \not =1$. Therefore, since $Q$ does not have covered inequalities of the form $x \leq 1$, neither does $Q'$.
        
        \item We did not add any non-covered inequalities to $Q$ to obtain $Q'$. Moreover, $Q'$ has the same free variables as $Q$. Therefore, all non-covered inequalities $y \leq c$ in $Q'$ continue to be such that $y$ is not free in $Q'$, unless $c=1$. Hence for all pairs $(x \leq c, y \leq d)$ of covered inequality $x \leq c$ and non-covered inequality $y \leq d$ with $d\neq 1$ in $Q'$ we have that $y$ is not free in $Q'$ as required for constant-disjointness.
    \end{itemize}
    Finally, we show that $\Eval{Q',\Voc}{\cK} \in \ConstLin$ (property P3), as claimed. We do this by reducing $\Eval{Q',\Voc}{\cK}$ to $\Eval{\mQ,\Sch}{\cK}$ as follows. Let $\db$ be a $\cK$-database over $\Voc$, input to $\Eval{Q',\Voc}{\cK}$. First, create the database $\db'$ that is equal to $\db$ except that for every relation symbol $R \in \Voc$ that is not mentioned in $Q$ we set $R^{\db'}$ to empty. This can clearly be done in linear time. Note that, if $R$ is not mentioned in $Q$, then the answer of $Q$ on $\db$ is independent of the contents of $R^\db$. Therefore, $\AnsEnum{Q'}{\db} = \AnsEnum{Q'}{\db'}$. Subsequently, compute $\db''$, the atomic reduction of $\db'$ w.r.t. $Q$. Computing $\db''$ can be done in time $\bigo(\size{\db'}) = \bigo(\size{\db})$.  By Claim~\ref{claim:equal-after-reduction} we have $\AnsEnum{Q'}{\db} = \AnsEnum{Q'}{\db'} = \AnsEnum{Q'}{\db''}$. Now verify the following claim: $\db''$ is consistent w.r.t. $\Mat = \Rel^{-1}$. This is because we have added, for each matrix symbol $\mA$ with $\Rel(\mA)$ occurring in $Q$, the inequalities that are required by consistency as atomic inequalities to $Q'$. (The matrix symbols $\mA$ with $\Rel(\mA)$ not occurring in $Q$ are empty in $\db'$ and hence vacuously consistent w.r.t. $\Mat$.) Since $\db''$ is consistent with $\Mat$ we have that $\Mat(\db'')$ is a matrix instance over $\Sch$ and hence a valid input to $\mQ$. Then, because $Q'$ simulates $\mQ$, we have $\AnsEnum{Q'}{\db} = \AnsEnum{Q'}{\db''} = \AnsEnum{\mQ}{\Mat(\db'')}$. The last equality holds because the head of $Q'$ is a binary relation since $\Rel$ is enumeration-suited for $\mQ$. %

    Because $\Eval{\mQ,\Sch}{\cK}$ is in $\EnumClass{\size{\I}}{1}$ it hence follows that with additional preprocessing of time $\bigo(\size{\Mat(\db'')}) = \bigo(\size{\db})$ we may enumerate $\AnsEnum{Q'}{\db''} = \AnsEnum{Q'}{\db}$ with constant delay, as desired.
\end{proof}

\section{Efficient evaluation of q-hierarhical queries}\label{sec:evaluation-qh}

This last section derives our algorithmic results for \qcq and \qhLang. \qcq is the subclass of CQ that allows efficient evaluation in a dynamic setting, where insertion and deletion of tuples are admitted. Then, similar to the previous section, we aim to lift these algorithmic results to \qhLang. 

We start by introducing the dynamic setting for query evaluation, and then study the upper and lower bounds.

\subsection{The dynamic evaluation setting}

We move now to the dynamic query evaluation both in the relational and matrix scenarios. Specifically, we consider the following set of updates. Recall that $\cK=(K, \ksum, \kprod, \kzero, \kone)$ is a semiring and $\Voc$ a vocabulary.
\begin{itemize}
    \item A single-tuple \emph{insertion} (over $\cK$ and $\Voc$) is an operation
    $u = \Insertion{R,\seq{d},k}$ with $R \in \Voc$, $\seq{d}$ a
    tuple of arity $\arity(R)$, and $k \in K$. When applied to a database
    $\db$ it induces the database $\db + u$ that is
    identical to $\db$, but $R^{\db + u}(\seq{d})=R^{\db}(\seq{d})\ksum k$. 
    \item A single-tuple \emph{deletion} (over $\cK$ and $\Voc$) is an expression
    $u=\Deletion{R,\seq{d}}$ with $R\in \Voc$ and $\seq{d}$ a
    tuple of arity $\arity(R)$.  When applied to a database $\db$ it induces the database $\db + u$ that is identical
    to $\db$, but $R^{\db + u}(\seq{d})=\kzero$. 
\end{itemize}
Notice that if every element in $\cK$ has an additive inverse (i.e., $\cK$ is a ring), one can simulate a deletion with an insertion. However, if this is not the case (e.g., $\bbB$ or $\bbN$), then a single-tuple deletion is a necessary operation.

As the reader may have noticed, updates allow to modify the contents of relations, but not of constant symbols. This is because an update to a constant translates as an update to the dimension value of a matrix within the equivalent linear algebra setting.  And updates in the linear algebra setting affect only entry values, not dimensions. An interesting line of future work is to consider dimension updates.

\paragraph{Dynamic enumeration problems}
A \emph{dynamic enumeration problem} is an enumeration problem $P$ together with a set $U$ of \emph{updates} on $P$'s inputs: a set of operations such that applying an update $u \in U$ to an input $I$ of $P$ yields a new input to $P$, denoted $I + u$.  An algorithm $A$ \emph{solves} a dynamic enumeration problem $D = (P,U)$ if it solves the enumeration problem $P$ (computing, for each input $I$ a data structure in a preprocessing phase from which $P(I)$ may be enumerated) and, moreover, it is possible, for every input $I$ and update $u$ to update the data structure that $A$ computed on $I$ during the update phase to a data structure for $I + u$. By definition, the latter updated data structure hence allows to enumerate $P(I + u)$. The preprocessing time and enumeration delay of $A$ are defined as for normal (static) enumeration problems. The \emph{update time} of $A$ is an upper bound on the time needed to update the data structure for $I$ into one for $I+u$.

In this document, we will consider two versions of the dynamic query evaluation problem for conjunctive query $Q$ over $\Voc$ under semiring $\cK$:
\begin{itemize}
    \item The \emph{unbounded} version, $\DynEval{Q,\Voc}{\cK}$ which is the dynamic enumeration problem $(\Eval{Q,\Voc}{\cK}, U)$ with $U$ the set of all single-tuple insertions and deletions over $\cK$ and $\Voc$.

    \item The \emph{bounded} version, denoted $\DynEval{Q,\Voc}{\cK,M}$, where $M \in \bbN_{\geq 0}$ is a constant positive natural number, and which is defined as follows.
    
    Define $\Eval{Q,\Voc}{\cK,M}$ to be the bounded version of the \emph{static} query evaluation problem: this is the collection of all pairs $(I,O)$ with
    $I = \db$ with $\db$ a $\cK$-database over $\Voc$ \emph{such that any data value $d$ occurring in $\db$ is at most $M$, i.e., $d \leq M$}\footnote{Recall that the set of data values that may appear in database tuples equals $\bbN_{\geq 0}$.}, and $O = \AnsEnum{Q}{\db}$. Hence, $\Eval{Q,\Voc}{\cK,M}$ is the evaluation problem of $Q$ on $\cK$-databases whose values in the active domain are bounded by $M$.

    Then the bounded dynamic evaluation problem $\DynEval{Q,\Voc}{\cK,M}$ is the dynamic enumeration problem $(\Eval{Q,\Voc}{\cK,M}, U)$ with $U$ the set of all single-tuple updates $\Insertion{R,\seq{d},k}$ or $\Deletion{R,\seq{d}}$ where every value $d \in \seq{d}$ satisfies $d \leq M$.
\end{itemize}
Clearly, one can check that $\DynEval{Q,\Voc}{\cK} = \bigcup_{M=1}^{\infty}\DynEval{Q,\Voc}{\cK,M}$.

In an analogous manner, in the matrix setting for a query $\mQ$ over $\Sch$ we define the dynamic query evaluation problem $\DynEval{\mQ,\Sch}{\cK}$ and $\DynEval{\mQ,\Sch}{\cK,M}$ of $\Eval{\mQ,\Sch}{\cK}$ and $\Eval{\mQ,\Sch}{\cK,M}$, respectively. The allowed updates in these problems are of the form $\Insertion{\mA,i,j,k}$, which sets $\mA_{i,j} := \mA_{i,j} \ksum k$, and $\Deletion{\mA,i,j}$ which sets $\mA_{i,j} := \kzero$.

\paragraph{Dynamic complexity classes} For functions $f, g$, and $h$ from the natural numbers to the positive reals we define $\DynClass{f}{g}{h}$ to be the class of dynamic enumeration problems $(P,U)$ for which there exists an enumeration algorithm $A$ such that for every input $I$ it holds that the preprocessing time is $\bigo\left( f(\size{I}) \right)$, the delay is $\bigo\left( g(\size{I}) \right)$, where $\size{I}$ denotes the size of $I$. Moreover, processing update $u$ on current input $I$ takes time $\bigo(\left( h(\size{I}, \size{u}) \right)$. Note that, since the inputs in a dynamic evaluation problem $\DynEval{Q,\Voc}{\cK}$ or $\DynEval{Q,\Voc}{\cK,M}$ consists only of the database and not the query, we hence measure complexity in \emph{data complexity} as is standard in the literature~\cite{DBLP:conf/csl/BaganDG07,DBLP:journals/vldb/IdrisUVVL20,DBLP:journals/siglog/BerkholzGS20,DBLP:conf/pods/BerkholzKS17}.
In what follows we set the size $\size{u}$ of a single-tuple update $u$ to $1$: each update is of constant size, given the fixed vocabulary $\Voc$.

Note that, since $\DynEval{Q,\Voc}{\cK} = \bigcup_{M=1}^{\infty}\DynEval{Q,\Voc}{\cK,M}$ if $\DynEval{Q,\Voc}{\cK} \in \Maintainable$ then also $\DynEval{Q,\Voc}{\cK,M} \in \Maintainable$, for every $M$. Conversely, if $\DynEval{Q,\Voc}{\cK,M} \not \in \Maintainable$ for some $M$, then also $\DynEval{Q,\Voc}{\cK} \not \in \Maintainable$. Specifically, we will use the bounded evaluation problem $\DynEval{Q,\Voc}{\cK,M}$ to obtain our lower bounds in the relational setting. Analogously, we focus on the bounded evaluation problem $\DynEval{\mQ,\Sch}{\cK,M}$ to obtain our lower bounds in the matrix setting.
\paragraph{Complexity hypothesis and known results.}
Like the authors of \cite{DBLP:conf/stoc/HenzingerKNS15,DBLP:conf/pods/BerkholzKS17} we use the following hypothesis concerning the hardness of dynamic problems as hypothesis for obtaining conditional lower bounds:
\begin{itemize}
\item The \emph{Online Boolean Matrix-Vector Multiplication (OMv) problem}:
given an $n\times n$ boolean matrix $A$, compute $A\cdot v_1,\ldots,A\cdot v_n$ sequentally
for $n\times 1$ vectors $v_1,\ldots,v_n$. It is required that the result of $A\cdot v_{i-1}$ must already be computed in order to access $v_i$.
The \emph{OMv conjecture} states that this problem cannot be solved in time $\bigo(n^{3-\varepsilon})$ for any $\varepsilon > 0$.
\end{itemize}

\begin{thmC}[\cite{DBLP:conf/pods/BerkholzKS17}]\label{theo:cqNotIneq-q-hierarchical-cde}
  Let $Q$ be a \cqNotIneq over $\Voc$.
  \begin{itemize}
    \item If $Q$ is q-hierarchical, then $\DynEval{Q,\Voc}{\bools}\in \Maintainable$.
    \item If $Q$ is a query without self joins and
      $\DynEval{Q,\Voc}{\bools,M}\in \Maintainable$ for every $M \in \posnat$,
      then $Q$ is q-hierarchical unless the $OMv$ conjecture is false.
  \end{itemize}
\end{thmC}

\subsection{Upper bounds for q-hierarchical queries}
Similar as for free-connex queries, we can provide dynamic evaluation algorithms for \qcq and \qhLang queries. However, for this dynamic setting, we require some additional algorithmic assumptions over the semiring. 
Let $\cK=(K, \ksum, \kprod, \kzero, \kone)$ be a semiring and $M$ be the set of all multisets of $K$.
For any $k\in K$ and $m\in M$, define $\MultisetInsert{k,m}$ and $\MultisetDelete{k,m}$ to be the multisets resulting from inserting or deleting $k$ from $m$, respectively.
Then we say that $\cK$ is \emph{sum-maintainable} if there exists a data structure $\cD$ to represent multisets of $K$ such that the empty set $\emptyset$ can be built in constant time, and if $\cD$ represents $m\in M$ then:
(1) the value $\bigksum_{k\in m} k$ can always be computed from $\cD$ in constant time;
(2) a data structure that represents $\MultisetInsert{k,m}$
can be obtained from $\cD$ in constant time; and
(3) a data structure that represents $\MultisetDelete{k,m}$
can be obtained from $\cD$ in constant time.
One can easily notice that if each element of $\cK$ has an additive inverse (i.e., $\cK$ is a ring), then $\cK$ is sum-maintainable, like $\bbR$. Other examples of sum-maintainable semirings (without additive inverses) are $\bbB$ and $\bbN$.

The main result of this subsection is the following.

\begin{thm}\label{theo:upper-bound-q-hierarchical-body}
    Let $\cK$ be a sum-maintainable semi-integral domain.
    For every q-hierarchical \cq $Q$, $\DynEval{Q,\Voc}{\cK}$ can be evaluated
    dynamically with constant-time update and constant-delay. In particular,
    $\DynEval{\mQ,\Sch}{\cK}$ can also be evaluated dynamically with constant-time
    update and constant-delay for every \qhLang $\mQ$ over $\Sch$.
  \end{thm}

We now prove Theorem~\ref{theo:upper-bound-q-hierarchical-body} in two steps. We first transfer the upper bound of \cqNotIneq queries over $\bools$-databases to \cqNotIneq queries over $\cK$-databases. We then generalize the latter to \cq queries over $\cK$-databases. 

\begin{prop}\label{prop:notIneq-qh-cq-over-sum-maintainable-and-zdf-semirings-is-maintainable}
    Let $Q$ be a q-hierarchical \cqNotIneq over $\Voc$ and $\cK$ a sum-maintainable semi-integral domain.
    Then $\DynEval{Q,\Voc}{\cK}\in \Maintainable$.
\end{prop}
\begin{proof}
    Note that since $Q$ is q-hierarchical it has a guarded query plan $(T,N)$ by
    Proposition~\ref{prop:q-hierarchical-iff-guarded-query-plan}. This implies in
    particular that $Q$ is also free-connex. By
    Proposition~\ref{prop:upper-bound-notIneq-for-semi-integral} we hence know
    that $\Eval{Q,\Voc}{\cK} \in \ConstLin$. As such, there is an algorithm $A$
    that, given $\db$ builds a data structure in linear time from which
    $\AnsEnum{Q}{\db}$ can be enumerated with constant delay. We will show that
    for any update $u$ to $\db$ this data structure can also be updated in
    $\bigo(1)$ into a data structure for $\AnsEnum{Q}{\db+u}$. Because, in the
    proof of Proposition~\ref{prop:upper-bound-notIneq-for-semi-integral} the
    algorithm (and data structure) depends on the shape of $Q$ and $(T,N)$ we make
    a corresponding case analysis here.
  
    \begin{enumerate}
        \item If $Q$ is a full-join, i.e., no variable is quantified, we reduce to the Boolean case as follows. Assume that $Q = H(\z) \gets R_1(\z_1)\wedge\cdots\wedge R_n(\z_n)$. In $\bigo(\size{\db})$ time construct $\Bool(\db)$, the $\bools$-database obtained from $\db$ defined as:
        \[
            R^{\Bool(\db)}\colon \seq{d}\mapsto
            \begin{cases}
                \btrue \text{ if }R^{\db}(\seq{d})\neq\kzero \\
                \bfalse \text{ otherwise, }
            \end{cases}
        \]
        for every $R\in\Voc$. Furthermore, preprocess $\db$ by creating lookup tables such that for every relation $R$ and tuple $\seq{a}$ of the correct arity we can retrieve the annotation $R^\db(\seq{a})$ in time $\bigo(\card{\seq{a}})$ time. Note that since $\Voc$ is fixed, $\card{\seq{a}}$ is constant. This retrieval is hence $\bigo(1)$ in data complexity. It is well-known that such lookup tables can be created in $\bigo(\size{\db})$ time in the RAM model.
        
        Finally, invoke the algorithm for $\DynEval{Q,\Voc}{\bools}$ on $\Bool(\db)$. By Theorem~\ref{theo:cqNotIneq-q-hierarchical-cde}, this algorithm has $\bigo(\size{\Bool(\db)}) = \bigo(\size{\db})$ preprocessing time, after which we may enumerate $\AnsEnum{Q}{\Bool(\db)}$ with $\bigo(1)$ delay and maintain this property under updates to $\Bool(\db)$ in $\bigo(1)$ time. We have shown in Proposition~\ref{prop:upper-bound-notIneq-for-semi-integral} that using the enumeration procedure for $\AnsEnum{Q}{\Bool(\db)}$ and the lookup tables, we may also enumerate $\AnsEnum{Q}{\db}$ with constant delay. We do not repeat this argument here.

        Given a single-tuple update $u$ we update the lookup tables and the data structure maintained by $\DynEval{Q,\Voc}{\bools}$ as follows.
        \begin{itemize}
            \item If $u=\Insertion{R,\seq{d},k}$ then lookup $\seq{d}$ in the lookup table for $R$ and retrieve its old annotation $\ell$ (if the lookup table does not contain $\ell$, set $\ell = \kzero$). There are two cases to consider:
                \begin{itemize}
                    \item If $k + \ell = \kzero$ then the tuple $\seq{d}$ is now deleted from $R^{\db+u}$ and we similarly remove it from $\Bool(\db+u)$ by issuing the update $u' = \Deletion{R, \seq{d}}$ to $\Bool(\db)$ (and $\DynEval{Q,\Voc}{\bools}$).
                    \item If $k + \ell \not = \kzero$ then the tuple is certainly present in $R^{\db + u}$ inserted and similarly make sure it is in $\Bool(\db+u)$ by issuing $u' = \Insertion{R,\seq{d}, \btrue}$ to $\Bool(\db)$ (and $\DynEval{Q,\Voc}{\bools}$). Note that if the tuple was already in $\Bool(\db)$ then $u'$ has no-effect since we are working in the boolean semiring there.
                \end{itemize}
            \item If $u=\Deletion{R,\seq{d}}$ then delete $\seq{d}$ from the lookup table and issue the update $u' = \Deletion{R, \seq{d}}$ to $\Bool(\db)$ and $\DynEval{Q,\Voc}{\bools}$.
        \end{itemize}
        Note that in all cases, after issuing $u'$ we obtain $\Bool(\db + u)$ as desired, so that after this update using the enumeration procedure for $\AnsEnum{Q}{\Bool(\db+u)}$ and the lookup tables, we will enumerate $\AnsEnum{Q}{\db+u}$ with constant delay.

        \item If the guarded query plan $(T,N)$ is such that $N$ consists of a single node $N = \{r\}$, then by Lemma~\ref{lemma:full-guarded-query-plan} we may assume without loss of generality that for every node $n$ with two children $c_1,c_2$ in $T$ we have $\var(n) = \var(c_1) = \var(c_2)$. Intuitively, the algorithm will work as follows. We show that for any database $\db$ we may compute $\AnsEnum{Q}{\db}$ in $\bigo(\size{\db})$ time. Because we can simply store this result, we may certainly enumerate it with constant delay. Then, by also storing the subresults computed during the computation of $\AnsEnum{Q}{\db}$ we show that we can also maintain $\AnsEnum{Q}{\db}$ and all subresults in $\bigo(1)$ time under updates.

        Formally the data structure used by $\DynEval{Q,\Voc}{\cK}$ will be the query result itself, plus some extra lookup tables. In particular, given input database $\db$ we:
            \begin{itemize}
                \item compute, for every node $n \in T$ the $\AnsEnum{\varphi[T,n]}{\db}$. We have shown in Proposition~\ref{prop:upper-bound-notIneq-for-semi-integral} that we may compute $\AnsEnum{\varphi[T,n]}{\db}$ in linear time, for every node~$n$. We do not repeat the argument here. Since $Q$ (and hence $T$) is fixed there are a constant number of nodes in $T$. So this step takes linear time overall.
                \item convene that we store $\AnsEnum{\varphi[T,n]}{\db}$ for every node $n$ by creating a lookup table such that for any tuple $(\seq{a},k) \in \AnsEnum{\varphi[T,n]}{\db}$ we can retrieve $k$ in $\bigo(1)$ time given $\seq{a}$. It is well-known that the lookup tables can be made such that may we enumerate the entries $(\seq{a},k)$ of this lookup table with constant delay. It is well-known that such lookup tables can be created in time linear in $\size{\AnsEnum{\varphi[T,n]}{\db}} = \bigo(\size{\db})$. So also this step takes linear time.
            \end{itemize}
        Note that by definition $Q \equiv \varphi[T,r]$; therefore the root represents $\AnsEnum{Q}{\db}$ as desired, and can be used to enumerate the result with constant delay.

        It remains to show that given an update $u$ to $\db$ we can update $\AnsEnum{\varphi[T,n]{\db}}$ into $\AnsEnum{\varphi[T,n]{\db+u}}$ in $\bigo(1)$ time, for every node $n$. In order to be able to show this, we store one auxiliary lookup table $L_n$ for all nodes $n$ that have a single child $c$. This auxiliary table is defined as follows. By Lemma~\ref{lemma:child-formula-projection} if $n$ has a single child $c$ then $\varphi[T,n] \equiv \exists \y. \varphi[T,c]$ with $\y = \var(c) \setminus \var(n)$. The lookup table $L_n$ stores, for each $(\seq{d},k)$ in $\AnsEnum{\varphi[T,n]}{\db}$ a pointer to a data structure $\cD_{\seq{d}}$ that represents the multiset
        \[ \{\!\{ k \mid (\seq{a},k) \in \AnsEnum{\varphi[T_c,c]}{\db}, \restr{\seq{a}}{\var(r)} = \seq{d}\}\!\}.\] Because $\cK$ is
        sum-maintainable, and because $\AnsEnum{\varphi[T_c,c]}{\db}$ is linear in $\db$, this extra lookup table can be computed in linear time.\footnote{Initially, the lookup table is empty. Loop over the entries $(\seq{a},k)$ of $\AnsEnum{\varphi[T_c,c]}{\db}$ one by one. For each such tuple, compute $\seq{d} = \restr{\seq{a}}{\var(r)}$ and lookup $\seq{d}$ in $L$. If it is present with data structure $\cD_{\seq{d}}$ then apply $\MultisetInsert{k,\cD_{\seq{d}}}$, hence updating $\cD_{\seq{d}}$. If it is not present, initialize $\cD_{\seq{d}}$ to empty; add $k$ to it, and insert $(\seq{d}, \cD_{\seq{d}})$ to the lookup table.}
  
        We now verify that for every single-tuple update $u$ to $\db$ and every node $n$ we can update $\AnsEnum{\varphi[T,n]{\db}}$ into $\AnsEnum{\varphi[T,n]{\db+u}}$ in $\bigo(1)$ time. The proof is by induction on the height of $n$ in $T$. (I.e., the length of longest path
        from $n$ to a leaf.)
        \begin{itemize}
            \item When $n$ is a leaf, it is an atom, say $R(\x)$. Upon update $u$, if $u$ is of the form $\Insertion{R,\seq{a},k}$ or $\Deletion{R,\seq{a}}$ then we update the lookup table that represents $\AnsEnum{\varphi[T,n]{\db}}$ in the obvious way. Such updates take $\bigo(1)$ in the RAM model. If $u$ pertains to a relation other then $R$, we simply ignore it.
            \item When $n$ is an interior node with a single child $c$ then by Lemma~\ref{lemma:child-formula-projection} we have $\varphi[T,n] \equiv \exists \y. \varphi[T,c]$ with $\y = \var(c) \setminus \var(r)$.  When we receive update $u$ to $\db$, by inductive hypothesis this yields a bounded number of single-tuple updates $u_c$ to $A = \AnsEnum{\varphi[T,c]}{\db}$ which we may compute in $\bigo(1)$ time. We translate these updates $u_c$ into updates $u'$ to apply to $\AnsEnum{\varphi[T,n]}{\db}$:
            \begin{itemize}
                \item If $u_c = \Insertion{\seq{a},k}$. Then apply $u_c$ to $A$ and update the auxiliary lookup table $L_n$ by applying $\MultisetInsert{k, \cD_{\seq{d}}}$ with $\seq{d} = \restr{\seq{a}}{\var(r)}$. Furthermore, update $\AnsEnum{\varphi[T,n]}{\db}$ by applying $u' = \Insertion{\seq{d},k}$ to the lookup table representing it.

                \item If $u_c = \Deletion{\seq{a}}$. Before applying $u_c$ to $A = \AnsEnum{\varphi[T,c]}{\db}$, look up the old annotation $k$ of $\seq{a}$ in $A$. Then apply $u_c$ to $A$ and update $L_n$ by applying $\MultisetDelete{k, \cD_{\seq{d}}}$ with $\seq{d} = \restr{\seq{a}}{\var(r)}$. If $\DSQuery(\cD_{\seq{d}}) = \kzero$, then $u$ effectively causes $\seq{d}$ to be removed from $\AnsEnum{\varphi[T,n]}{\db}$. Hence, the update $u'$ to apply to (the lookup table representing) $\AnsEnum{\varphi[T,n]}{\db}$ is $u' =\Deletion{\seq{d}}$. Otherwise, $\DSQuery(\cD_{\seq{d}}) \not = \kzero$ and the update $u'$ to apply is $u' = \Insertion{\seq{d}, k}$.
            \end{itemize}
  
            \item Node $n$ has two children $c_1$ and $c_2$. By Lemma~\ref{lemma:children-formulas-conjunction} we have $\varphi[T,n] \equiv \varphi[T,c_1] \wedge \varphi[T,c_2]$. Note that this denotes an intersection since $\var(n)=\var(c_1)=\var(c_2)$. When we receive update $u$ to $\db$, by inductive hypothesis this yields a bounded number of single-tuple updates to $A_1 = \AnsEnum{\varphi[T,c_1]}{\db}$ and $A_2 = \AnsEnum{\varphi[T,c_1]}{\db}$ which we may compute in $\bigo(1)$ time. We translate these updates to $A_1$ and $A_2$ into updates to $\AnsEnum{\varphi[T,n]{\db}}$ as follows. We only show the reasoning for updates to $A_1$. The reasoning for updates to $A_2$ is similar. Let $u_1$ be update to $A_1$. We translate this into update $u'$ to apply to $\AnsEnum{\varphi[T,n]}{\db}$ as follows:
            \begin{itemize}
                \item if $u_1 = \Insertion{\seq{d},k_1}$ then use the lookup table on $A_2$ to find the annotation $\ell$ of $\seq{d}$ in $A_2$. (If $\seq{d}$ does not appear in $A_2$ then take $\ell = \kzero$). If $k_1 \kprod \ell \not = \kzero$ then apply $u' = \Insertion{d, k_1 \kprod \ell}$ to $\AnsEnum{\varphi[T,n]}{\db}$; otherwise there is no update $u'$ to~apply.
                \item if $u_1 = \Deletion{\seq{d}}$ then apply $u' = \Deletion{d}$ to $\AnsEnum{\varphi[T,n]}{\db}$.
            \end{itemize}
        \end{itemize}

        \item When $Q$ is not full and $N$ has more than one node we reduce to the previous two cases because $\varphi[T,N]\equiv \varphi[T_1, n_1] \wedge \dots \wedge \varphi[T_l, n_l]$ where $F=\{ n_1, \ldots, n_l \}$ is the frontier of $N$, i.e, the nodes without children in $N$, and $T_1,\dots, T_{l}$ are the subtrees of $T$ rooted at $n_1,\dots, n_l$ respectively. Hence, we may enumerate $\AnsEnum{\varphi[T,N]}{\db}$ with constant delay by first fully computing the full join $\varphi[T_1, n_1] \wedge \dots \wedge \varphi[T_l, n_l]$ on these subresults. By item (2) above, each $A_i := \AnsEnum{\varphi[T_{i}, n_i]}{\db}$ can be computed in linear time from $\db$. Furthermore, each $\AnsEnum{\varphi[T_{i}, n_i]}{\db}$ can be maintained under updates in $\bigo(1)$ time by item (2) above. By item (1) the result of the final full join can be enumerated with constant delay after linear time processing on $A_i$. In addition, we can maintain this property under updates in $\bigo(1)$ time by item (1) above, since this is a full join and $\varphi[T,N]$ is q-hierarchical. Therefore, the entire result $\AnsEnum{\varphi[T,N]}{\db}$ can be enumerated with constant delay after linear time processing, and be maintained under updates in $\bigo(1)$ time. \qedhere
    \end{enumerate}
\end{proof}

Next, we show how to extend Proposition~\ref{prop:notIneq-qh-cq-over-sum-maintainable-and-zdf-semirings-is-maintainable} when queries have inequalities.
Towards this goal, note that the conditions for a \cq to be q-hierarchical are imposed over relational atoms. Hence, the following result is straightforward (recall the definition of $\Split(Q)$ given in Section~\ref{sec:evaluation}).
\begin{cor}
	\label{cor:query-with-ineq-q-hierarhical-if-rel-is}
	Let $Q$ be a \cq and $\Split(Q)=(\Qrel, \Qineq)$.
	Then $Q$ is q-hierarchical if and only if $\Qrel$ is q-hierarchical.
\end{cor}

Then, we can exploit the split of $Q$ between $\Qrel$ and $\Qineq$ to evaluate efficiently any $q$-hierarchical query $Q$ as follows.

\begin{prop}\label{prop:relational-maintain-implies-withIneq-maintain}
    Let $Q$ be a \cq over $\Voc$ and $\Split(Q)=(\Qrel, \Qineq)$ and $\cK$ a
    semi-integral domain.  If $\DynEval{\Qrel,\Voc}{\cK}\in \Maintainable$ then
    $\DynEval{Q,\Voc}{\cK}\in \Maintainable$.
\end{prop}
\begin{proof}
    Let $\Voc$ be the vocabulary of $Q$ and $\db$ be an arbitrary $\cK$-database over $\Voc$. Since $\DynEval{\Qrel,\Voc}{\cK}\in \Maintainable$ we have $\Eval{\Qrel,\Voc}{\cK}\in\ConstLin$ and consequently $\Eval{Q,\Voc}{\cK}\in\ConstLin$, by Proposition~\ref{prop:relational-enum-implies-withIneq-enum}.
  
    Note that Proposition~\ref{prop:relational-enum-implies-withIneq-enum} shows that $\Eval{Q,\Voc}{\cK}\in\ConstLin$ by first making input database $\db$ atomically consistent with $Q$ and then invoking $\Eval{\Qrel,\Voc}{\cK}$ on the atomically reduced database $\db'$. It is then possible to enumerate $\ValEnum{Q}{\db'}=\ValEnum{Q}{\db}$ with constant delay, by enumerating $\ValEnum{\Qrel}{\db'}=\ValEnum{\Qrel}{\db}$ with constant delay. In other words, the data structure of $\Eval{Q,\Voc}{\cK}$ is simply that of $\Eval{\Qrel,\Voc}{\cK}$.
  
    Hence, to obtain the proposition, it suffices to translate a given update $u$ to $\db$ into an update $u'$ to $\db'$ such that $\db'+u'$ is the atomic reduction of $\db + u$. This allows us to update the data structure $\Eval{\Qrel,\Voc}{\cK}$ in $\bigo(1)$ time by assumption, and still be able to enumerate $\ValEnum{\Qrel}{\db'+u'}=\ValEnum{\Qrel}{\db+u}$. We translate the update as follows
  
    \begin{itemize}
        \item If $u=\Insertion{R,\seq{d},k}$ then for each atom $R(\x)$ in $Q$ with  $\seq{d} \models \x$ check that the inequality constraints imposed by $Q[R(\x)]$ are satisfied. If this is the case, apply $u' = u$ to $\db'$. Otherwise, no update needs to be done to $\db'$.

        \item If $u=\Deletion{R,\seq{d}}$ then apply $u'= u$ to $\db'$. \qedhere
    \end{itemize}
\end{proof}

We finally have all the machinery to prove Theorem~\ref{theo:upper-bound-q-hierarchical-body}.

\begin{proof}[Proof of Theorem~\ref{theo:upper-bound-q-hierarchical-body}]
    Let $\Split(Q)=(\Qrel,\Qineq)$. Since $Q$ is q-hierarchical, we have that $\Qrel$ is q-hierarchical by Corollary~\ref{cor:query-with-ineq-q-hierarhical-if-rel-is}. Then, by Proposition~\ref{prop:notIneq-qh-cq-over-sum-maintainable-and-zdf-semirings-is-maintainable}, $\DynEval{\Qrel,\Voc}{\cK} \in\Maintainable$ since $\cK$ is a sum-maintainable semi-integral domain. Hence we get that
     $\DynEval{Q,\Voc}{\cK}\in\Maintainable$ because of Proposition~\ref{prop:relational-maintain-implies-withIneq-maintain}.

    We proceed to prove the second part of the result, i.e., achieve the same processing complexity for the matrix evaluation problem $\DynEval{\mQ,\Sch}{\cK}$. Let $\mQ$ be a \qhLang query over $\Sch$ and $\I$ and instance over $\Sch$. Fix a matrix-to-relational schema encoding $\Rel$ over $\Sch$ such that the head atom of $Q$ is binary. Since in particular $\mQ$ is also a \fcLang query, by Theorem~\ref{theo:free-connex-upper-bound-body} $\sem{\mQ}{\I}$ can be enumerated with constant delay after a preprocessing that runs in time $\bigo(\size{\I})$.
 	Now, use Theorem~\ref{theo:qhLang-equiv-q-hierarchical-cq} to compute the q-hierarchical relational query $Q$ over $\Voc$ that is equivalent to $\mQ$ under $\Rel$. This is in constant time in data complexity. Next, let $u$ be a update $u$ to $\I$. We define a series of single tuple updates $u_1,\ldots,u_n$ to $\Rel(\I)$ such that $\Rel(\I)+u_1+\ldots+u_n=\Rel(\I)+u=\Rel(\I_u)$. This can be done in $\bigo(\size{u})$ time.
 	Because $Q$ is q-hierarchical, $\DynEval{Q,\Voc}{\cK}$ can be evaluated dynamically with constant-time update and constant-delay, and hence we can enumerate each $\AnsEnum{Q}{\Rel(\I)}$, $\AnsEnum{Q}{\Rel(\I)+u_1}$ up to $\AnsEnum{Q}{\Rel(\I)+u_1+\ldots+u_n}$ with constant delay. This holds because processing each update $u_i$ one at a time takes time $\bigo(1)$. After $\bigo(\size{u})$ time, the former yields constant delay enumeration for $\AnsEnum{\mQ}{\I_u}$.
\end{proof}

\subsection{Lower bounds for q-hierarchical queries}
Similar as for free-connex \cq, we can extend the lower bound in~\cite{DBLP:conf/pods/BerkholzKS17}
when the subsemiring generated by $\kzero_{\cK}$ and $\kone_{\cK}$ is zero-sum free,
by assuming the Online Boolean Matrix-Vector Multiplication (OMv) conjecture~\cite{DBLP:conf/stoc/HenzingerKNS15}. For this, recall the definition of constant-disjoint introduced in~Section~\ref{sec:evaluation}.

\begin{thm}\label{theo:lower-bound-q-hierarchical}
    Let $Q$ be a \cq over $\Voc$ without self-joins and constant-disjoint. Let 
    $\cK$ be a semiring such that the subsemiring generated by $\kzero_{\cK}$ and
    $\kone_{\cK}$ is zero-sum free.  
    If $\DynEval{Q,\Voc}{\cK}$ can be evaluated dynamically with constant-time update and constant-delay, then $Q$ is q-hierarchical, unless the OMv conjecture is false.
\end{thm}

To prove Theorem~\ref{theo:lower-bound-q-hierarchical} it suffices to show that the statement holds in the bounded problem $\DynEval{Q,\Voc}{\bools,M}$, for every $M\in \posnat$. %
Towards that goal, we first reduce the $\DynEval{Q,\Voc}{\bools,M}$ problem into the $\DynEval{\Qrel,\Voc}{\bools,M}$ problem assuming that $Q$ is constant-disjoint. Subsequently, we reduce  $\DynEval{Q,\Voc}{\cK,M}$ to $\DynEval{Q,\Voc}{\bools,M}$.

\begin{prop}\label{prop:ineq-maintain-implies-relational-maintain}
    Let $M \in \posnat$ and let $Q$ be a \cq over $\Voc$ with
    $\Split(Q)=(\Qrel, \Qineq)$.  If $Q$ is constant-disjoint and
    $\DynEval{Q,\Voc}{\bools,M}\in \Maintainable$ then
    $\DynEval{\Qrel,\Voc}{\allowbreak\bools,M}\in \Maintainable$.
\end{prop}
\begin{proof}
    The proof uses the same ideas as the proof of
    Proposition~\ref{prop:ineq-enum-implies-relational-enum}, but specialized to the dynamic setting, and utilizing that $M$ is a bound on the active domain of the databases that we will evaluate $\Qrel$ on.
  
    Specifically, we reduce $\DynEval{\Qrel,\Vocrel}{\bools,M}$ to
    $\DynEval{Q,\Voc}{\bools,M}$. Denote the free variables of $Q$, $\Qrel$ and
    $\Qineq$ by $\x$, $\x_1$ and $\x_2$, respectively. Note that $\x_1$ and $\x_2$
    are disjoint and $\var(\x) = \var(\x_1) \cup \var(\x_2)$.  Consider an
    arbitrary database $\db_{\text{rel}}$ over $\Voc$, input to
    $\DynEval{\Qrel,\Voc}{\bools,M}$.  In particular, all values in the active domain of $\db$ are bounded by $M$. Construct the database $\db$, input to
    $\DynEval{Q,\Voc}{\bools,M}$ as follows.
    \begin{itemize}
    \item $R^{\db} = R^{\dbrel}$ for each relation symbol $R \in \Voc$;
    \item $db(c) := M$ for all constant symbols $c \in \Voc$ that occur in a
      covered inequality in $Q$. Note that $c \not = 1$ because $Q$ is
      constant-disjoint;
    \item $db(c) := 1$ for all other constant symbols $c \in \Voc$.
    \end{itemize}
    This is well-defined because $Q$ is constant-disjoint. The computation of
    $\db$ is clearly in $\bigo(\size{\dbrel})$ and $\size{\db} = \size{\dbrel}$.
    We call $\db$ the \emph{extension} of $\dbrel$.
  
    Because $M$ is an upper bound on the domain values occurring in $\dbrel$ we have shown in Proposition~\ref{prop:ineq-enum-implies-relational-enum} that $\AnsEnum{\Qrel}{\dbrel} = \AnsEnum{\Qrel}{\db}$ and, that moreover, we may enumerate $\AnsEnum{\Qrel}{\db}$ with constant delay using the constant-delay enumeration for $\AnsEnum{Q}{\db}$, which exists because $\DynEval{Q,\Voc}{\bools,M}\in \Maintainable$.
  
    Hence, to obtain the proposition, it suffices to show that we can translate
    any given update $u$ to $\dbRel$ into an update $u'$ to $\db$ such that
    $\db+u'$ is the extension of $\dbRel + u$. Because
    $\DynEval{Q,\Voc}{\bools,M}\in \Maintainable$ this allows us to update the
    data structure of $\Eval{\Qrel,\Voc}{\bools,M}$ in $\bigo(1)$, and still be
    able to enumerate $\AnsEnum{Q}{\db+u'}$, which yields the enumeration of
    $\AnsEnum{Q}{\dbrel+u}$ with constant delay. We reason as follows.  Let $u$ be
    a single-tuple update to $\dbRel$ such that all data values occurring in $u$
    are bounded by $M$. Then simply apply $u$ to $\db$. Clearly, $\db+u$ is the
    extension of $\dbRel + u$.
\end{proof}

\begin{prop}\label{prop:maintenance-of-bool-using-zsf-semiring}
    Let $Q$ be a \cq over $\Voc$ and $\cK$ a semiring such that the subsemiring
    generated by $\kzero_{\cK}$ and $\kone_{\cK}$ is non-trivial and zero-sum
    free. Let $M \in \posnat$.   If $\DynEval{Q,\Voc}{\cK,M}\in \Maintainable$ then
    $\DynEval{Q,\Voc}{\bools,M}\in \Maintainable$.
\end{prop}
\begin{proof}
    Let $\cK=(K, \ksum, \kprod, \kzero_{\cK}, \kone_{\cK})$ be a a semiring such
    that the subsemiring generated by $\kzero_{\cK}$ and $\kone_{\cK}$ is
    non-trivial and zero-sum free.  The proof uses the same ideas as the proof
    Proposition~\ref{prop:enumeration-of-bool-using-zsf-semiring}, but
    specialized to the dynamic setting, with bounded evaluation.
  
    Specifically, we reduce $\DynEval{Q,\Voc}{\bools,M}$ to
    $\DynEval{Q,\Voc}{\cK,M}$ as follows. Let $\db_\bools$ be a $\bools$-database
    over $\Voc$, input to $\DynEval{Q,\Voc}{\bools,M}$.  Construct the
    $\cK$-database $\db_\cK$, input to $\DynEval{Q,\Voc}{\cK,M}$ by setting
    \begin{itemize}
        \item for every relation symbol $R\in\Voc$
        \[
            \fssem{R}{\db_\cK}\colon \seq{d}\mapsto
            \begin{cases}
                \kone_{\cK} \text{ if }R^{\db_\bools}(\seq{d})=\btrue \\
                \kzero_{\cK} \text{ otherwise. }
            \end{cases}
        \]

        \item $\db_\cK(c) := \db_\bools(c)$ for every constant symbol $c$.
    \end{itemize}
    We call $\db_\cK$ the $\cK$ version of $\db_\bools$
    Note that this takes time $\bigo(\size{\db_\bools})$. Then run
    $\DynEval{Q,\Voc}{\cK,M}$ on $\db_\cK$, which takes linear time. We have
    shown in the proof of
    Proposition~\ref{prop:enumeration-of-bool-using-zsf-semiring} that we may
    then enumerate $\AnsEnum{Q}{\db_\bools}$ with constant delay using the
    constant-delay enumeration procedure for $\AnsEnum{Q}{\db_\cK}$ provided by
    $\DynEval{Q,\Voc}{\cK,M}$.
    By assumption, $\Eval{Q,\Voc}{\cK}\in\ConstLin$ thus the set
    $\ValEnum{Q}{\db_\cK}$ can be enumerated with constant delay after a linear
    time preprocessing.

    Hence, to obtain the proposition, it suffices to show that we can translate
    any given update $u$ to $\db_\bools$ into an update $u'$ to $\db_{\cK}$ such
    that $\db_{\cK} + u'$ yields the $\cK$-version of $\db_\bools + u$.  Because
    $\DynEval{Q,\Voc}{\cK,M}\in \Maintainable$ this allows us to update the
    data structure of $\DynEval{Q,\Voc}{\bools,M}$ in $\bigo(1)$, and still be
    able to enumerate $\AnsEnum{Q}{\db_\cK+u'}$, which yields the enumeration of
    $\AnsEnum{Q}{\dbrel+u}$ with constant delay.

    We reason as follows.  Let $u$ be a single-tuple update to $\db_\bools$ such
    that all data values occurring in $u$ are bounded by $M$.
    \begin{itemize}
        \item If $u = \Insertion{R,\seq{d},\btrue}$ and $\seq{d}$ does not occur in $R^{\db_{\cK}}$ (meaning that $\seq{d}$ did not occur in $\db_\bools$ before), then set $u' = \Insertion{R, \seq{d}, \kone_\cK}$.
        \item If $u =\Insertion{R,\seq{d},\bfalse}$ then $\db+u$ is the same as $\db$, and we hence issue no update $u'$ to $\db_{\cK}$.
        \item If $u = \Deletion{R,\seq{d}}$ then the update $u' = \Deletion{R,\seq{d}}$. \qedhere
    \end{itemize}
\end{proof}

As argued, the next theorem clearly implies Theorem~\ref{theo:lower-bound-q-hierarchical}.

\begin{thm}
    \label{thm:qh-cq-semiring-lower-bound}
    Let $Q$ be a \cq over $\Voc$ without self-joins and constant-disjoint and
    $\cK$ a semiring such that the subsemiring generated by $\kzero_{\cK}$ and
    $\kone_{\cK}$ is non-trivial and zero-sum free.
    $\DynEval{Q,\Voc}{\cK,M}\in\Maintainable$ for every $M \in \posnat$, then $Q$
    is q-hierarchical, unless the $OMv$ conjecture is false.
\end{thm}
\begin{proof}
    Let $\Split(Q)=(\Qrel,\Qineq)$.  Since
    $\DynEval{Q,\Voc}{\cK,M}\in\Maintainable$ for every $M$, by
    Proposition~\ref{prop:maintenance-of-bool-using-zsf-semiring} we have that
    $\DynEval{Q,\Voc}{\bools,M}\in\Maintainable$ for every $M$, because the
    subsemiring generated by $\kzero_{\cK}$ and $\kone_{\cK}$ is non-trivial
    zero-sum free.
    Proposition~\ref{prop:ineq-maintain-implies-relational-maintain} states then
    that $\DynEval{\Qrel,\Voc}{\bools,M}\in\Maintainable$ for every $M$, since $Q$
    is constant disjoint. Because $\Qrel \in \cqNotIneq$, by
    Theorem~\ref{theo:cqNotIneq-q-hierarchical-cde}, $\Qrel$ is q-hierarchical
    unless the $OMv$ conjecture is false.  Given that $\Qrel$ is q-hierarchical if
    and only if $Q$ is q-hierarchical, the former directly implies that $Q$ is
    q-hierarchical unless the $OMv$ conjecture is false.
\end{proof}

\subsection{Lower bounds for \qhLang queries}

Similarly to the lower bound in the free-connex case, we show that Theorem~\ref{thm:qh-cq-semiring-lower-bound} transfers to $\conjLang$.

\begin{cor}\label{cor:lower-bound-qh-matrix}
    Let $\mQ$ be a \conjLang query over $\Sch$ such that $\mQ$ does not repeat matrix symbols and $\mQ$ is constant-disjoint. Let $\cK$ be a semiring such that the subsemiring generated by $\kzero_{\cK}$ and $\kone_{\cK}$ is zero-sum free. If $\DynEval{\mQ,\Sch}{\cK}$ can be evaluated dynamically with constant-time update and constant-delay, then $\mQ$ is equivalent to a \qhLang query, unless the OMv conjecture is false.
\end{cor}

Similarly as in the relational case, we rely on the fact that the following theorem clearly implies Corollary~\ref{cor:lower-bound-qh-matrix}: indeed, if $\DynEval{\mQ,\Sch}{\cK} \in \EnumClass{\size{\I}}{1}$ then trivially $\DynEval{\mQ,\Sch}{\cK,M} \in \EnumClass{\size{\I}}{1}$ for every $M$, in which case the following proposition yields the claim of Corollary~\ref{cor:lower-bound-qh-matrix}.

\begin{thm}\label{theo:lower-bound-q-hierarchical-matrix}
    Let $\mQ$ be a \conjLang query over $\Sch$ such that $\mQ$ does not repeat matrix symbols and $\mQ$ is constant-disjoint. Let $\cK$ be a semiring such that the subsemiring generated by $\kzero_{\cK}$ and $\kone_{\cK}$ is non-trivial and zero-sum free. If $\DynEval{\mQ,\Sch}{\cK,M} \in \EnumClass{\size{\I}}{1}$ for every $M \in \posnat$, then $\mQ$ is equivalent to a \qhLang query, unless the OMv conjecture is false.
\end{thm}
\begin{proof}
    Let $\Rel$ be a matrix-to-relational encoding scheme over $\Sch$ such that it is enumeration-suited for $\mQ$ (see Section~\ref{subsec:fc-lowerbound}). Let $\Mat = \Rel^{-1}$ be its inverse relational-to-matrix encoding.
  
    We apply Proposition~\ref{prop:simulating-cq-is-constant-disjoint} on $\mQ$ to obtain binary \cq $Q$ over $\Voc = \Rel(\Sch)$ that simulates $\mQ$ w.r.t. $\Rel$. We know that $Q$ is binary, self-join free and constant-disjoint.  We next modify $Q$ into a query $Q'$ over $\Voc$ such that the following properties hold:
    \begin{enumerate}[(P1)]
        \item $Q'$ continues to simulate $\mQ$ w.r.t. $\Rel$;
        \item $Q'$ continues to be constant-disjoint;
        \item $\DynEval{Q',\Voc}{\cK,M} \in \Maintainable$, for every $M \in \posnat$.
    \end{enumerate}
    This suffices to prove the proposition. Indeed, by Theorem~\ref{thm:qh-cq-semiring-lower-bound} $Q'$ is necessarily q-hierarchical unless the OMv conjecture fails. If $Q'$ is q-hierarhical, then by Corollary~\ref{cor:fcLang-equiv-free-connex-cq} there exists a \fcLang query $\mQ'$ that simulates $Q'$ w.r.t $\Mat = \Rel^{-1}$. Consequently, $\mQ$ and $\mQ'$ are then equivalent.

    Intuitively, the modification of $Q$ into $Q'$ is necessary to obtain property (P3): the fact that $\DynEval{\mQ,\Sch}{\cK,M} \in \EnumClass{\size{\I}}{1}$ only implies that we may dynamically evaluate $Q$ in $\Maintainable$ \emph{on databases that encode some matrix instance}. To apply Theorem~\ref{thm:qh-cq-semiring-lower-bound}, by contrast we need to show that we can dynamically evaluate $Q$ on \emph{arbitrary databases}. The crux will be that we obtain $Q'$ by adding extra inequalities to $Q$, such that evaluating $Q'$ on an arbitrary database yields the same result as evaluating it (and $Q$, $\mQ$) on some matrix instance.

    The definition of $Q'$ is as follows. Assume $Q$ is of the form $Q: H(\y) \gets \exists \z. \psi$ with $\psi$ quantifier free. For every relational atom $\alpha = R(\x)$ of $\psi$ we define a formula $\varphi_\alpha$ as follows. This formula is a conjunction of inequalities:
    \begin{itemize}
        \item If $\alpha = R(x_1,x_2)$ with $R$ binary and $\Mat(R)\colon (\beta,\gamma)$ then $\varphi_\alpha := x_1 \leq \Rel(\beta) \wedge x_2 \leq \Rel(\gamma)$. Note that, by definition of $\Rel$, $\Rel(\alpha) \not = 1 \not = \Rel(\beta)$.
        \item If $\alpha = R(x_1)$ with $R$ unary then by definition of $\Rel$ either $\Mat(R)\colon (\beta,1)$ or $\Mat(R)\colon (1,\beta)$ for some size symbol $\beta \not = 1$. Then take $\varphi_\alpha := x_1 \leq \Rel(\beta)$. Note that by definition of matrix-to-relational encoding schemes, $\Rel(\beta) \not =1$ since $\beta \not = 1$.
        \item If $\alpha= R()$ with $R$ nullary then by definition of $\Rel$ we have $\Mat(R)\colon (1,1)$ and we take $\varphi_\alpha$ the empty formula (which is equivalent to \btrue).
    \end{itemize}
    Then we define $Q'$ to be the CQ
    \[ Q'\colon H(\y) \gets \exists z. (\psi \wedge \bigwedge_{\alpha \text{ rel. atom in } \psi} \varphi_\alpha).\]

    Note that $Q'$ continues to simulate $\mQ$ over $\Rel$ (property P1). Indeed, for any matrix instance $\I$ over $\Sch$ we know that $\AnsEnum{\mQ}{\I} = \AnsEnum{Q}{\Rel(\I)}$. By definition, $\Rel(\I)$ is consistent with $\Rel^{-1} = \Mat$, and therefore the tuples in $\Rel(\I)$ vacuously satisfy all the extra inequalities that we have dded to $Q$ to obtain $Q'$. As such, $\AnsEnum{Q}{\Rel(\I)} = \AnsEnum{Q'}{\Rel(\I)}$. Thus, $Q'$ continues to simulate $\mQ$.

    Also note that $Q'$ is constant-disjoint (property P2). Indeed:
    \begin{itemize}
        \item We have only added covered inequalities to $Q$ to obtain $Q'$. Any such covered inequality that we have added was of the form $x \leq c$ with $c \not =1$. Therefore, since $Q$ does not have covered inequalities of the form $x \leq 1$, neither does $Q'$.

        \item We did not add any uncovered inequalities to $Q$ to obtain $Q'$. Moreover, $Q'$ has the same free variables as $Q$. Therefore, all uncovered inequalities $y \leq c$ in $Q'$ continue to be such that $y$ is not free in $Q'$, unless $c=1$. Hence for all pairs $(x \leq c, y \leq d)$ of covered inequality $x \leq c$ and non-covered inequality $y \leq d$ with $d\neq 1$ in $Q'$ we have that $y$ is not free in $Q'$ as required for constant-disjointness.
    \end{itemize}

    Finally, we show that $\DynEval{Q',\Voc}{\cK,M} \in \Maintainable$ for every $M \in \posnat$ (property P3), as claimed. Fix $M \in \posnat$ arbitrarily. We reduce $\DynEval{Q',\Voc}{\cK,M}$ to $\DynEval{\mQ,\Sch}{\cK}$ (which is an \emph{unbounded} evaluation problem) as follows. Let $\db$ be a $\cK$-database over $\Voc$, input to $\DynEval{Q',\Voc}{\cK,M}$. First, create the database $\db'$ that is equal to $\db$ except that for every relation symbol $R \in \Voc$ that is not mentioned in $Q$ we set $R^{\db'}$ to empty. This can clearly be done in linear time. Note that, if $R$ is not mentioned in $Q$, then it is also not mentioned in $Q'$, and hence of $Q'$ on $\db$ is independent of the contents of $R^\db$. Therefore, $\AnsEnum{Q'}{\db} = \AnsEnum{Q'}{\db'}$. Subsequently, compute $\db''$, the atomic reduction of $\db'$ w.r.t. $Q$. Computing $\db''$ can be done in time $\bigo(\size{\db'}) = \bigo(\size{\db})$. %
    By Claim~\ref{claim:equal-after-reduction} we have $\AnsEnum{Q'}{\db} = \AnsEnum{Q'}{\db'} = \AnsEnum{Q'}{\db''}$. Now verify the following claim: $\db''$ is consistent w.r.t. $\Mat = \Rel^{-1}$. This is because we have added, for each matrix symbol $\mA$ with $\Rel(\mA)$ occurring in $Q$, the inequalities that are required by consistency as atomic inequalities to $Q'$. (The matrix symbols $\mA$ with $\Rel(\mA)$ not occurring in $Q$ are empty in $\db'$ and hence vacuously consistent w.r.t. $\Mat$.) Since $\db''$ is consistent with $\Mat$ we have that $\Mat(\db'')$ is a matrix instance over $\Sch$ and hence a valid input to $\mQ$. Then, because $Q'$ simulates $\mQ$, we have $\AnsEnum{Q'}{\db} = \AnsEnum{Q'}{\db''} = \AnsEnum{\mQ}{\Mat(\db'')}$. Because $\DynEval{\mQ,\Sch}{\cK}$ is in $\EnumClass{\size{\I}}{1}$ it hence follows that with additional preprocessing of time $\bigo(\size{\Mat(\db'')}) = \bigo(\size{\db})$ we may enumerate $\AnsEnum{Q'}{\db''} = \AnsEnum{Q'}{\db}$ with constant delay, as desired.

    In essence, therefore, the data structure needed to enumerate $\AnsEnum{Q'}{\db}$ with constant delay is the data structure computed by $\DynEval{\mQ,\Sch}{\cK}$ when run on $\Mat(\db'')$. We next show how to maintain this data structure under updates $u$ to $\db$. 

    Specifically, we next show how we may translate in $\bigo(1)$ time an update $u$ to $\db$ into an update $u''$ to $\db''$ and an update $u_M$ to $\Mat(\db'')$ so that (1) $\db'' + u''$ yields the same database as when we would do the above procedure starting from $\db +u$ instead of $\db$: $(\db+u)'' = \db''+ u''$ and (2) applying $u_M$ to $\Mat(\db'')$ yields $\Mat(\db'' + u'')$.  Therefore, applying $u_M$ to $\DynEval{\mQ,\Sch}{\cK}$ will give us the same data structure as when it was run on $\Mat(\db'' + u'')$. From this we can again enumerate $\AnsEnum{Q'}{\db'' + u''} = \AnsEnum{Q'}{\db+u}$ with constant delay. This is because the output relation is binary since $\Rel$ is enumeration-suited for $\mQ$. We reason as follows.
    \begin{itemize}
        \item If $u = \Insertion{R, \seq{d},k}$ with $R$ not occurring in $Q'$ then we ignore $u$, there are no updates $u''$ and $u_M$ to be done.

        \item If $u = \Insertion{R, \seq{d},k}$ with $R$ occurring in $Q'$ then we check whether $\seq{d}$ is atomically consistent with $Q'[R(\x)]$ for every atom $R(\x)$ in $Q'$. This can be done in constant time, because $Q'$ (and hence the number of inequalities to check) is fixed. If it is atomically consistent then $u' = u$ and $u_M$ inserts $k$ in $\Mat(R)$ at the coordinates specified by $\seq{d}$.

        \item If $u = \Deletion{R, \seq{d},k}$ with $R$ not occurring in $Q'$ then we ignore $u$, there are no updates $u''$ and $u_M$ to be done.

        \item If $u = \Deletion{R,\seq{d},k}$ with $R$ occurring in $Q'$ then we check whether $\seq{d}$ is atomically consistent with $Q'[R(\x)]$ for every atom $R(\x)$ in $Q'$. This can be done in constant time, because $Q'$ is fixed. If it is atomically consistent then $u' = u$ and $u_M$ sets the entry in $\Mat(R)$ at the coordinates specified by $\seq{d}$ to $\kzero$. \qedhere
    \end{itemize}
\end{proof}

\section{Conclusions and future work}\label{sec:conclusions}

In this work, we isolated the subfragments of \conjLang that admit efficient evaluation in both static and dynamic scenarios. We found these algorithms by making the correspondence between \cq and \lang, extending the evaluation algorithms for free-connex and q-hierarchical \cq, and then translating these algorithms to the corresponding subfragments, namely, \fcLang and \qhLang. To the best of our knowledge, this is the first work that characterizes subfragments of linear algebra query languages that admit efficient evaluation. Moreover, this correspondence improves our understanding of its expressibility.

Regarding future work, a relevant direction is to extend \fcLang and \qhLang with disjunction, namely, matrix summation. This direction is still an open problem even for \cq with union~\cite{DBLP:journals/tods/CarmeliK21}. Another natural extension is to add point-wise functions and understand how they affect expressibility and efficient evaluation. Finally, improving the lower bounds to queries without self-join would be interesting, which is also an open problem for \cq~\cite{DBLP:journals/siglog/BerkholzGS20,DBLP:conf/pods/CarmeliS23}.
 
%% Bibliography
\bibliographystyle{alphaurl}% the mandatory bibstyle
\bibliography{biblio}

\appendix

\section{Omitted proofs from Section~\ref{sec:preliminaries}}\label{sec:app-preliminaries}

\subsection*{Proof of Proposition~\ref{prop:foplus-safeness}}

\begin{proof}
    Let $\varphi$ be a safe \foplus formula. First observe the following:  every subformula $\psi$ of $\varphi$ that is itself not an equality atom, is also safe. This is because safety is defined ``universally'' on $\varphi$. 
    Fix a database $\db$. We prove that $\fssem{\varphi}{\db}$ has finite support
    by induction on $\varphi$.
  
    \begin{itemize}
      
        \item Both $R(\x)$ and $x\leq c$  have finite support by definition.
        \item $x=y$ by itself is not safe; this case hence cannot occur.
        \item If $\varphi = \varphi_1 \wedge \varphi_2$, we discern the following cases.
        \begin{itemize}
            \item Neither $\varphi_1$ nor $\varphi_2$ are an equality atom $x=y$. Then
                both $\varphi_1$ and $\varphi_2$ are safe, and the result follows by
                induction hypothesis because an annotation computed by $\varphi$ can only
                be non-zero if is obtained by multiplying non-zero annotations originating
                from $\varphi_1$ and $\varphi_2$, which both have finite support.
            \item One of $\varphi_1$ or $\varphi_2$ is of the form $x=y$. Assume for the
                sake of presentation that $\varphi_2 = (x=y)$. Then, by definition of
                safety, at least one of $x,y$ is in $\rr(\varphi_1)$. In particular, from
                the definition of $\rr(\cdot)$ it follows that $\varphi_1$ itself cannot
                be an equality atom. Therefore, $\varphi_1$ is safe and the induction
                hypothesis applies to it.  We discern two further cases.
                \begin{itemize}
                    \item Both $x,y \in \rr(\varphi_1)$. Since
                    $\rr(\varphi_1) \subseteq \free(\varphi_1)$, it follows that
                    $\free(\varphi_1) = \free(\varphi)$. The result follows directly from
                    the induction hypothesis, since it follows that the support of
                    $\fssem{\varphi}{\db}$ is a subset of the support of
                    $\fssem{\varphi_1}{\db}$, which is finite by induction hypothesis. 
                    \item Only one of $x,y \in \rr(\varphi_1)$.  Assume w.l.o.g. that
                    $x \in \rr(\varphi_1)$ but $y \not \in \rr(\varphi_1)$. By induction
                    hypothesis, $\fssem{\varphi_1}{\db}$ has finite support. Now consider a
                    valuation $\nu\colon \free(\varphi)$ such that
                    $\fssem{\varphi}{\db}(\nu) \not = \kzero$ and define
                    $\nu_1 = \restr{\nu}{\free(\varphi_1)}$. By definition of the semantics
                    of $\wedge$ we have also $\fssem{\varphi_1}{\db}(\nu_1) \not =
                    \kzero$. Now note that $\nu_1$ completely specifies $\nu$: the only
                    variable that is potentially in the domain of $\nu$ but not in the
                    domain of $\nu_1$ is $y$, but $\varphi$ requires $x = y$. So, knowing
                    $\nu(x)$, which is given by $\nu_1(x)$, we also know $\nu(y)$. We
                    conclude that if there were an infinite number of distinct
                    $\nu\colon \free(\varphi)$ such that
                    $\fssem{\varphi}{\db}(\nu) \not = \kzero$ then there are also an
                    infinite number of $\nu_1\colon \free(\varphi_1)$ such that
                    $\fssem{\varphi}{\db}(\nu) \not = \kzero$, which cannot happen by
                    induction hypothesis.
                \end{itemize}
        \end{itemize}
        \item If $\varphi = \varphi_1\vee \varphi_2$, then neither $\varphi_1$ nor $\varphi_2$ can themselves be equality atoms. (If they were it would violate condition (1) of safety.). The result then straightforwardly follows from the fact that $\free(\varphi_1) = \free(\varphi_2) = \free(\varphi)$ and from the induction hypothesis.
        \item If $\varphi = \exists y. \varphi'$ then $\varphi'$ itself cannot be an equality atom: if it were it would violate condition (1) of safety. The result then straightforwardly follows from  the induction hypothesis. \qedhere
    \end{itemize}
\end{proof} 
\section{Omitted proofs from Section~\ref{sec:sumlang-and-fo}}\label{sec:app-sumlang-and-fo}

\subsection*{Proof of Lemma~\ref{lemma:conjlang-prenex-normal-form}}

\begin{proof}
	We prove the lemma by induction on $e$.
	\begin{itemize}
		\item If $e=\mA\colon\left(\alpha,\beta\right)$ then fix $\mx\colon\left(\alpha, 1\right)$ and $\my\colon\left(\beta,1\right)$. Clearly, $e \equiv \Sigma \mx,\my.\hspace{1ex}\left( \mx^T \cdot \mA \cdot \my \right) \times \mx\cdot\my^T$.%
		
		\item If $e=\mw\colon (\alpha,1)$ with $\mw$ a vector variable, then fix $\mx\colon\left(\alpha, 1\right)$ and $\my\colon\left(1,1\right)$. Clearly, $e \equiv \Sigma \mx, \my.\left( \mx^T \cdot \mw\right) \times \mx\cdot\my^T$.
		
		\item If $e=(e_1)^T\colon(\beta,\alpha)$ with $e_1\colon (\alpha,\beta)$. By inductive hypothesis we can assume that $e_1 \equiv \Sigma \mx,\my,\mv_1,\ldots,\mv_k. \, \ms \times \mx\cdot \my^T$. Hence, $e \equiv \Sigma \my,\mx,\mv_1,\ldots,\mv_k. \, \ms \times \my\cdot \mx^T$.
		\item If $e=e_1 \kprod e_2$ with  $e_1\colon(\alpha,\beta)$ and $e_2\colon(\alpha,\beta)$, then:
		\[e_1\kprod e_2 = \Sigma \mx,\my. \left( \left(\mx^T\cdot e_1\cdot\my\right) \times \left(\mx^T\cdot e_2\cdot\my\right) \right)\times \left(\mx\cdot\my^T\right)\] with $\mx\colon (\alpha, 1)$ and $\my\colon(\beta,1)$, and we can reduce $e_1 \kprod e_2$ to the other cases. 
		
		\item If $e=e_1\times e_2\colon (\alpha,\beta)$. By inductive hypothesis we have
		\begin{align*}
			e_1& \equiv \Sigma \mx_1,\my_1,\mv_1,\ldots,\mv_k. \hspace{1ex} \ms \times \mx_1\cdot \my_1^T \\ e_2&\equiv \Sigma \mx_2,\my_2,\mw_1,\ldots,\mw_l. \hspace{1ex} \mt \times \mx_2\cdot \my_2^T. 
		\end{align*}
		Note that $\mx_1\colon(1,1)$ and $\my_1\colon(1,1)$. Therefore,
		\[ e \equiv \Sigma
		\mx_2,\my_2,\mv_1,\ldots,\mv_k,\mw_1,\ldots,\mw_l,\mx_1,\my_1. \hspace{1ex}\left(\ms \times
		\mt \right) \times \mx_2\cdot \my_2^T.\]
		
		\item If $e=e_1\cdot e_2\colon (\alpha,\beta)$. By inductive hypothesis we have 
		\begin{align*}
			e_1&\equiv \Sigma \mx_1,\my_1,\mv_1,\ldots,\mv_k. \hspace{1ex} \ms \times \mx_1\cdot \my_1^T\\
			e_2&\equiv \Sigma \mx_2,\my_2,\mw_1,\ldots,\mw_l. \hspace{1ex} \mt \times \mx_2\cdot \my_2^T.
		\end{align*}
		Note that in this case, $\mx_1\colon(\alpha,1)$, $\my_1\colon(\gamma,1)$, $\mx_2\colon(\gamma,1)$ and $\my_2\colon(\beta,1)$. Therefore, \[e \equiv \Sigma \mx_1,\my_2,\mv_1,\ldots,\mv_k,\mw_1,\ldots,\mw_l,\my_1,\mx_2. \hspace{1ex}\left(\ms \times \mt \times \my_1^T\cdot \mx_2\right) \times \mx_1\cdot \my_2^T.\] 
		\item If $e=\Sigma \mv. e_1 \colon (\alpha,\beta)$. By inductive hypothesis we have \[ e_1 \equiv \Sigma \mx,\my,\mv_1,\ldots,\mv_k. \hspace{1ex} \ms \times \mx\cdot \my^T.\] Therefore, $e \equiv \Sigma \mx,\my,\mv_1,\ldots,\mv_k, \mv. \hspace{1ex}\ms  \times \mx\cdot \my^T$. \qedhere
	\end{itemize}
      \end{proof}

\section{Omitted proofs from Section~\ref{sec:free-connex-matlang}}\label{sec:app-free-connex-matlang}

\subsection*{Proof of Lemma~\ref{lem:fotwo-unify}}

\begin{proof}
	We first observe:
	\begin{enumerate}
		\item If $\free(\varphi) = \{x\}$ then essentially $\psi$ expresses a renaming
		of variable $x$ into variable $y$, in the sense that for all $\db$ and all
		$\nu\colon \{y\}$ we have
		$\fssem{\psi}{\db}(\nu) = \fssem{\varphi}{\db}(x\mapsto \nu(y))$. In such a
		case, it hence suffices to take $\psi' := \varphi[x \leftrightarrow y]$, the
		$\foTwo$ formula obtained from $\varphi$ by simultaneously replacing all
		occurrences (free and bound) of $x$ in $\varphi$ by $y$, and all occurrences
		of $y$ in $\varphi$ by $x$.
		\item 
		If $\free(\varphi) = \{y\}$ then observe that $\psi \equiv \varphi$. In this case it hence suffices to take $\psi' := \varphi$.
	\end{enumerate}

	It hence remains to show the result when $\free(\varphi) = \{x,y\}$.  The proof
	for this case is by well-founded induction on the length of $\foTwo$ formulas.  
	
	\begin{itemize}
		\item When $\varphi$ is an atom, it must be an atom of the form $A(\z)$ since
		$\free(\varphi) = \{x,y\}$. Then $\psi$ expresses that $x$ and $y$ must be
		bound to the same values in $A$, and $x$ dropped from the resulting
		valuation. It hence suffices to take $\psi' := \varphi[x \to y]$, which is
		the atom obtained from $A(\z)$ where all occurrences of $x$ are replaced by
		$y$.
		
		\item The case where $\varphi = \exists z. \varphi'$ cannot happen, as this
		necessarily has only one free variable.
		
		\item When $\varphi = \varphi_1 \wedge \varphi_2$ we make a further case
		analysis.
		\begin{itemize}
			\item If $\free(\varphi_1) = \{x\}$ and $\free(\varphi_2) = \{y\}$ then $\psi \equiv \exists x. (\varphi_1[x \leftrightarrow y] \wedge \varphi_2)\equiv \varphi_1[x \leftrightarrow y] \wedge \varphi_2$ where $\varphi_1[x \leftrightarrow y]$ is the formula obtained by simultaneously replacing all occurrences (free and bound) of $x$ by $y$ and all occurrences of $y$ by $x$. Hence we take $\psi' := \varphi_1[x \leftrightarrow y] \wedge \varphi_2$.
			\item If $\free(\varphi_1) = \{x,y\}$ and $\free(\varphi_2) = \{x,y\}$ then $\psi \equiv (\exists x. \varphi_1 \wedge x=y) \wedge (\exists x. \varphi_2 \wedge x=y)$. The formula $\psi'$ is then obtained by applying the induction hypothesis on $\varphi_1$ and $\varphi_2$, and taking the conjunction of the resulting formulas.
			\item If $\free(\varphi_1) = \{x,y\}$ and $\free(\varphi_2) = \{y\}$ then $\psi \equiv (\exists x. \varphi_1 \wedge x = y) \wedge \varphi_2$. The formula $\psi'$ is obtained by applying the induction hypothesis on $\varphi_1$ and taking the conjunction of the result with $\varphi_2$.
			\item If $\free(\varphi_1) = \{x,y\}$ and $\free(\varphi_2) = \{x\}$ then we observe that \[ \psi \equiv \exists x. \left( (\exists y. \varphi_1 \wedge x = y) \wedge \varphi_2  \wedge x= y\right).\] The latter formula first selects from $\varphi_1$ all valuations where $x=y$, projects the result on $x$ and then multiplies this result with $\varphi_2$. By induction hypothesis there exists $\psi'_1 \in \foTwo$  equivalent to $\exists y. (\varphi_1 \wedge x=y)$. Let $\psi'_2 := \psi'_1 \wedge \varphi$. Then $\psi \equiv \exists x. (\psi'_1 \wedge x= y)$. Note that $\psi'_1$ has only one free variable, $x$, and hence we can apply the reasoning of (1) above to obtain our desired formula $\psi$.
			\item If $\free(\varphi_1) = \{x,y\}$ and $\free(\varphi_2) = \emptyset$
			then $\psi \equiv (\exists x. \varphi_1 \wedge x= y) \wedge \varphi_2$. The formula $\psi'$ is obtained by applying the induction hypothesis on $\varphi_1$ and taking the conjunction of the result with $\varphi_2$.
			\item All other cases are symmetrical variants of those already seen. Note that the cases where $\free(\varphi_1) \cup \free(\varphi_2) \not = \{x,y\}$ cannot occur. \qedhere
		\end{itemize}
	\end{itemize}
	
\end{proof}

\subsection*{Proof of Lemma~\ref{lem:fotwoeq-equiv-equality-toplevel}}

\begin{proof}
	The proof is by induction on $\psi$. We assume w.l.o.g. that the only
	variables that occur in $\psi$ are $\{x,y\}$. The case where $\psi$ is an atom
	is immediate, and the case where $\psi$ is a conjunction follows immediately
	from the induction hypothesis. Hence, assume $\psi$ is of the form
	$\psi = \exists x. \psi_1$. (The case $\exists y. \psi_1$ is analogous.) By
	induction hypothesis, there exists $\psi'_1$ equivalent to $\psi_1$ in which
	equality atoms only occur at the top level. If no equality atom occurs in
	$\psi'_1$ then it clearly suffices to take $\psi' = \exists
	x. \psi'_1$. Hence, assume that some equality atom occurs in $\psi'_1$. We may
	assume w.l.o.g. that only the atom $x=y$ occurs; atoms of the form $x = x$ or
	$y=y$ are always valid and hence can be easily be removed. Also, because $x$
	and $y$ are the only variables ever used, there are no other
	possibilities. Then, $\psi'_1$ is of the form $\varphi \wedge x= y$ with
	$\varphi$ an $\foTwo$ formula. The result then follows by
	Lemma~\ref{lem:fotwo-unify}.
\end{proof}
 
\section{Omitted proofs from Section~\ref{sec:q-hierarchical-matlang}}\label{sec:app-q-hierarchical-matlang}

\subsection*{Proof of Lemma~\ref{lemma:foTwoH-unify}}

\begin{proof}
	The case when $\varphi\in\foTwoSimp$ is entirely analogous to the proof of Lemma~\ref{lem:fotwo-unify} with fewer cases involved in the conjunction, since only conjunction between formulas with the same free variables is allowed.
	
	Hence we focus when $\varphi=\varphi_1\wedge\cdots\wedge\varphi_k$ with $\varphi_i\in\foTwoSimp$ for every $i$. Consider the set $F = \{ \free(\varphi_i) \mid 1 \leq i \leq k\}$. Note that $F \not = \emptyset$. Furthermore, because $\phi$ is a hierarchical conjunction, $\{ \{x,y\}, \{x\}, \{y\}\} \not \subseteq F$. There are hence $2^{2^2}$-3 = 13 possibilities for $F$. We construct $e$ by case analysis on $F$. We only illustrate three important cases; the other cases can be easily derived from the ones presented next. We adopt the following notation: for each $S \in F$ the formula $\varphi_S$ denotes the conjunction of all $\varphi_i$ with $\free(\varphi_i) = S$.
	\begin{itemize}
		\item Assume $F = \{ \{ x \}, \{ y \}, \emptyset \}$. Then $\varphi \equiv \varphi_{x}\wedge\varphi_{y}\wedge\varphi_{\emptyset}$. To express $\exists x. \varphi \wedge x = y$ it suffices to take $\psi'\coloneqq \varphi_{x}[x \leftrightarrow y] \wedge \varphi_{y}\wedge\varphi_{\emptyset}$, where $\varphi_{y}[x \leftrightarrow y]$ is the formula obtained by simultaneously replacing all occurrences (free and bound) of $x$ by $y$ and all occurrences of $y$ by $x$.
		\item Assume $F = \{ \{x,y\}, \{y\}, \emptyset \}$. Then $\varphi \equiv \varphi_{x,y}\wedge\varphi_{y}\wedge\varphi_{\emptyset}$. To express $\exists x. \varphi \wedge x = y$ it suffices to take $\psi'\coloneqq (\exists x. \varphi_{x,y} \wedge x = y) \wedge \varphi_{y}\wedge\varphi_{\emptyset}$, where $\exists x. \varphi_{x,y} \wedge x = y$ can be expressed in $\foTwoH$ since $\varphi_{x,y}\in\foTwoSimp$.
		\item Assume $F = \{ \{x,y\}, \{x\}, \emptyset \}$. Then $\varphi \equiv \varphi_{x,y}\wedge\varphi_{x}\wedge\varphi_{\emptyset}$. To express $\exists x. \varphi \wedge x = y$ it suffices to take $\psi' \coloneqq \exists x. \left( (\exists y. \varphi_{x,y} \wedge x = y) \wedge \varphi_{x}\wedge\varphi_{\emptyset}\right)$, where the latter formula first selects from $\varphi_{x,y}$ all valuations where $x=y$, projects the result on $x$ and then multiplies this result with $\varphi_{x}$ (and $\varphi_{\emptyset}$). Note that here we also rely on the fact that $\exists y. \varphi_{x,y} \wedge x = y$ can be expressed in $\foTwoH$ since $\varphi_{x,y}\in\foTwoSimp$. \qedhere
	\end{itemize}
\end{proof}

\subsection*{Proof of Lemma~\ref{lem:foTwoH-force-equal}}

\begin{proof}
	For the proof, assume w.l.o.g. that $z = y$.  Let $\psi = \exists x. \varphi \wedge x = y$. Then for all $\val\colon\{z\}$ and $\db$ we have $\fssem{\psi}{\db}(\val) = \fssem{\varphi\wedge x=y}{\db}(x \mapsto \val(z), y \mapsto \val(z))$. It hence suffices to show that we can express $\psi(z)$ in $\foTwoH$. The result is then implied by Lemma~\ref{lemma:foTwoH-unify}.
\end{proof} 
\section{Related Work in Detail}

\subsection{On treewidth and finite-variable logics}\label{subsec:finite-var-logics}

Geerts and Reutter~\cite{DBLP:conf/iclr/GeertsR22} introduce a tensor logic \tl\xspace over binary relations and show that conjunctive expressions in this language that have treewidth $k$ can be expressed in $\tl_{k+1}$, the $k$-variable fragment of $\tl$. In essence, $\tl$ is a form of $\foplus$ evaluated over $K$-relations.

Geerts and Reutter~\cite{DBLP:conf/iclr/GeertsR22} do take free variables into account when defining treewidth. Specifically, a formula with $\ell$ free variables has treewidth at least $\ell$. Therefore, the treewidth of the following free-connex conjunctive query,
\[ \varphi(x,y) = \exists z,u,v. R(x,y), R(x,z), R(z,u), R(z,v) \]
has treewidth at least $2$. For this query, therefore, the result by Geerts and Reutter only implies expressibility in $\focon_3$, not $\focon_2$ as we show in this paper.

\end{document}